\documentclass[10pt,prd,preprintnumbers,floatfix,aps,nofootinbib,showpacs]{revtex4} 
\usepackage{epsfig}
\usepackage{url}
\usepackage{hyperref}
\usepackage[normalem]{ulem} 
\usepackage{latexsym}
\usepackage{epsfig}
\usepackage{amsmath}
\usepackage{amssymb}
\usepackage{graphicx}
\usepackage{verbatim}
\usepackage{enumerate,mdwlist}%lists
\usepackage[titletoc]{appendix}
\usepackage{amsfonts}
\usepackage{tikz} %diagrams

\usepackage[all]{xy} %array diagrams
\usepackage{amsfonts}
\usepackage{multirow}%tables
\allowdisplaybreaks[1]%eqs break accross pages, larger numbers -> easier breaks
\newcommand{\lp}{\left(}
\newcommand{\rp}{\right)}

\newcommand{\ba}{\begin{eqnarray}}
\newcommand{\ea}{\end{eqnarray}}
\newcommand{\be}{\begin{equation}}
\newcommand{\ee}{\end{equation}}

\newcommand{\D}{{\mathcal{D}}}%derivative

\newcommand{\R}{\mathcal{R}}
\newcommand{\Lag}{\mathcal{L}}
\newcommand{\A}{\mathcal{A}}
\newcommand{\F}{\mathcal{F}}

\newcommand{\T}[1]{\tilde{#1}}%Tilde
\newcommand{\B}[1]{\bar{#1}}%Bar
\newcommand{\J}[1]{\hat{#1}}%Hat
\newcommand{\I}[1]{\check{#1}}%Check
\newcommand{\pP}{\langle\Phi\rangle}
\newcommand{\pPP}{\langle\Phi^{2}\rangle}
\newcommand{\pPPP}{\langle\Phi^{3}\rangle}
\newcommand{\tP}{[\Phi]}
\newcommand{\tPP}{[\Phi^{2}]}
\newcommand{\tPPP}{[\Phi^{3}]}
\newcommand{\vO}{\varOmega} %old \xi_{0}
\newcommand{\vU}{\varUpsilon} %old \xi_{3}
\newcommand{\vOP}{\varOmega_{\phi}} %old \xi_{1}
\newcommand{\vUP}{\varUpsilon_{\phi}}
\newcommand{\vOX}{\varOmega_{X}} %old \xi_{2}
\newcommand{\vUX}{\varUpsilon_{X}} %old \xi_{4}
\newcommand{\gC}{\gamma_{\T{C}}} %old \gamma_{2}
\newcommand{\gD}{\gamma_{\T{D}}} %old \gamma_{1}
\newcommand{\gX}{\gamma_{\T{X}}} %old \gamma_{3}
\newcommand{\vOp}{\varOmega_{,\phi}} 
\newcommand{\vUp}{\varUpsilon_{,\phi}} 
\newcommand{\vOPp}{\varOmega_{\phi,\phi}}

\newcommand{\vOx}{\varOmega_{,X}} 
\newcommand{\vUx}{\varUpsilon_{,X}} 
\newcommand{\vOPx}{\varOmega_{\phi,X}}

\newcommand{\lPh}{\lambda_{\Phi}}
\newcommand{\lT}{\lambda_{\theta}}
\newcommand{\lPs}{\lambda_{\Psi}}
\newcommand{\lBPh}{\lambda_{\B{\Phi}}}
\newcommand{\lBPs}{\lambda_{\B{\Psi}}}
\newcommand{\BPh}{\B{\Phi}}
\newcommand{\BPs}{\B{\Psi}}
\newcommand{\HPs}{\J{\Psi}}
\newcommand{\IPs}{\I{\Psi}}
\definecolor{grey}{rgb}{0.4,0.4,0.4}
\definecolor{dullmagenta}{rgb}{0.4,0,0.4}
\definecolor{darkblue}{rgb}{0,0,0.4}
\definecolor{midblue}{rgb}{0,0,0.5}
\definecolor{midred}{rgb}{0.5,0,0}
\definecolor{orange}{rgb}{1,0.5,0}
\definecolor{lightbrown}{rgb}{0.75,0.5,0.25}
\definecolor{tan}{cmyk}{0.14,0.42,0.56,0}
\definecolor{djunglegreen}{cmyk}{0.99,0,0.52,0}
\definecolor{lightgreen}{rgb}{0,1,0}
\definecolor{olivegreen}{cmyk}{0.64,0,0.95,0.40}
\definecolor{midgreen}{rgb}{0.0,0.675,0.0}
\definecolor{darkgreen}{rgb}{0,0.5,0}
\begin{document} 
%TITLE
\title{Field redefinitions in theories beyond Einstein gravity \\ using the language of differential forms}
%AFFILIATIONS
\author{Jose Mar\'ia Ezquiaga}
\email{jose.ezquiaga@uam.es}
\affiliation{Instituto de F\'isica Te\'orica UAM/CSIC, Universidad Aut\'onoma de Madrid, \\ C/ Nicol\'as Cabrera 13-15, Cantoblanco, Madrid 28049, Spain}
\author{Juan Garc\'ia-Bellido}
\email{juan.garciabellido@uam.es}
\affiliation{Instituto de F\'isica Te\'orica UAM/CSIC, Universidad Aut\'onoma de Madrid, \\ C/ Nicol\'as Cabrera 13-15, Cantoblanco, Madrid 28049, Spain}
\author{Miguel Zumalac\'arregui}
\email{miguelzuma@berkeley.edu}
\affiliation{Nordita, KTH Royal Institute of Technology and Stockholm University, \\
Roslagstullsbacken 23, SE-106 91 Stockholm, Sweden}
\affiliation{Berkeley Center for Cosmological Physics and University of California at Berkeley, \\
Berkeley, California 94720, USA}
%-----
%ABSTRACT
\begin{abstract}
We study the role of field redefinitions in general scalar-tensor theories. In particular, we first focus on the class of field redefinitions linear in the spin-2 field and involving derivatives of the spin-0 mode, generically known as disformal transformations. We start by defining the action of a disformal transformation in the tangent space. Then, we take advantage of the great economy of means of the language of differential forms to compute the full transformation of Horndeski's theory under general disformal transformations. We obtain that Horndeski's action maps onto itself modulo a reduced set of non-Horndeski Lagrangians. These new Lagrangians are found to be invariant under disformal transformation that depend only in the first derivatives of the scalar. Moreover, these combinations of Lagrangians precisely appear when expressing in our basis the constraints of the recently proposed Extended Scalar-Tensor (EST) theories. These results allow us to classify the different orbits of scalar-tensor theories invariant under particular disformal transformations, namely the special disformal, kinetic disformal and disformal Horndeski orbits. In addition, we consider generalizations of this framework. We find that there are possible well-defined extended disformal transformations that have not been considered in the literature. However, they generically cannot link Horndeski theory with EST theories. Finally, we study further generalizations in which extra fields with different spin are included. These field redefinitions can be used to connect different gravity theories such as multi-scalar-tensor theories, generalized Proca theories and bi-gravity. We discuss how the  formalism of differential forms could be useful for future developments in these lines.
\end{abstract}
%----
\date{\today}


\maketitle

%-------
%SEC. I: INTRODUCTION
%-------
\section{Introduction}
\label{sec:Introduction}

General Relativity (GR) is a rather unique theory. On the observational side, it provides an excellent description of gravitational phenomena ranging from laboratory up to Solar System scales \cite{Will:2014kxa}, being also the base of our concordance cosmological model \cite{Ade:2015xua}. On the theoretical side, it is the only theory for a spin-2 field in four dimensions with covariant second order equations of motion (EoM) \cite{Lovelock:1971yv}. However, at the same time, this simplicity has the drawback that very different approaches become equivalent in the GR limit, yielding to a plethora of alternative theories of gravity \cite{Clifton:2011jh,Joyce:2014kja}. In this sense, it is crucial to establish which of these theories are theoretically consistent and distinct. This is a basic requirement in order to test these models against the new data from local experiments \cite{Will:2014kxa}, astrophysics \cite{Berti:2015itd} and cosmology \cite{Koyama:2015vza}.

When considering alternatives to Einstein's theory of gravity, some of its fundamental principles need to be violated. This can be effectively parametrized by the inclusion of extra physical degrees of freedom (DoF). In this perspective, the simplest modification of GR consist of adding a single scalar DoF. If one then tries to systematically build the most general scalar-tensor (ST) interactions, one soon faces with Ostrogradski's theorem \cite{ostrogradski1850member}, which limits the possible derivative interactions without instabilities in the Hamiltonian (see \cite{Woodard:2015zca} for a recent review). 

Along the years, this fatal Ostrogradski's instability has been surpassed in several manners that can be classified in three generations of ST theories. The first one is characterized by Jordan-Brans-Dicke (JBD) theories \cite{Brans:1961sx} in which there are not derivative interactions of the scalar. The second one allows for derivative interactions at the price of having very complex Lagrangians that prevent the EoM to be higher than second order. This corresponds to Horndeski theory \cite{Horndeski:1974wa}. The third generation, in active research nowadays, is composed of theories with higher than second order EoM but with additional constraints killing the instabilities \cite{Motohashi:2016ftl,Klein:2016aiq}. This generation was initiated with the so-called beyond-Horndeski theories \cite{Zumalacarregui:2013pma,Gleyzes:2014dya} of which completions encompassing more general degenerate theories, named Extended Scalar-Tensor, have been found \cite{Langlois:2015cwa,Crisostomi:2016czh,BenAchour:2016fzp}. Recently, a new formulation for ST theories was presented based on the language of differential forms \cite{Ezquiaga:2016nqo}. This new approach allows for a natural description of these complicated, general ST Lagrangians. 

All these theories are described with classical field theory. Within this framework, a basic tool to disentangle the physical DoF is to use field redefinitions \cite{deRham:2016wji}. If a field redefinition is well-defined, meaning non-singular and invertible, then, two theories related by this transformation contain the same number of DoF. This simple statement can be very useful to determine if a novel proposal is theoretically viable and different from previous theories. In fact, this type of arguments has been extensively applied to the three generations of ST theories, always trying to find a convenient reformulation (or frame) simplifying the analysis of the dynamics of the theory.

In the first generation of ST theories, non-minimal couplings of the scalar with the metric can arise. However, since there are not derivative interactions, only couplings of functions of the scalar field $f(\phi)$ to the Ricci scalar are allowed. This is nothing but a \emph{conformal} rescaling of the curvature, where the conformal factor is parametrized by the scalar field $\phi(x)$. Consequently, the non-minimal coupling can be eliminated choosing an appropriate parametrization and the difficulties of the analysis are moved entirely to the scalar and matter sectors, see e.g. \cite{Dicke:1961gz,Wetterich:1987fm}. For this reason, conformal transformations play a central role in this class of theories.

In the second generation, there are derivative couplings to the metric. This introduces derivative interactions with both the Ricci scalar and tensor, as well as intricate functions of second derivatives of the scalar. Therefore, it is necessary a field redefinition more complex than a conformal one to decouple both sectors. This can be achieved if the redefined metric is \emph{disformal} \cite{Bekenstein:1992pj}, i.e. if it does not share the same causal structure of the original metric. In other words, if the new metric contains a tensorial part constructed with derivatives of the scalar. For the case of Horndeski theory, the relevant transformations are those including first derivatives of the scalar in the tensorial part alone \cite{Bettoni:2013diz}. This is because the form of the action remains invariant and just the four free functions of the theory change. In some cases, one can use these transformations to erase the non-minimal couplings \cite{Bettoni:2015wta}. 

Nevertheless, when one considers more general transformation with the coefficients of the conformal and disformal terms depending also on first derivatives of the scalar, one encounters well-defined theories outside of Horndeski's domain \cite{Zumalacarregui:2013pma}, thus inaugurating the third generation of ST theories. Interestingly, there are other theories in this third generation that cannot be disformally related with the second one \cite{Achour:2016rkg}. It is still an open question if they can be generated through a more general redefinition, i.e. through an \emph{extended} disformal transformation \cite{Zumalacarregui:2013pma}. In this work, we will study this possibility. 

In this context, it is important to emphasize that two gravitational theories related by a field redefinition are only directly equivalent in the absence of matter. When matter is present, the metric to which this sector is minimally coupled should also be transformed for the two theories to be equivalent \cite{Casas:1991ky,Domenech:2016yxd,Domenech:2015hka}. In cosmology, the former situation is a good approximation in the early Universe when studying inflation, while the latter is the typical case in the late Universe when considering dark energy models. This is the explanation of why investigating new couplings to matter can be equivalent to survey novel gravitational theories.

In this work, we are going to study field redefinition in ST theories using the formalism of differential forms \cite{Ezquiaga:2016nqo}. This novel approach will allow us to present a global picture of the current status of the problem and extend its scope. This study will also serve as an excellent example of the great economy of means in the calculations and the analysis of the results that exhibit this new formalism. We will begin by presenting how to define disformal transformations in the language of differential forms in Sec. \ref{sec:Disformal}. The great advantage is that once the transformation of every building block of the theory is found, the process of transforming a given Lagrangian becomes very clear and systematic. This knowledge will permit us to compute in a transparent manner the disformal transformation of different theories. For instance, in Sec. \ref{sec:DisformalTheories}, we compute the most general disformal transformation of the full Horndeski theory. We also show explicitly how the Gauss-Bonnet Lagrangian disformally transforms onto itself plus total derivatives. Subsequently, in Sec. \ref{sec:Orbits}, we classify the different orbits of ST theories invariant under different classes of disformal transformations. Afterwards, we study generalizations of disformal transformation. We start by considering field redefinitions with higher derivatives of the scalar in Sec. \ref{sec:ExtendedDisformal}, discussing its relation with the Extended Scalar-Tensor theories. Then, in Sec. \ref{sec:ConnectingMG}, we investigate generalizations in which other fields with different spin are included. This will allow us to connect different gravity theories such as multi-scalar-tensor theories \cite{Damour:1992we}, generalized Proca theory \cite{Heisenberg:2014rta} or bi-gravity \cite{Hassan:2011zd}. Lastly, we conclude in Sec. \ref{sec:Discussion} with a discussion of the main results and future prospects of this work.

%-------
%SEC. II: DISFORMAL TRANSFORMATIONS
%-------
\section{Disformal Transformations in the Tangent Space}
\label{sec:Disformal}

Gravity can be easily formulated in the tangent space. In this context, the usual diffeomorphism (Diff) invariance becomes an invariance under local Lorentz Transformations (LLT), resembling in a clear manner the similarities with gauge theories. Moreover, the geometry of the manifold, which was previously determined by the space-time components of the metric tensor $g_{\mu\nu}$, is now contained in the basis elements of the cotangent space $\theta^{a}$. The two objects are directly connected through the definition of the metric tensor $\mathfrak{g}=g_{\mu\nu}dx^{\mu}\otimes dx^{\nu}=\eta_{ab}\theta^{a}\otimes\theta^{b}$, where $\eta_{ab}$ is the Minkowski metric, given by $\eta_{ab}=\mathrm{diag}(-1,1,1,1)$. Subsequently, if one wants to transform the metric while maintaining its causal structure, we simply need to rescale the basis elements $\theta^{a}$ by $\T{\theta}^{a}=\T{C}(x)\theta^{a}$, which is nothing but a \emph{conformal transformation}. In the case of JBD theories \cite{Brans:1961sx}, one could choose the conformal factor to depend on the scalar field $\phi$ in order to eliminate the non-minimal coupling with the Ricci scalar. This transformation, in the usual component notation, takes the well-known form $\T{g}_{\mu\nu}=\T{C}^{2}g_{\mu\nu}$.

However, when there are derivative couplings of the scalar field to the curvature, such as in Horndeski's theory \cite{Horndeski:1974wa}, conformal transformations are not enough to erase the non-minimal couplings. Nonetheless, one could take advantage of the results of Ref. \cite{Ezquiaga:2016nqo}, where it was shown that scalar-tensor theories can be naturally built using the language of differential forms. Apart from the curvature 2-form describing the geometry, one just needs two 1-forms, $\Psi^{a}\equiv\nabla^{a}\phi\nabla_{b}\phi\theta^{b}$ and $\Phi^{a}\equiv\nabla^{a}\nabla_{b}\phi\theta^{b}$, encoding respectively first and second derivatives of the scalar field. Thus, one could perform a redefinition of the vielbein $\theta^{a}$ that includes first derivatives of the scalar field by applying
\begin{equation}
\T{\theta}^{a}=\T{C}(\phi,X)\theta^{a}+\T{D}(\phi,X)\Psi^{a},
\label{eq:disfT}
\end{equation}
where $X$ is the scalar kinetic term $-2X=\nabla_{\mu}\phi\nabla^{\mu}\phi$. This kind of field redefinition is known as a \emph{disformal transformation} \cite{Bekenstein:1992pj}. In order for this transformation to be well behaved, it must have a non-vanishing determinant, so that it can be inverted. In the language of differential forms, the determinant of the transformation can be very easily computed via the volume element, which encodes it naturally with a square root. Recalling that it is given by $\eta\equiv\frac{1}{D!}\epsilon_{a_{1}\cdots a_{D}}\theta^{a_{1}}\wedge\cdots\wedge\theta^{a_{D}}=\sqrt{-g}dx^{D}$, we obtain that it transforms as
\begin{equation}
\T{\eta}=\frac{1}{D!}\epsilon_{a_{1}\cdots a_{D}}\T{\theta}^{a_{1}}\wedge\cdots\wedge\T{\theta}^{a_{D}}=\T{C}^{D-1}\left(\T{C}-2X\T{D}\right)\eta,
\label{eq:disfDet}
\end{equation}
where $\epsilon_{a_{1}\cdots a_{D}}$ is the totally antisymmetric symbol\footnote{This relation can be trivially obtained using that $\Psi^{a}\wedge\Psi^{b}=0$, which is a consequence of the antisymmetry of the exterior product. For more details, one can see Ref. \cite{Ezquiaga:2016nqo} where all this notation is explained in detail.}. Therefore, to prevent the disformal volume element to become a complex number, the disformal coefficients must satisfy that $\T{C}>0$ and $\T{C}>2X\T{D}$, where the transformed determinant arises from $\T{\eta}=\sqrt{-\T{g}}\eta$. Nevertheless, this mathematical consistency requirement can be relaxed because it has been pointed out in the literature that  cosmological solutions may maintain the second condition only dynamically \cite{Zumalacarregui:2010wj,Koivisto:2012za,Zumalacarregui:2012us} (see also \cite{Sakstein:2014aca,Sakstein:2015jca}). 

These disformal transformations in the tangent space can be traced back to the conventional component notation. Starting from the definition of the disformal metric, we find that
\begin{equation}
\T{\mathfrak{g}}=\eta_{ab}\T{\theta}^{a}\otimes\T{\theta}^{b}=\T{g}_{\mu\nu}dx^{\mu}\otimes dx^{\nu}=\lp \T{C}^{2}g_{\mu\nu}+2\T{D}\lp\T{C}-X\T{D}\rp\nabla_{\mu}\phi\nabla_{\nu}\phi\rp dx^{\mu}\otimes dx^{\nu}.
\label{eq:disfMetric}
\end{equation}
Subsequently, one can recover the original formulation of Bekenstein \cite{Bekenstein:1992pj} by appropriately redefining the disformal coefficients, cf. Eq. (\ref{eq:cJ}) and (\ref{eq:dJ}) for the precise expression. Moreover, the above conditions on the disformal coefficients $\T{C}$ and $\T{D}$ can be directly related with the ones discussed first in Ref. \cite{Bekenstein:1992pj} and more recently in \cite{Bettoni:2013diz}.

The benefits of applying disformal transformations in the tangent space are considerable. First of all, it naturally introduces the 1-form $\Psi^{a}$, which was one of the basic building blocks used to construct scalar-tensor theories in differential forms language \cite{Ezquiaga:2016nqo}. Thus, it connects in an interesting and fundamental manner this new formulation for ST theories with disformal transformations. Secondly, the building blocks of the ST theories will transform in a very transparent way. From the transformed frame field $\tilde{\theta}^{a}$, the rest of geometrical quantities, the connection 1-form $\tilde{\omega}^{ab}$ and the 2-form curvature $\tilde{\R}^{ab}$, can be constructed. For the scalar field building blocks, as it will be shown in the next subsections, the transformation also follows directly. Therefore, this method provides us with a way to compute the disformal building blocks $\T{\R}^{ab}$, $\T{\Psi}^{a}$ and $\T{\Phi}^{a}$ in terms of the original ones $\Psi^{a}$, $\Phi^{a}$ and $\R^{ab}$. This fact will simplify enormously the computations because any disformally transformed Lagrangian could be expressed as a linear combination of Lagrangians with the same building blocks, which we already know from \cite{Ezquiaga:2016nqo}. Moreover, it will become straightforward to elucidate how each new Lagrangian is generated through the specific dependence of the disformal coefficients on $\phi$ and $X$.

In the following, we are going to present explicitly how general ST theories transform under disformal transformations (\ref{eq:disfT}). Since we have defined these field redefinitions in the tangent space, we will be working with the  formalism of differential forms for ST gravity \cite{Ezquiaga:2016nqo}. In this language, a general basis of Lagrangians reads
\begin{equation}
\Lag_{(lmn)}=\bigwedge_{i=1}^{l}\mathcal{R}^{a_{i}b_{i}}\wedge\bigwedge_{j=1}^{m}\Phi^{c_{j}}\wedge\bigwedge_{k=1}^{n}\Psi^{d_{k}}\wedge\theta^{\star}_{~a_{1}b_{1}\cdots a_{l}b_{l}c_{1}\cdots c_{m}d_{1}\cdots d_{n}}\,,
\label{eq:L}
\end{equation}
where $\bigwedge$ is an abbreviation for a set of consecutive wedge products, $\theta^{\star}_{a_{1}\cdots a_{k}}$ is the Hodge dual basis defined by $\theta^{\star}_{~a_{1}\cdots a_{k}}=\frac{1}{(D-k)!}\epsilon_{a_{1}\cdots a_{k}a_{k+1}\cdots a_{D}}\theta^{a_{k+1}}\wedge\cdots\wedge\theta^{a_{D}}$, and the subindices satisfy $l,m,n\in\mathbb{N}$ and $2l+m+n\leq D$. Our objective will be to determine if disformal transformations map the general basis $\Lag_{(lmn)}$ to itself or not, and if not what is the structure behind the new terms. This is a major question in trying to understand Horndeski's theory \cite{Horndeski:1974wa} and its generalizations \cite{Zumalacarregui:2013pma,Gleyzes:2014dya,Langlois:2015cwa,Crisostomi:2016czh,Ezquiaga:2016nqo,BenAchour:2016fzp}. As we will see, the kinetic dependence of the disformal coefficients $\T{C}$ and $\T{D}$ will introduce new building blocks, which we refer as the \emph{extended basis}. Thus, let us now study the effect of a disformal transformation in the basis elements.

%-------
%DISFORMAL BUILDING BLOCKS
%-------
\subsection{Disformal Building Blocks}
\label{subsec:DisformalBasis}

First, we are going to investigate how disformal transformations modify the building blocks of the theory, i.e. $\theta^{\star}_{~a_{1}\cdots a_{k}}$, $\omega^{ab}$, $\Psi^{a}$, $\Phi^{a}$ and $\R^{ab}$. We will focus on the origin of each terms, paying special attention to the terms that extend the basis. Generically, the new terms will correspond to higher order contractions of second derivatives with gradient fields. Subsequently, we proceed to analyze each building block one by one.
%--
\begin{enumerate}[(i)] 
%-Hodge Dual- 
\item \emph{Disformal Hodge Dual Basis:}

We have already defined how $\theta^{a}$ transforms in (\ref{eq:disfT}). Consequently, we only need to apply this transformation to the definition of the Hodge dual basis given before, after Eq. (\ref{eq:L}). We obtain
\begin{equation}
\T{\theta}^{\star}_{a_{1}\cdots a_{k}}=\T{C}^{D-k-1}\lp\T{C}\theta^{\star}_{a_{1}\cdots a_{k}}+\T{D}\Psi^{b}\wedge\theta^{\star}_{a_{1}\cdots a_{k}b}\rp\,.
\label{eq:disfHodge}
\end{equation}
One should notice that this expression is a generalization of the volume element introduced in (\ref{eq:disfDet}), which can be denoted also as $\T{\theta}^{\star}=\T{\eta}$. In fact, we recover this result for $k=0$ recalling that $\Psi^{a}\wedge\theta^{\star}_{~a}=-2X\theta^{\star}$.
%-Connection-
\item \emph{Disformal connection 1-form:}

The next step is to compute how the connection 1-form transforms, because it encodes how covariant derivatives act. To do so, we impose the torsionless and metricity conditions on the disformal connection, i.e. $\tilde{T}^{a}=0$ and $\tilde{\omega}_{ab}=-\tilde{\omega}_{ba}$. We postulate that the disformal connection takes the form $\tilde{\omega}^{a}_{~b}=\omega^{a}_{~b}+X^{a}_{~b}$, where $X^{a}_{~b}$ must satisfy $X_{ab}=-X_{ba}$. Then, it can be determined from
\begin{equation}
\begin{split}
\tilde{T}^{a}=& \tilde{\D}\tilde{\theta}^{a}=d\tilde{\theta}^{a}+\tilde{\omega}^{a}_{~b}\wedge\tilde{\theta}^{b}=\D\T{\theta}^{a}+X^{a}_{\  b}\wedge\T{\theta}^{b} \\
=&\tilde{C}\D\theta^{a}+\tilde{D}\wedge \D\Psi^{a}+ \D(\tilde{C})\wedge\theta^{a}+\D(\tilde{D})\wedge\Psi^{a}+X^{a}_{~b}\wedge\tilde{C}\theta^{b}+X^{a}_{~b}\wedge\tilde{D}\Psi^{b} \\
=&\tilde{C}T^{a}+\tilde{D}\Phi^{a}\wedge \D\phi+\tilde{C}_{,\phi}\D\phi\wedge\theta^{a}-\tilde{C}_{,X}\nabla_{b}\phi\Phi^{b}\wedge\theta^{a}-\tilde{D}_{,X}\nabla_{b}\phi\Phi^{b}\wedge\Psi^{a}+X^{a}_{~b}\wedge\tilde{C}\theta^{b}+X^{a}_{~b}\wedge\tilde{D}\Psi^{b}\,,
\end{split}
\end{equation}
where $\D$ represents an exterior, covariant derivative \cite{Ezquiaga:2016nqo}. Interestingly, the scalar dependence of the disformal coefficient $\tilde{D}_{,\phi}$ does not affect the connection since $d\phi\wedge\Psi^{a}=0$ by antisymmetry. Consequently, after implementing the torsionless condition, the 1-form connection follows
\begin{equation}
\tilde{\omega}^{ab}=\omega^{ab}-(\vOP+\frac{1}{2}\pP\vOX\vU)\cdot\theta^{[a}\nabla^{b]}\phi+\vOX\cdot\theta^{[a}\nabla^{b]}\nabla_{c}\phi\nabla^{c}\phi+\vU\cdot\Phi^{[a}\nabla^{b]}\phi+\vUX\cdot\Psi^{[a}\nabla^{b]}\nabla_{c}\phi\nabla^{c}\phi\,.
\label{eq:disfConnection}
\end{equation}
Here, we have introduced the coefficients $\vO_{i}$ and $\vU_{i}$, which encodes the field dependence of the disformal coefficients $\T{C}$ and $\T{D}$ respectively. The subindex $_{i}$ indicates if the coefficient is generated through the derivative of $\phi$ or $X$. When there is no subindex, it means that $\vO$ or $\vU$ are sourced directly by the disformal coefficients. The complete set is given by
\begin{align}
\vO&=\gD\T{C}\,, \label{eq:x0} \\
\vOP&=2\gD\T{C}_{,\phi}\,, \label{eq:x1} \\
\vOX&=2\gC\T{C}_{,X}\,, \label{eq:x2} \\
\vU&=2\gD\T{D}\,, \label{eq:x3} \\
\vUP&=2\gD\T{D}_{,\phi}\,, \label{eq:x3phi} \\
\vUX&=\gC(2\T{D}_{,X}-\vU\T{D})\,, \label{eq:x4}
\end{align}
where we have defined the fractions $\gC\equiv\T{C}^{-1}$ and $\gD\equiv(\T{C}-2X\T{D})^{-1}$. For later convenience, we introduce an extra fraction $\gX$ given by $\gX\equiv(\vO-X(\vO_{X}-2X\vU_{X}))^{-1}$ that appears when computing the disformal transformation of the scalar kinetic term $\T{X}$. As it was mentioned, the disformal connection is not affected by the $\phi$ dependence of $\T{D}$. This becomes explicit by the absence of $\vUP$ in (\ref{eq:disfConnection}). Also, the torsionless condition makes the transformed connection insensitive to the conformal factor alone. Thus, $\vO$ does not appear either in (\ref{eq:disfConnection}).

Noticeably, all coefficients are functions of the scalar field and its first derivatives, i.e. $\phi$ and $X$ only. However, in the second term of the disformal 1-form connection (\ref{eq:disfConnection}), there is a $\pP$ factor that already hints that there will be new terms beyond the basis $\Lag_{(lmn)}$. This factor contains second derivatives of the scalar field contracted with first ones $\pP\equiv\nabla^{a}\phi\nabla_{a}\nabla_{b}\phi\nabla^{b}\phi$. We will study such new terms in the following Sec. \ref{subsec:ExtendedBasis}. Remarkably, the dependence of $\T{C}$ and $\T{D}$ in $\phi$ and $X$ is described separately by each coefficient $\vO_{i}$ and $\vU_{i}$. This fact will simplify the analysis of disformal theories because it will become clear at first sight which new Lagrangian arises from which part of the disformal transformation. In particular, it will be transparent that all higher order terms will be generated through the kinetic dependence of the coefficients $\T{C}$ and $\T{D}$, i.e. through $\vOX$ and $\vUX$. With this result, we can obtain the disformal transformation of the rest of the building blocks.

%-Scalar 1-forms- 
\item \emph{Disformal Scalar 1-forms:}

Using the previous results, we can easily see how the 1-forms describing the first and second derivatives of the scalar field transform. It is just necessary to recall that they are defined by $\T{\Psi}^{a}=\T{\nabla}^{a}\phi\T{\D}\phi$ and $\T{\Phi}^{a}=\T{\D}(\T{\nabla}^{a}\phi)$. We find that they correspond to
\begin{align}
\tilde{\Psi}^{a}&=\gD\Psi^{a}\,, \label{eq:DisformalPsi} \\
\tilde{\Phi}^{a}&=\lPh[\vO]\Phi^{a}+\lT[\vOP,\vOX]\theta^{a}+\lPs[\vOP,\vOX,\vUP,\vUX]\Psi^{a}+\lBPh[\vOX,\vU,\vUX]\BPh^{a}+\lBPs[\vOX,\vUX]\BPs^{a}\,, \label{eq:DisfPhi} 
\end{align}
where we have defined for shortness $\BPh^{a}=\nabla^{a}\phi\nabla_{z}\phi\Phi^{z}$ and $\BPs=\nabla^{a}\nabla_{z}\phi\Psi^{z}$. This notation can be generalized to an arbitrary number of contractions, as it will be discussed in Sec. \ref{subsec:ExtendedBasis} and \ref{subsec:FullExtendedBasis}, and summarized in Tab. \ref{tab:notation}. Noticeably, the coefficients $\lambda_{i}$ in front of each term, named with a subindex referring to it, can be expressed in terms of $\vO_{i}$ and $\vU_{i}$. To make this explicit, we have include in brackets the main dependence of each coefficient. As before, it will appear a $\pP$ term triggered by the kinetic dependence of the disformal coefficients. Such a term will only arise in $\lT$ and $\lPs$ so that the whole expression is of the same order of second derivatives for all the pieces in the transformed 1-form (\ref{eq:DisfPhi}). The specific definitions of these $\lambda_{i}$ coefficients can be found in Eqs. (\ref{eq:l0}-\ref{eq:l4}) of Appendix \ref{app:Computations}.

%-Curvature- 
\item \emph{Disformal Curvature 2-form:}

Finally, the disformal 2-form curvature can be computed through its definition
\begin{equation}
\begin{split} 
\tilde{\R}^{ab}=&\tilde{\D}\tilde{\omega}^{ab}=d\omega^{ab}+\omega^{a}_{~c}\wedge\omega^{cb}+dX^{ab}+\omega^{a}_{~c}\wedge X^{cb}+X^{a}_{~c}\wedge\omega^{cb}+X^{a}_{~c}\wedge X^{cb} \\
=&\R^{ab}+\D X^{ab}+X^{a}_{~c}\wedge X^{cb}
\end{split} \label{eq:TransfCurvature}
\end{equation}
once $X^{ab}=\T{\omega}^{ab}-\omega^{ab}$ is known. Remarkably, this simple expression is valid for any field redefinition of the vielbein. In tensorial notation, the analog of this expression corresponds to Eq. (36) of Ref. \cite{Zumalacarregui:2013pma}. Particularizing for disformal transformations, we obtain
\begin{equation}
\begin{split}
\tilde{\R}^{ab}=&\R^{ab}+\alpha_{\B{\R}}[\vU]\nabla_{z}\phi\R^{z[b}\nabla^{a]}\phi+\alpha_{\Phi\Phi}[\vU]\Phi^{a}\wedge\Phi^{b}+\alpha_{\BPh\Phi}[\vU,\vUX]\BPh^{[a}\wedge\Phi^{b]}+\alpha_{\BPs\Phi}[\vOX,\vUX]\BPs^{[a}\wedge\Phi^{b]} \\
+&\alpha_{\Phi\Psi}[\vOP,\vOX,\vUP,\vUX]\Phi^{[a}\wedge\Psi^{b]}+\alpha_{\Phi\theta}[\vOP,\vOX]\Phi^{[a}\wedge\theta^{b]}+\alpha_{\BPh\theta}[\vOP,\vOX]\BPh^{[a}\wedge\theta^{b]} \\
+&\alpha_{\BPs\theta}[\vOX,\vUX]\BPs^{[a}\wedge\theta^{b]}+\alpha_{\Psi\theta}[\vOP,\vOX,\vUX]\Psi^{[a}\wedge\theta^{b]}+\alpha_{\theta\theta}[\vOP,\vOX,\vU]\theta^{a}\wedge\theta^{b} \\
+&\alpha_{\HPs\theta}[\vOX]\HPs^{[a}\wedge\theta^{b]} +\alpha_{\HPs\Psi}[\vOX,\vUX]\HPs^{[a}\wedge\Psi^{b]}+\alpha_{\vec{\Phi}\theta}[\vOX,\vU]\vec{\Phi}^{[a}\wedge\theta^{b]}+\alpha_{\BPs\BPh}[\vU,\vUX]\BPs^{[a}\wedge\BPh^{b]} \\
+&\D(\B{\alpha}_{\Phi\theta}[\vOX]\nabla^{z}\phi\nabla_{z}\nabla^{[a}\phi)\wedge\theta^{b]}+\D(\B{\alpha}_{\Phi\Psi}[\vUX]\nabla^{z}\phi\nabla_{z}\nabla^{[a}\phi)\wedge\Psi^{b]}+\D(\B{\alpha}_{\pP\theta}[\vOX,\vU]\pP\nabla^{[a}\phi)\wedge\theta^{b]}\,,
\label{eq:DisfCurvature}
\end{split}
\end{equation}
where we have introduced two additional elements $\HPs^{a}=\nabla^{a}\nabla_{z}\phi\nabla^{z}\phi\nabla_{y}\phi\Phi^{y}$ and $\vec{\Phi}^{a}=\nabla^{a}\phi\nabla_{z}\phi\nabla^{z}\nabla_{y}\phi\Phi^{y}$, which, again, will be discussed in more detail in Sec. \ref{subsec:ExtendedBasis} and whose summary can be found in Tab. \ref{tab:notation}. Clearly, the outcome of the transformation (\ref{eq:DisfCurvature}) has become more cumbersome. Nevertheless, the discussion is very similar. We obtain all possible combinations of the building blocks that constitute a 2-form. From which every term containing elements beyond the original basis will be proportional to $\vOX$ or $\vUX$, which are the coefficients sensitive to a kinetic dependence. In addition, there are exterior derivatives appearing in the last line. Once a Lagrangian is formed, these derivatives can be eliminated by introducing total derivatives. Technically, one needs to apply the graded Leibniz rules for exterior derivatives and Stoke's theorem. Lastly, the particular form of the $\alpha_{i}$ coefficients is presented in (\ref{eq:a1}-\ref{eq:a17}). They are functions of $\vO_{i}$ and $\vU_{i}$, and include contractions of second derivatives, $\pP$ factors, when possible, so that all the terms are of the same order in second derivatives. Similarly to the previous case of second derivatives, we have named each $\alpha_{i}$ coefficient with a subindex referring to the term after it. We also show the dependence on $\vO_{i}$ and $\vU_{i}$ in brackets. The last three coefficients has an over-bar to indicate that they differ from the rest since they are inside an exterior derivative.
%--
\end{enumerate}
%--
%-------
%EXTENDED BASIS
%-------
\subsection{Extended Basis}
\label{subsec:ExtendedBasis}
%
%--
%TABLE NOTATION
%--
\bgroup
\def\arraystretch{1.5}
\begin{table}
\begin{tabular}{ l || c}
& Scalar Field Derivatives \\ \hline\hline
\emph{First derivatives:} & $\nabla_{a}\phi=\partial_{a}\phi=\phi_{,a}$ \\ 
\emph{Second derivatives:} & $\nabla_{a}\nabla_{b}\phi=\phi_{;ab}=\phi_{;ba}$ \\ \hline\hline
& Contractions of Second Derivatives \\ \hline\hline
\emph{$n$-th power:} & $\Phi^{n}_{\  ab}=\phi_{az_{1}}\phi^{;z_{1}}_{\  \  \  ;z_{2}}\cdots\phi^{;z_{n-1}}_{\  \  \  \  \  \  ;b}$ \\ 
\emph{With the metric:} & $[\Phi^{n}]=\Phi^{n}_{\  ab}\eta^{ab}$\\ 
\emph{With first derivatives:} & $\langle\Phi^{n}\rangle=\phi^{,a}\Phi^{n}_{\  ab}\phi^{,b}$ \\ \hline\hline
& 1-forms with Scalar Field Derivatives \\ \hline\hline
\emph{First derivatives:} & $\Psi^{a}=\nabla^{a}\phi\D\phi=\nabla^{a}\phi\nabla_{b}\phi\cdot\theta^{b}$ \\ 
\emph{Second derivatives:} & $\Phi^{a}=\D\nabla^{a}\phi=\nabla^{a}\nabla_{b}\phi\cdot\theta^{b}$ \\ \hline\hline
& 1-forms with Contractions of Second Derivatives \\ \hline\hline
\emph{$n$-th power:} & $\lp\Phi^{n}\rp^{a}=\left.\Phi^{n}\right.^{a}_{\  b}\cdot\theta^{b}$ \\ 
\emph{With first derivatives:} & $\lp\Psi^{mn}\rp^{a}=\left.\Phi^{m}\right.^{a}_{\  b}\phi^{,b}\phi_{,c}\lp\Phi^{n}\rp^{c}=\left.\Phi^{m}\right.^{a}_{\  b}\phi^{,b}\phi_{,c}\left.\Phi^{n}\right.^{c}_{\  d}\cdot\theta^{d}$ \\ \hline\hline
& Extended Basis \\ \hline\hline
\emph{Original elements:} & $\Psi^{a}$, $\Phi^{a}$ \\ 
\emph{Additional elements:} & $\  \  \BPh^{a}=\lp\Psi^{01}\rp^{a}$, $\BPs^{a}=\lp\Psi^{10}\rp^{a}$, $\HPs^{a}=\lp\Psi^{11}\rp^{a}$, $\vec{\Phi}^{a}=\lp\Psi^{02}\rp^{a}$, $\IPs^{a}=\pP\HPs^{a}\  \  $ \\ \hline\hline
\end{tabular}
\caption{Summary of the notation used throughout the text and necessary for constructing the extended basis. Any higher order basis element can be constructed in terms of $\lp\Phi^{n}\rp^{a}$, $\lp\Psi^{mn}\rp^{a}$ and factors of $[\Phi^{n}]$ and $\langle\Phi^{n}\rangle$, as discussed in Sec. \ref{subsec:FullExtendedBasis}. However, for the case of disformal transformations, we will only need a reduced set of them that we define, for shortness, in the last row of the table as $\BPh^{a},\  \BPs^{a},\  \HPs^{a},\  \vec{\Phi}^{a}$ and $\IPs^{a}$. The square and angle bracket notation is based on Ref. \cite{Zumalacarregui:2013pma}.}\label{tab:notation}
\end{table}
%-----
In the previous section, we have learned that a disformal transformation introduces additional building blocks. At linear order in second derivatives, there have appeared two 1-forms $\BPh^{a}\equiv\nabla^{a}\phi\nabla_{z}\phi\Phi^{z}$ and $\BPs\equiv\nabla^{a}\nabla_{z}\phi\Psi^{z}$. When contracted with the Hodge dual basis $\theta^{\star}_{\  a_{i}\cdots a_{k}}$, these two terms become equivalent, since there is a symmetry between the free index and the one contracted to $\theta^{a}$. Moreover, they correspond to a contraction of $\Phi^{a}$ with first derivatives. Lagrangians with this kind of terms were studied in Ref. \cite{Ezquiaga:2016nqo}. They were named with an over-bar $\Lag_{(l\B{m}n)}$ to indicate that there was one such contraction (thus the present notation). It was shown that, after the use of algebraic identities and total derivatives, they are equivalent to the original basis $\Lag_{(lmn)}$. Therefore, we do not consider them here.

Furthermore, we have seen that the kinetic dependence of the disformal coefficients introduces an additional 1-form that is quadratic in second derivatives. We have defined it as  $\J{\Psi}^{a}\equiv\lp\Psi^{11}\rp^{a}=\nabla^{a}\nabla_{z}\phi\nabla^{z}\phi\nabla_{y}\phi\Phi^{y}$, where $\lp\Psi^{11}\rp^{a}$ is a particular realization of the notation to denote higher order 1-form $\lp\Psi^{mn}\rp^{a}$ presented in Tab. \ref{tab:notation}. One can use the same rules as before to construct a general Lagrangian such as (\ref{eq:L}). Whenever there is a term of this type, we will denote it with an over-hat. Accordingly, an extended basis of Lagrangians can be constructed as 
\begin{equation}
\Lag_{(l\J{m}n)}=\bigwedge_{i=1}^{l}\mathcal{R}^{a_{i}b_{i}}\wedge\J{\Psi}^{c_{1}}\wedge\bigwedge_{j=2}^{m}\Phi^{c_{j}}\wedge\bigwedge_{k=1}^{n}\Psi^{d_{k}}\wedge\theta^{\star}_{~a_{1}b_{1}\cdots a_{l}b_{l}c_{1}\cdots c_{m}d_{1}\cdots d_{n}}\,.
\label{eq:Lhat}
\end{equation}
In the case of disformal Horndeski theory, we will find in the next section that only terms up to $m=2$ and $l=0$ arise, which is a consequence of the fact that a disformal transformation does not change the power of second derivatives of a given theory and that Horndeski theory is at most cubic in second derivatives\footnote{This statement can be proved by analyzing how each building block transforms under a disformal redefinition. We have exactly performed this study in Sec. \ref{subsec:DisformalBasis} and one can observe that every disformal element maintains the same order in second derivatives.}. In Appendix \ref{subapp:Lag}, one can find the component notation of these Lagrangians that will emerge in the next sections. Also, the relations of this extended basis, (\ref{eq:Lhat}), with the original one, (\ref{eq:L}), which could be either algebraic identities or exact forms, are placed respectively in Appendix \ref{subapp:AntisymId} and \ref{subapp:ExactForms}.

In addition, there will be another extended building block arising in the calculation of the disformal Horndeski theory. This element will be a consequence of the dependence on scalars with second derivatives, e.g. $\pP$, of the coefficients $\lambda_{i}$ and $\alpha_{i}$ of the disformal transformation of $\Phi^{a}$ and $\R^{ab}$ respectively (cf. Appendix \ref{app:Computations} for their definition). This additional element corresponds to a contraction with second derivatives of one case of the generalized 1-form $\lp\Psi^{mn}\rp^{a}$. We define it as $\I{\Psi}^{a}\equiv\nabla^{a}\nabla_{z}\phi\nabla^{z}\phi\nabla_{y}\phi\nabla^{y}\nabla_{x}\phi\lp\Psi^{01}\rp^{x}=\pP\J{\Psi}^{a}$. Consequently, one only has to substitute $\I{\Psi}^{c_{1}}$ for $\J{\Psi}^{c_{1}}$ in the Lagrangian (\ref{eq:Lhat}) to obtain a new set of Lagrangians $\Lag_{(l\I{m}n)}$. Now, they will be characterized by a Czech accent instead of an over-hat. Just the two lowest order Lagrangians of this class will show up in the computations and their component expressions are given in Appendix \ref{subapp:Lag}. Altogether, with these extended basis Lagrangians $\Lag_{(l\J{m}n)}$ and $\Lag_{(l\I{m}n)}$, one has all the necessary ingredients to construct the disformal transformation of any theory build up with the basis elements $\Lag_{(lmn)}$. This corresponds to any \textquotedblleft Galileon-like\textquotedblright theory.
%-------
%FULL EXTENDED BASIS
%-------
\subsection{Towards a Complete Basis (of Differential Forms)}
\label{subsec:FullExtendedBasis}
Aside from the disformal transformation of the scalar-tensor theories $\Lag_{(lmn)}$, one may wonder what is the maximal capability of this extended basis. Let us imagine that we want to construct an action for a theory with arbitrary powers of second derivatives of the scalar field up to maximum order $T$. In component notation, one could define the Lagrangian density of the theory $S=\int L\sqrt{-g}d^{D}x$ as a sum of all possible monomials\footnote{Here, a square bracket represents the contraction of two free indices, e.g. $[t_{\mu\nu}]\equiv t^{\mu}_{~\mu}$, and an angle bracket the contraction with partial derivatives of the scalar field, e.g. $\langle t_{\mu\nu}\rangle\equiv\phi^{,\mu}t_{\mu\nu}\phi^{,\nu}$ \cite{Zumalacarregui:2013pma}. This notation is summarized in Tab. \ref{tab:notation}.} 
\begin{equation} \label{eq:Lcomponents}
L(\phi,X,\tP,\cdots,[\Phi^{M}],\pP,\cdots,\langle\Phi^{N}\rangle)=\sum_{i_{1}\cdots i_{M}}\sum_{j_{1}\cdots j_{N}}g_{i_{1}\cdots i_{M}j_{1}\cdots j_{N}}(\phi,X)\prod_{k=1}^{M}[\Phi^{k}]^{i_{k}}\cdot\prod_{l=1}^{N}\langle\Phi^{l}\rangle^{j_{l}}\,.
\end{equation}
satisfying that their total order of the $M+N$ factors is bounded, specifically $\sum_{k}k\cdot i_{k}+\sum_{l}l\cdot j_{l}<T$. If we wish to construct a similar theory using the  formalism of differential forms, we will encounter several difficulties. A fundamental limitation is related to the fact that a general Lagrangian can accommodate at most $D$ wedge products of 1-forms, where $D$ is the dimension of the space-time manifold. Thus, in principle, one could not get more than $D$ products in the final theory. Moreover, some of the combinations may lead to trivial outcomes. For instance, whenever we have two general building blocks $\lp\Psi^{m_{1}n_{1}}\rp^{a}$ and $\lp\Psi^{m_{2}n_{2}}\rp^{a}$ with $m_{1}=m_{2}$, $n_{1}=n_{2}$ or both, the wedge product $\lp\Psi^{m_{1}n_{1}}\rp^{a}\wedge\lp\Psi^{m_{2}n_{2}}\rp^{b}$ vanishes due to antisymmetry. These two issues constrain the representation of arbitrary theories in a simple basis of differential forms.

However, one can manage to describe general theories of the form of (\ref{eq:Lcomponents}) in the differential form formalism enlarging the set of building blocks. When $T\leq D$, the basic blocks are the generalized 1-forms $\lp\Phi^{m}\rp^{a}$ and $\lp\Psi^{mn}\rp^{a}$, which represent contractions of second derivatives with the metric and with first derivatives respectively. A formal definition of them can be found in Tab. \ref{tab:notation}. In this case, since the maximum power $T$ does not exceed the number of dimensions, there are no problems with the required number of products. Nevertheless, in order to have a complete basis, and due to the antisymmetric redundancies aforementioned, one has also to allow for arbitrary contractions of $\lp\Psi^{mn}\rp^{a}$ with second derivatives. This will consist of a generalization of the second extended building block $\I{\Psi}^{a}=\pP\J{\Psi}^{a}$ introduced in Sec. \ref{subsec:ExtendedBasis}. Effectively, one would end up with additional building blocks of the form of $\langle\Phi^{p}\rangle\lp\Psi^{mn}\rp^{a}$. In the case of $T> D$, one cannot invoke further contractions. The way out is to enable new terms of the form $[\Phi^{p}]\lp\Phi^{m}\rp^{a}$. In conclusion, this larger set of building blocks allows for constructing theories with arbitrary powers of second derivatives. The shortcoming of this generalization is that the new building blocks contains scalar functions such as $\pP$ or $\tP$ that will not be affected by the antisymmetric structure. Therefore, they will not benefit from all the advantages of this formalism. Furthermore, not all of the possible Lagrangians will be independent. There will be relations among them that limit their number. Such links between different theories could be either antisymmetric identities or exact forms. We include some examples of them in Appendix \ref{subapp:AntisymId} and \ref{subapp:ExactForms} respectively.
%-------
%SEC. III: DISFORMAL THEORIES
%-------
\section{A tour through Different Disformal Theories}
\label{sec:DisformalTheories}

After presenting how the fundamental building blocks transform with a disformal vielbein, we are going to study particular disformal theories. For this task, we only need to glue together our previous results and work out the coefficients, which will be all functions of $\vO_{i}$ and $\vU_{i}$. Along with the computations, there will be extra factors corresponding to contractions of second derivatives such as $\pP$ or $\pPP$. We have already seen that they are originated in the disformal curvature and second derivatives of the scalar. Applying algebraic identities among Lagrangians, one can rewrite those terms as Lagrangians of the basis $\Lag_{(lmn)}$ (or its higher order extensions $\Lag_{(l\J{m}n)}$ and $\Lag_{(l\I{m}n)}$). We summarize the relevant relations in Appendix \ref{subapp:AntisymId}. Moreover, when curvature terms are present, one can implement exact forms (equivalent to total derivatives) to express the results with second derivatives only. The set of exact forms relating different Lagrangians of the extended basis can be found in Appendix \ref{subapp:ExactForms}.\footnote{Here, and in the rest of the text, we will assume that the spacetime is asymptotically flat so that the fields vanish at infinity and total derivative terms can be neglected. In spacetimes with boundaries, such de Sitter, these terms might be relevant when setting equivalences between theories and boundary conditions might not be preserved in different frames.}

It is important to emphasis that all the Lagrangians of the extended basis will be proportional to either $\vOX$ or $\vUX$, which respectively encode the dependence of $\T{C}$ and $\T{D}$ in $X$. This stresses the fact, which was first shown in Ref. \cite{Bettoni:2013diz}, that Horndeski theory maps onto itself for special disformal transformations, i.e. for transformation coefficients depending only in $\phi$. In our context, this special case translates into $\vOX=\vUX=0$. However, in Ref. \cite{Bettoni:2013diz} the new non-Horndeski terms were not computed. Here we compute them. It is relevant to remember that these new terms appearing represent healthy beyond-Horndeski theories as it was first realized in Ref. \cite{Zumalacarregui:2013pma}. In this section, we will concentrate on computing the disformal version of Horndeski theory $\Lag^{H}=\sum_{i=2}^{5}\Lag_{i}^{H}$, which in our formalism takes the form
\begin{align}
\Lag_{2}^{H}[G_{2}]=&G_{2}\Lag_{(000)}\,, \label{eq:LH2} \\
\Lag_{3}^{H}[G_{3}]=&G_{3}\Lag_{(010)}\,, \label{eq:LH3} \\
\Lag_{4}^{H}[G_{4}]=&G_{4}\Lag_{(100)}+G_{4,X}\Lag_{(020)}\,, \label{eq:LH4} \\
\Lag_{5}^{H}[G_{5}]=&G_{5}\Lag_{(110)}+\frac{1}{3}G_{5,X}\Lag_{(030)}\,. \label{eq:LH5} 
\end{align}
One should recall that there are other combinations of Lagrangians $\Lag_{i}^{NH}$ which are equivalent, up to total derivatives, to the above ones \cite{Ezquiaga:2016nqo}. For completeness, we include them in the Appendix \ref{app:DisformalHorndeski}.

Some results that we are going to present partially overlap with previous work. First, the special disformal transformation of Horndeski theory was computed in \cite{Bettoni:2013diz}, i.e. $\vOX=\vUX=0$ for $\T{\Lag}^{H}$. Then, the kinetic disformal Einstein-Hilbert action was obtained in \cite{Zumalacarregui:2013pma}, i.e. $\vOP=\vUP=0$ for $\T{\Lag}_{(100)}$ (see Ref. \cite{Brax:2016kin} for cosmological implications). Moreover, in Ref. \cite{Gleyzes:2014qga}, it was shown that a purely kinetic disformal transformation of quartic and quintic Horndeski Lagrangians introduce the quartic and quintic beyond Horndeski Lagrangians respectively, i.e. $\vOP=\vOX=0$ for $\Lag_{4}^{H}$ and $\Lag_{5}^{H}$. Afterwards, the kinetic disformal quadratic degenerate Lagrangian terms were computed in \cite{Achour:2016rkg}. Later, the kinetic disformal transformation of kinetic quartic and kinetic quintic Horndeski were computed respectively in \cite{Crisostomi:2016czh} and \cite{BenAchour:2016fzp}, i.e. $\vOP=\vUP=0$ for $\Lag_{4}^{H}[G_{4}(X)]$ and $\Lag_{5}^{H}[G_{5}(X)]$. The different $\vO_{i}$ and $\vU_{i}$ coefficients of each theory can be found in Table \ref{tab:xi}.

Nevertheless, our calculation is still valuable for two main reasons. Firstly, it is based in a completely new approach that benefits from a systematic and much more clear structure. This novel formalism allows for an easy extrapolation to other  gravity theories that we will discuss in Sec. \ref{sec:ConnectingMG}. Secondly, we will be presenting for the first time the \emph{full calculation of disformal Horndeski theory}, i.e. the most general dependence of the disformal coefficients $\vO_{i},\vU_{i}\neq0$ for all Horndeski Lagrangians $\T{\Lag}^{H}$. We will present in detail how this result is obtained. Moreover, for the future benefit of the community, we include the full result with all the coefficients in Appendix \ref{app:DisformalHorndeski}, so that it can be easily used for further computations.

Thus, in the following, we start a journey through different disformal theories. As a warm up exercise, we will begin with a conformal transformation, particularizing for a kinetic dependence only. This will be useful for visualizing the type of new terms and the method of computation. This will allow us to jump to the calculation of the full disformal Horndeski theory. Later, we compute the disformal transformation of beyond-Horndeski Lagrangians. Finally, as a bonus, we investigate the effect of disformal transformations in Lovelock theory.
%--
%TABLE DISFORMAL COEFFICIENTS
%--
\bgroup
\def\arraystretch{1.5}
\begin{table}
\begin{tabular}{l l || c | c | c | c | c | c |}
& & $\  \vO\  $ & $\  \vOP\  $ & $\  \vOX\  $ & $\  \vU\  $ & $\  \vUP\  $  & $\  \vUX\  $ \\ \hline\hline
\emph{Special Conformal:} & $\T{C}=\T{C}(\phi)$, $\T{D}=0$ & $\  1\  $ & $\  \neq0\  $ & $0$ & $0$ & $0$ & $0$ \\ \hline
\emph{Kinetic Conformal:} & $\T{C}=\T{C}(X)$, $\T{D}=0$ & $\  1\  $ & $0$ & $\  \neq0\  $ & $0$ & $0$ & $0$ \\ \hline
\emph{(General) Conformal:} & $\T{C}=\T{C}(\phi,X)$, $\T{D}=0$ & $\  1\  $ & $\  \neq0\  $ & $\  \neq0\  $ & $0$ & $0$ & $0$ \\ \hline
\emph{Constant Disformal:} & $\T{C}=const.$, $\T{D}=const.$ & $\  \neq0\  $ & $0$ & $0$ & $\  \neq0\  $ & $0$ & $0$ \\ \hline
\emph{Special Disformal:} & $\T{C}=\T{C}(\phi)$, $\T{D}=\T{D}(\phi)$ & $\neq0$ & $\neq0$ & $0$ & $\  \neq0\  $ & $\  \neq0\  $ & $0$ \\ \hline
\emph{Purely Kinetic Disformal:} & $\T{C}=const.$, $\T{D}=\T{D}(X)$ & $\neq0$ & $0$ & $0$ & $\neq0$ & $0$ & $\  \neq0\  $ \\ \hline
\emph{Kinetic Disformal:} & $\T{C}=\T{C}(X)$, $\T{D}=\T{D}(X)$ & $\neq0$ & $0$ & $\neq0$ & $\neq0$ & $0$ & $\  \neq0\  $ \\ \hline
\emph{(General) Disformal:} & $\T{C}=\T{C}(\phi,X)$, $\T{D}=\T{D}(\phi,X)$ & $\  \neq0\  $ & $\  \neq0\  $ & $\  \neq0\  $ & $\   \neq0\   $ & $\  \neq0\  $ & $\  \neq0\  $ \\ \hline
\end{tabular}
 \caption{Summary of different disformal transformations in terms of $\vO_{i}$ and $\vU_{i}$, which are functions of $\T{C}$ and $\T{D}$ and are defined in Eqs. (\ref{eq:x0}-\ref{eq:x4}).}\label{tab:xi}
 \end{table}
%-------
%CONFORMAL HORNDESKI THEORY
%-------
\subsection{Conformal Horndeski Theory}
\label{subsec:ConformalHorndeski}

Before going to the full disformal Horndeski theory, we are going to analyze the conformal Horndeski theory, focusing in the kinetic conformal transformations since it is the one triggering the new terms beyond Horndeski. For this task, one should notice that a kinetic conformal transformation, i.e. $\T{C}=\T{C}(X)$ and $\T{D}=0$, is characterized by $\vO=1$, $\vOP=\vU=\vUP=\vUX=0$ and $\vOX=\vOX(X)$, as summarized in Table \ref{tab:xi}. In order to obtain the transformed theory, one needs to transform each building block, e.g. $\Phi^{a}\rightarrow\T{\Phi}^{a}$. Furthermore, one should remember to transform appropriately the coefficient in front of each Lagrangian. For shortness, it can be defined $\T{G}_{i}(\phi,X)\equiv G_{i}(\phi,\T{X}(\phi,X))$. In the case that the coefficient has a partial derivative, $\T{G}_{i,\T{X}}(\phi,X)$, the following factor has to be included
\begin{equation}
\frac{\partial \T{X}}{\partial X}=\gD^{2}\vO(\vO-X(\vOX-2X\vUX))=\gD^{2}\gX^{-1}\vO\,,
\label{eq:Xcoef}
\end{equation}
where we have introduced the fractions $\gamma_{i}$ defined in Sec. \ref{subsec:DisformalBasis} as $\gD\equiv(\T{C}-2X\T{D})^{-1}$ and $\gX\equiv(\vO-X(\vOX-2X\vUX))^{-1}$. These terms are important because they account for the exact cancellation of the higher derivative terms in the equations of motion. When dealing with extra non-Horndeski Lagrangians, our aim will be to group the terms into as many Horndeski Lagrangians as possible to isolate properly the remaining part.

Starting with the simple $\Lag_{2}^{H}$, we only need to rescale the volume element 
\begin{equation}
\T{\Lag}_{2}^{H}[G_{2}]=G_{2}(\phi,\T{X})\T{\Lag}_{(000)}=G_{2}(\phi,\T{X})\wedge\T{\theta}^{\star}=\T{G}_{2}(\phi,X)\T{C}^{4}\wedge\theta^{\star}=\Lag_{2}^{H}[\T{G}_{2}(\phi,X)\T{C}^{4}]\,.
\label{eq:ConfH2}
\end{equation}
Clearly, this new Lagrangian belongs to Horndeski theory since only the coefficient in front of $\Lag_{(000)}=\theta^{\star}$ has changed. In the case of $\Lag_{3}^{H}$, we also need to transform the building block $\Phi^{a}$. It is easy to obtain
\begin{equation}
\begin{split}
\T{\Lag}_{3}^{H}[G_{3}]=&G_{3}(\phi,\T{X})\T{\Lag}_{(010)}=\T{G}_{3}\wedge\T{\Phi}^{a}\wedge\T{\theta}^{\star}_{\  a} \\
=&\T{G}_{3}(\lambda_{\Phi}\Phi^{a}+\lambda_{\theta}\theta^{a}+\lambda_{\BPh}\BPh^{a}+\lambda_{\BPs}\BPs^{a})\wedge\T{C}^{3}\theta^{\star}_{\  a} \\
=&\T{G}_{3}\T{C}^{2}(\Lag_{(010)}+\vOX\Lag_{(0\B{1}0)})=\Lag_{3}^{H}[\T{G}_{3}\T{C}^{2}(1+2X\vOX)]+\Lag_{3}^{NH}[\T{G}_{3}\T{C}^{2}\vOX]\,,
\label{eq:ConfH3}
\end{split}
\end{equation}
where we have used the explicit definition of the $\lambda_{i}$ coefficients in the second line and one algebraic identity in the third one to relate different Lagrangians, i.e. $\Lag_{(0\B{1}0)}=-\Lag_{(011)}-2X\Lag_{(010)}$. In the last line, we have made clear that these Lagrangians belong to Horndeski theory, since it was found in \cite{Ezquiaga:2016nqo} that $\Lag_{3}^{NH}$, given in Appendix \ref{app:DisformalHorndeski}, is part of Horndeski theory upon exact forms (total derivatives). 

For the quartic kinetic conformal Horndeski Lagrangian, there are two Lagrangians to transform and we need to include the aforementioned kinetic coefficient (\ref{eq:Xcoef}). We find
\begin{equation}
\begin{split}
\T{\Lag}_{4}^{H}[G_{4}]=& G_{4}(\phi,\T{X})\T{\Lag}_{(100)}+G_{4,\T{X}}\T{\Lag}_{(020)}=\T{G}_{4}\T{\Lag}_{(100)}+\T{G}_{4,X}\frac{\partial X}{\partial\T{X}}\T{\Lag}_{(020)} \\
=&\Lag_{4}^{H}[\T{C}^{2}\T{G}_{4}]+\Lag_{3}^{H}[6X\T{C}^{2}\T{G}_{4,\phi}\vOX]+\Lag_{3}^{NH}[3\T{C}^{2}\T{G}_{4,\phi}\vOX] \\
-&\frac{1}{2}\T{C}^{2}\vOX(\T{G}_{4}-\frac{2X\T{G}_{4,X}}{1-X\vOX})(2\Lag_{(020)}-3\vOX\Lag_{(0\J{1}0)})+\frac{1}{2}\frac{\T{C}^{2}\T{G}_{4,X}\vOX}{1-X\vOX}(\Lag_{(021)}-\vOX\Lag_{(0\J{1}1)})\,, 
\end{split}
\label{eq:ConfH4}
\end{equation}
where we present the result directly in terms of the known Lagrangians $\Lag_{i}^{H}$ and $\Lag_{i}^{NH}$. In order to get this result, one only needs to repeat each step performed in the previous calculation: $i)$ substitute every disformal building block, then $ii)$ identify known Lagrangians, and finally $iii)$ apply identities to rewrite everything in term of Horndeski Lagrangians. Here, the coefficient $\T{G}_{4,\phi}$ enters when applying an exact form to eliminate the exterior derivatives in the disformal curvature $\T{\R}^{ab}$, cf. (\ref{eq:DisfCurvature}). For the first time in the calculations of this section, there are terms appearing in the last line that cannot be recast in the Horndeski structure. These terms will be generated by the kinetic disformal dependence via $\vOX$ and $\vUX$. 

In order to compute the quintic kinetic conformal Horndeski Lagrangian, one could apply the same procedure as before. However, there will be terms in which there are third derivatives of the scalar that cannot be eliminated. This is because, when trying to eliminate the exterior derivatives of the disformal curvature $\T{\R}^{ab}$, we move the exterior derivative to terms in the disformal second derivative 1-form $\T{\Phi}^{a}$ that cannot be reduced to only second derivatives. To avoid this unwanted situation, one could apply the following trick. It consists of applying an exact form in the first step of the calculation. Since these terms appear through $\T{\Lag}_{(110)}$, we focus only in this part first. The trick reads
\begin{equation}
\begin{split}
\T{G}_{5}\T{\Lag}_{(110)}=& \T{G}_{5}\T{\R}^{ab}\wedge\T{\Phi}^{c}\wedge\T{\theta}^{\star}_{\  abc}=\T{G}_{5}\T{\R}^{ab}\wedge\T{\D}\lp\T{\nabla}^{c}\phi\rp\wedge\T{\theta}^{\star}_{\  abc} \\
 =&-\T{\D}(\T{G}_{5})\T{\nabla}^{c}\phi\wedge\T{\R}^{ab}\wedge\T{\theta}^{\star}_{\  abc}-\T{G}_{5}\T{\nabla}^{c}\phi\wedge\T{\D}\T{\R}^{ab}\wedge\T{\theta}^{\star}_{\  abc}+\T{G}_{5}\T{\nabla}^{c}\phi\wedge\T{\R}^{ab}\wedge\T{\D}\T{\theta}^{\star}_{\  abc}+\T{\D}\T{\Lag}_{(100)}^{D-1} \\
 =&-\T{\D}(\T{G}_{5})\T{\nabla}^{c}\phi\wedge\T{\R}^{ab}\wedge\T{\theta}^{\star}_{\  abc}+\T{\D}\T{\Lag}_{(100)}^{D-1}\,,
\end{split}
\label{eq:ConfL110}
\end{equation}
where we have included the exact forms used for completeness, $\T{\D}\T{\Lag}_{(100)}^{D-1}=\T{\D}(\T{G}_{5}\T{\nabla}^{c}\phi\wedge\T{\R}^{ab}\wedge\T{\theta}^{\star}_{\  abc})$. Notice that we have taken advantage of the fact that $\D\R^{ab}=0$ and $\D\theta^{\star}_{\  a_{1}\cdots a_{k}}=0$ (and thus $\T{\D}\T{\R}^{ab}=0$ and $\T{\D}\T{\theta}^{\star}_{\  a_{1}\cdots a_{k}}=0$). Once we are in the last line, we are safe since the action of the exterior derivatives of $\T{\R}^{ab}$ on $\D\T{G}_{5}$ is second order.\footnote{This can be easily proven since $\D G_{i}(\phi,X)=G_{i,\phi}\D\phi-G_{i,X}\nabla_{z}\phi\Phi^{z}$. Thus, only the last term can lead to third derivatives. Nevertheless, applying an exterior derivative, one finds $\nabla_{z}\phi\D\Phi^{z}=\nabla_{z}\phi\R^{zy}\nabla_{y}\phi=0$, due to the antisymmetry of the curvature indices.}

Consequently, we can use the above result to compute $\T{\Lag}_{5}^{H}$. Neglecting the exact form terms, which can be eliminated applying Stoke's theorem, we obtain
\begin{equation}
\begin{split}
\T{\Lag}_{5}^{H}[G_{5}]=& G_{5}(\phi,\T{X})\T{\Lag}_{(110)}+\frac{1}{3}G_{5,\T{X}}\T{\Lag}_{(030)}=-\T{\D}(\T{G}_{5})\T{\nabla}^{c}\phi\wedge\T{\R}^{ab}\wedge\T{\theta}^{\star}_{\  abc}+\frac{1}{3}\T{G}_{5,X}\frac{\partial X}{\partial\T{X}}\T{\Lag}_{(030)} \\
=&\Lag_{5}^{H}[\T{G}_{5}]+\frac{X\vOX\T{G}_{5,X}}{6(1-X\vOX)}(2\Lag_{(030)}-6\vOX\Lag_{(0\J{2}0)}+3\vOX^{2}\Lag_{(0\I{1}0)}) \\
+&X\vOX\T{G}_{5,\phi}(2\Lag_{(020)}-3\vOX\Lag_{(0\J{1}0)})+\vOX\T{G}_{5,\phi}(\Lag_{(021)}-\vOX\Lag_{(0\J{1}1)})\,.
\end{split}
\label{eq:ConfH5}
\end{equation}
Therefore, in addition to the non-Horndeski combinations of Lagrangians discovered in (\ref{eq:ConfH4}), we find another one, namely the second term in the second line of the above equation. Interestingly, all these new sets of Lagrangians have a clear structure that repeats in both quartic and quintic, kinetic conformal Horndeski Lagrangians. This is consistent with the relations existing between a quartic Lagrangian depending on $X$ and a quintic Lagrangian depending on $\phi$. Moreover, it will be a reflection of the structure of the healthy theories constructed with the extended basis. We will discuss how these combinations relate to degenerate scalar-tensor theories in the next section.

Before moving to the general disformal analysis, it is important to remark that these results have been checked with previous sub-cases considered in the literature \cite{Zumalacarregui:2013pma,Achour:2016rkg,Crisostomi:2016czh,BenAchour:2016fzp}.

%-------
%DISFORMAL HORNDESKI THEORY
%-------
\subsection{Disformal Horndeski Theory}
\label{subsec:DisformalHorndeski}

Subsequently, we are going to generalize the method presented in the previous section to full disformal transformations. We have found so far that the kinetic dependence of the conformal factor already introduces linear combinations of Lagrangians that do not belong to Horndeski theory, starting with the quartic Lagrangian. Moreover, we have been able to evade the appearance of third derivatives in the quintic Lagrangian applying appropriate exact forms, which does not modify the dynamics of the theory. Along this section, we will focus on searching for new non-Horndeski Lagrangians and on determining the origin of each term. With the purpose of facilitating the use of our results in future research, we devote Appendix \ref{app:DisformalHorndeski} to present the complete calculation of general disformal Horndeski theory altogether. In this section, we will investigate each disformal Horndeski Lagrangian one by one.

To begin with, the first disformal Horndeski Lagrangian simply yields
\begin{equation}
\T{\Lag}_{2}^{H}[G_{2}]=G_{2}(\phi,\T{X})\T{\Lag}_{(000)}=\T{G}_{2}(\phi,X)\gD^{-1}\T{C}^{3}\Lag_{(000)}=\Lag_{2}^{H}[\T{G}_{2}(\phi,X)\T{C}^{3}(\T{C}-2X\T{D})]\,.
\label{eq:DisfH2}
\end{equation}
Clearly, this term maps onto itself, i.e. $\T{\Lag}_{2}^{H}\subset\Lag_{2}^{H}$, and there is no difference, apart from the actual form of the coefficient, with the conformal case. The second Lagrangian becomes
\begin{equation}
\begin{split}
\T{\Lag}_{3}^{H}[G_{3}]=&G_{3}(\phi,\T{X})\T{\Lag}_{(010)}=\T{C}^{2}\T{G}_{3}\vO(\vO(\vO-X(\vOX-2X\vUX))+3X\vOX)\Lag_{(010)} \\
+&\frac{1}{2}\T{C}^{2}\T{G}_{3}\vO(3\vOX+(1+\vO)\vU-\vO(\vOX-2X\vUX))\Lag_{(011)}-2X\T{C}^{2}\T{G}_{3}(\vOP+X\vUp)\Lag_{(000)} \\
=&\Lag_{3}^{H}[\B{G}_{3}]+\Lag_{3}^{NH}[\B{E}_{3}]+\Lag_{2}^{H}[\B{G}_{2}]\,,
\label{eq:DisfH3}
\end{split}
\end{equation}
where $\B{G}_{3}$, $\B{E}_{3}$ and $\B{G}_{2}$ can be read directly from the first two lines of (\ref{eq:DisfH3}) since $\Lag_{3}^{H}[\B{G}_{3}]=\B{G}_{3}\Lag_{(010)}$, $\Lag_{3}^{NH}[\B{E}_{3}]=\B{E}_{3}\Lag_{(011)}$ and $\Lag_{2}^{H}[\B{G}_{2}]=\B{G}_{2}\Lag_{(000)}$. Thus, as in the conformal case, this result means that the disformal cubic Horndeski Lagrangian belongs to Horndeski theory. The only difference with respect to the previous case is that, since we are allowing the disformal coefficients to depend on $\phi$, we obtain one Horndeski Lagrangian of a lower order, so that $\T{\Lag}_{3}^{H}\subset\Lag_{3}^{H},\Lag_{2}^{H}$.

For the quartic disformal Lagrangian, the approach is equivalent, with a clear structure, but the coefficients involve more parameters. For that reason, we present them separately. The different Lagrangians appearing read
\begin{equation}
\begin{split}
\T{\Lag}_{4}^{H}[G_{4}]=& G_{4}(\phi,\T{X})\T{\Lag}_{(100)}+G_{4,\T{X}}\T{\Lag}_{(020)}=\T{G}_{4}\T{\Lag}_{(100)}+\T{G}_{4,X}\frac{\partial X}{\partial\T{X}}\T{\Lag}_{(020)} \\
=&\Lag_{4}^{H}[\B{G}_{4}]+\Lag_{4}^{NH}[\B{E}_{4}]+\Lag_{3}^{H}[\B{G}_{3}]+\Lag_{3}^{NH}[\B{E}_{3}]+\Lag_{2}^{H}[\B{G}_{2}]+\Lag_{2}^{NH}[\B{E}_{2}] \\
+&H_{4}(2\Lag_{(020)}-3\vOX\Lag_{(0\J{1}0)})+I_{4}(\Lag_{(021)}-\vOX\Lag_{(0\J{1}1)})+F_{4}\Lag_{(021)}\, 
\end{split}
\label{eq:DisfH4}
\end{equation}
whose corresponding coefficients are
\begin{align}
\B{G}_{4}=&\T{C}^{2}\T{G}_{4}\vO \,, \\
\B{E}_{4}=&\frac{1}{2}\T{C}\T{G}_{4}(2\T{D}+\T{C}\vU) \,, \\
\B{G}_{3}=&-3\T{C}^{2}\T{G}_{4}(\vOP+2X\vOPx)-2X\T{C}^{2}(\T{G}_{4,\phi}\vO\vOX-\T{G}_{4,X}(\vOP+{X\gX\vO^{-1}\vOX(\vOP+2\vOp)})) \,, \\
\B{E}_{3}=&\T{C}^{2}(\T{G}_{4}(\vUp-3\vOPx)+3\T{G}_{4,\phi}\vO\vOX-\T{G}_{4,X}\gX\vO^{-1}(\vOP(3\vO^{2}-1)-2\vOp+6X^{2}\vO\vOP\vUX \\
+&3X\vOX(2\vOp-X\vOP\vU))) \,,  \nonumber \\
\B{G}_{2}=&3\T{C}^{2}X\vOP(\T{G}_{4}\vOP+{4X\T{G}_{4,X}\gX\vO^{-2}\vOp})\,, \\
\B{E}_{2}=&-3\T{C}\T{G}_{4}(\T{C}\vOPp-\T{D}X\vOP^{2}) \,, \\
H_{4}=&-\frac{1}{2}\T{C}^{2}\vOX(\T{G}_{4}\vO-{2X\gX\T{G}_{4,X}}) \,, \\
I_{4}=&\frac{1}{2}\T{C}(\T{G}_{4}(X\T{D}\vOX\vU-2\T{C}(\vOX\vU+\vUX))+\T{G}_{4,X}(4X\T{D}\vU+\T{C}\gX{((2\vO-1)\vOX-2X\vU^{2}-4X(\vO-2)\vUX)})\,, \\
F_{4}=&\frac{1}{4}\T{C}(\T{C}\T{G}_{4}(\vOX\vU+2\vUX)-2\T{G}_{4,X}(2\T{D}(2\vO-1)+{\T{C}\gX(2X\vOX\vU+(1-2\vO)\vU+(3-2\vO)2X\vUX)})\,.
\end{align}
Therefore, we obtain a set of Horndeski Lagrangians, second line of (\ref{eq:DisfH4}), and another of non-Horndeski ones, third line of (\ref{eq:DisfH4}). The difference with the conformal case is that we obtain more lower order Horndeski Lagrangians, due to the scalar field dependence of the disformal coefficients via $\vOP$ and $\vUP$. Moreover, in addition to the two linear combinations of non-Horndeski Lagrangians $(2\Lag_{(020)}-3\vOX\Lag_{(0\J{1}0)})$ and $(\Lag_{(021)}-\vOX\Lag_{(0\J{1}1)})$, we obtain an isolated $F_{4}\Lag_{(021)}$ which vanish for the conformal case.\footnote{One should notice that this grouping of the terms containing $\Lag_{(021)}$ is not unique. However, we have chosen this particular division to highlight the conformal/disformal origin of each term.} This term is nothing but the quartic beyond-Horndeski Lagrangian.
 
For the quintic disformal Horndeski Lagrangian, we follow the same procedure as in the previous case. One only needs to remember to apply an initial exact form to avoid the appearance of third derivative terms in the action, as explained for the conformal calculation, cf. (\ref{eq:ConfL110}). Splitting the Lagrangians from their coefficients, we find
\begin{equation}
\begin{split}
\T{\Lag}_{5}^{H}[G_{5}]=& G_{5}(\phi,\T{X})\T{\Lag}_{(110)}+\frac{1}{3}G_{5,\T{X}}\T{\Lag}_{(030)}=-\T{\D}(\T{G}_{5})\T{\nabla}^{c}\phi\wedge\T{\R}^{ab}\wedge\T{\theta}^{\star}_{\  abc}+\frac{1}{3}\T{G}_{5,X}\frac{\partial X}{\partial\T{X}}\T{\Lag}_{(030)} \\
=&\Lag_{5}^{H}[\B{G}_{5}]+\Lag_{4}^{NH}[\B{E}_{4}]+\Lag_{3}^{H}[\B{G}_{3}]+\Lag_{3}^{NH}[\B{E}_{3}]+\Lag_{2}^{H}[\B{G}_{2}] \\
+&H_{5}(2\Lag_{(030)}-6\vOX\Lag_{(0\J{2}0)}+3\vOX^{2}\Lag_{(0\I{1}0)})+I_{5}(2\Lag_{(031)}-6\vOX\Lag_{(0\J{2}1)}+3\vOX^{2}\Lag_{(0\I{1}1)}) \\
+&H_{4}(2\Lag_{(020)}-3\vOX\Lag_{(0\J{1}0)})+I_{4}(\Lag_{(021)}-\vOX\Lag_{(0\J{1}1)})+F_{4}\Lag_{(021)}\,.
\end{split}
\label{eq:DisfH5}
\end{equation}
where
\begin{align}
\B{G}_{5}=&\vO\T{G}_{5}-\int\T{G}_{5}\vOx dX \,, \\
\B{E}_{4}=&\vOp\T{G}_{5}-(\int\T{G}_{5}\vOx dX)_{,\phi} \,, \\
\B{G}_{3}=&6X^{2}\vOP\vOX(2\vO^{2}\T{G}_{5,\phi}+{X\T{G}_{5,X}\gX(\vOP-4X\vUp)}) \,, \\
\B{E}_{3}=&\vOP(3X\vOX-1)(2\vO^{2}\T{G}_{5,\phi}+{X\T{G}_{5,X}\gX(\vOP-4X\vUp)}) \,, \\
\B{G}_{2}=&2X^{2}\vOP^{2}(3\vO\T{G}_{5,\phi}+{2X\T{G}_{5,X}\vO^{-1}\gX(\vOP-3X\vUp)}) \,, \\
H_{5}=&\frac{1}{6}{X\gX\vO^{2}\vOX\T{G}_{5,X}} \,, \\
I_{5}=&\frac{1}{12}{\gX\vO(X\vOX\vU+2X\vUX)\T{G}_{5,X}} \,, \\
H_{4}=&X\vO\vOX(\vO^{2}\T{G}_{5,\phi}-{2X^{2}\gX\vUp\T{G}_{5,X}}) \,, \\
I_{4}=&\frac{1}{2}(\vO((1+\vO(3\vO-2))\vOX+4X\vO\vUX)\T{G}_{5,\phi}-{2X^{2}\gX((3\vO-2)\vOX+4X\vUX)\vUp\T{G}_{5,X}}) \,, \\
F_{4}=&-\frac{1}{2}X(\vOX\vU+2\vUX)(\vO^{2}\T{G}_{5,\phi}-{2X^{2}\gX\vUp\T{G}_{5,X}}) \,.
\end{align}
Similarly to the latter case, we obtain a set of Horndeski and non-Horndeski Lagrangians in (\ref{eq:DisfH5}). In addition to the conformal calculation (\ref{eq:ConfH5}) and the disformal $\Lag_{4}^{H}$ (\ref{eq:DisfH4}), we obtain an additional combination of Lagrangians, that of the $I_{5}$ coefficient, which is only generated through the disformal coefficient $\T{D}$. Another new feature of this result is that there are integral terms in the first two coefficients. This is because in order to obtain the proper Horndeski Lagrangians one needs to add one exact form.\footnote{In particular, we have used the relation $\D\Lag_{(100)}^{D-1}[\alpha]=\alpha_{,\phi}\Lag_{(101)}-\alpha_{,X}\Lag_{(1\B{1}0)}+\alpha\Lag_{(110)}$, where $\alpha=\vO\T{G}_{5,X}$ was derived in \cite{Ezquiaga:2016nqo}.} One can notice that, when there is only kinetic dependence in the disformal sector, $\vOX=0$ and $\vUX\neq0$, one recovers Horndeski theory plus the beyond-Horndeski Lagrangians $\Lag_{(021)}$ and $\Lag_{(031)}$.

In conclusion, the main result is similar to the conformal case, i.e. disformal Horndeski maps onto Horndeski plus some particular combination of Lagrangians, which in the disformal case are more numerous. These Lagrangians can be related with degenerate scalar-tensor theories \cite{Langlois:2015cwa,Crisostomi:2016czh,BenAchour:2016fzp}. In these works, they write the Lagrangians in terms of monomials, as in (\ref{eq:Lcomponents}), up to quadratic and cubic order. Nevertheless, there is a direct correspondence with our approach. We present in Appendix \ref{subapp:EST} how the quadratic and cubic Lagrangians can be completely characterized using the differential form formalism. The difference is that in our basis Lagrangians are polynomials instead of monomials. This has the advantage that general theories can be represented directly in this language and that these combinations naturally arise in the computations. To illustrate this point, one can check how the definition of the different classes of degenerate theories greatly simplifies in our formalism. In the class of theories disformally related to Horndeski, the aforementioned combinations of Lagrangians are directly present. For the others, one can find Lagrangians of the extended basis that cannot be related to the original basis with a disformal transformation.

Finally, let us add that, as before, our results agree with previous sub-cases considered in the literature \cite{Zumalacarregui:2013pma,Achour:2016rkg,Crisostomi:2016czh,BenAchour:2016fzp}. In order to summarize this section, we present in Appendix \ref{app:DisformalHorndeski} the complete disformal transformation of all Horndeski Lagrangians at once. With this full result, one can obtain any particular subclass of disformal transformations simply by choosing the appropriate characteristic coefficients $\vO_{i}$ and $\vU_{i}$, cf. Table \ref{tab:xi} for a list of every possibility.

%-------
%DISFORMAL BEYOND-HORNDESKI LAGRANGIANS
%-------
\subsection{Disformal beyond-Horndeski Lagrangians}
\label{subsec:DisformalBeyond}
As an extension of the analysis of disformal Horndeski theory, one could think of applying these same techniques to other Lagrangians. In this subsection we study the disformal $G^{3}$ Lagrangians \cite{Gleyzes:2014dya}. This calculation has been done partially in different places \cite{Gleyzes:2014qga,Crisostomi:2016czh,Achour:2016rkg}. Here, we include the most general dependence of the disformal coefficients. Beginning with the quartic beyond-Horndeski Lagrangian, i.e. $\T{\Lag}_{4}^{BH}=F_{4}(\phi,\T{X})\T{\Lag}_{(021)}$ and focusing in the transformed Lagrangian without the coefficient, we obtain 
\begin{equation}
\begin{split}
\T{\Lag}_{(021)}=&\T{\Phi}^{a}\wedge\T{\Phi}^{b}\wedge\T{\Psi}^{c}\wedge\T{\theta}^{\star}_{\  abc} \\
=&\gD^{2}\vO\big(\vO^{2}(X\vOX-1)\Lag_{(021)}+\vO^{2}(2-3X\vOX)(\Lag_{(021)}-\vOX\Lag_{(0\J{1}1)})-2X^{2}\vO^{2}\vOX(2\Lag_{(020)}-3\vOX\Lag_{(0\J{1}0)}) \\
+&24\Lag_{3}^{H}[X^{3}\vO\vOP\vOX]-4\Lag_{3}^{NH}[X\vO\vOP(1-3X\vOX)]+6\Lag_{2}^{NH}[X^{2}\vOP^{2}]\big)\,.
\end{split}
\label{eq:DisfBH4}
\end{equation}
Interestingly, this result is insensitive to the disformal part of the field redefinition, thus mapping onto itself $\T{\Lag}_{(021)}\rightarrow \Lag_{(021)}$ when the conformal factor is constant, i.e. $\vOP,\vOX=0$. This can be easily understood in the language of differential forms since the original Lagrangian contains a term $\Psi^{a}$. Then, by antisymmetry, any wedge product of $\Psi^{a}\wedge\Psi^{b}$, $\Psi^{a}\wedge\BPs^{b}$ or $\Psi^{a}\wedge\BPh^{b}\wedge\theta^{\star}_{\  ab\cdots a_{k}}$ will vanish. Consequently, as one can verify from the transformation of $\T{\theta}^{\star}_{\  a_{1}\cdots a_{k}}$ and $\T{\Phi}^{a}$ in (\ref{eq:disfHodge}) and (\ref{eq:DisfPhi}) respectively, all the disformal dependence disappears. Thus, there are no $\vU_{i}$ coefficients in (\ref{eq:DisfBH4}). When $\vOP\neq0$ and $\vOX=0$, this Lagrangian maps onto itself plus Horndeski Lagrangians, i.e. $\T{\Lag}_{(021)}\subset \Lag_{(021)},\Lag_{3}^{H},\Lag_{2}^{H}$. In general transformations, when $\vOP,\vOX\neq0$, we obtain the same new combinations of Lagrangians as in the quartic, disformal Horndeski Lagrangian (\ref{eq:DisfH4}).

Regarding the disformal, quintic beyond-Horndeski Lagrangian, $\T{\Lag}_{5}^{BH}=F_{5}(\phi,\T{X})\T{\Lag}_{(031)}$, we obtain that
\begin{equation}
\begin{split}
\T{\Lag}_{(031)}=&\T{\Phi}^{a}\wedge\T{\Phi}^{b}\wedge\T{\Phi}^{c}\wedge\T{\Psi}^{d}\wedge\T{\theta}^{\star}_{\  abcd} \\
=&\gD^{4}\big(\frac{1}{2}\vO^{3}((1-X\vOX)(2\Lag_{(031)}-6\vOX\Lag_{(0\J{2}1)}+3\vOX^{2}\Lag_{(0\I{1}1)})-2X^{2}\vOX(2\Lag_{(030)}-6\vOX\Lag_{(0\J{2}0)}+3\vOX^{2}\Lag_{(0\I{1}0)})) \\
-&3X\vO^{2}\vOP((X\vOX-1)\Lag_{(021)}+(2-3X\vOX)(\Lag_{(021)}-\vOX\Lag_{(0\J{1}1)}))+6X^{3}\vO^{2}\vOP\vOX(2\Lag_{(020)}-3\vOX\Lag_{(0\J{1}0)}) \\
-&36\Lag_{3}^{H}[X^{2}\vO\vOP^{2}\vOX]+6\Lag_{3}^{NH}[X^{2}\vO\vOP^{2}(1-3X\vOX)]-6\Lag_{2}^{NH}[X^{3}\vOP^{3}]\big)\,.
\end{split}
\label{eq:DisfBH5}
\end{equation}
In the same fashion, this Lagrangian maps onto itself when the conformal factor is constant. Moreover, when $\T{C}=\T{C}(\phi)$, then $\T{\Lag}_{(031)}$ maps onto itself plus the other beyond Horndeski term $\Lag_{(021)}$ plus Horndeski Lagrangians, i.e. $\T{\Lag}_{(031)}\subset \Lag_{(031)},\Lag_{(021)},\Lag_{3}^{H},\Lag_{2}^{H}$. In addition, for a full disformal transformation, there appear extra combinations of Lagrangians, which are equal to the ones of the quintic, disformal Horndeski theory (\ref{eq:DisfH5}).

%-------
%DISFORMAL LOVELOCK
%-------
\subsection{From Lovelock to Horndeski}
\label{subsec:Lovelock}

In addition to the previous disformal transformations of scalar-tensor theories, one could think of applying them to other kinds of gravity theories. At this point, the best option is to consider Lovelock theory \cite{Lovelock:1971yv} because it is already contained in the general basis of Lagrangians (\ref{eq:L}) by simply setting $m=n=0$ and fixing $\alpha_{lmn}$ to a constant.

In four dimensions, Lovelock theory $\Lag^{L}$ is characterized by three pieces: the cosmological constant (or volume element) $\Lag_{0}^{L}=\Lag_{(000)}=\theta^{\star}$, the Einstein-Hilbert action $\Lag_{1}^{L}=\Lag_{(100)}=\R^{ab}\wedge\theta^{\star}_{ab}$ and the Gauss-Bonnet $\Lag_{2}^{L}=\Lag_{(200)}=\R^{ab}\wedge\R^{cd}\wedge\theta^{\star}_{abcd}$. We have already computed the disformal volume element and Einstein-Hilbert with the calculation of $\T{\Lag}_{2}^{H}$ in (\ref{eq:DisfH2}) and $\T{\Lag}_{4}^{H}$ in (\ref{eq:DisfH4}). We only need to particularize setting the coefficients $\T{G}_{2}$ and $\T{G}_{4}$ to constants. The result is the same set of Lagrangians but with simplified coefficients. At this level, one gets at most to the quartic Horndeski Lagrangian. Therefore, in order to complete the transformation of Lovelock theory, we only need to analyze the disformal Gauss-Bonnet. Of course, it is well-known that such Lagrangian is indeed a topological term, thus being insensitive to any change in the local geometry. Accordingly, we do not expect to obtain any dynamics from the field redefinition of the vielbein. One can show this in general by noting that the disformal Gauss-Bonnet action
\begin{equation}
\T{S}_{GB}=\int\T{\R}^{ab}\wedge\T{\R}^{cd}\wedge\T{\theta}^{\star}_{abcd}=\int\T{\D}\T{\omega}^{ab}\wedge\T{\D}\T{\omega}^{cd}\wedge\T{\theta}^{\star}_{abcd}
\label{eq:DisfGB}
\end{equation}
shares the same properties than the original one. Namely, it does not depend on the disformal vielbein $\T{\theta}^{a}$ since $\T{\R}^{ab}$ depends only on the connection and $\T{\theta}^{\star}_{\  abcd}=\theta^{\star}_{\  abcd}=\epsilon_{abcd}$. This means that the Lagrangian is not affected by the local geometry. Then, recalling that we are working with a metric-compatible and torsionless disformal theory, one finds that the equations of motion identically vanish using $\T{D}{\R}^{ab}=\T{\D}\T{\theta}^{\star}_{\  abcd}=0$.

Alternatively, one can show that the disformal Gauss-Bonnet remains a topological term explicitly, computing the transformed Lagrangians. Such a calculation is very lengthy. However, with the differential form approach, it becomes doable. As an example, we consider the special disformal transformation of the Gauss-Bonnet Lagrangian, i.e. $\vOX=\vUX=0$. Going directly to the results, we obtain that
\begin{equation}
\begin{split}
\T{\Lag}_{(200)}=&\Lag_{6}^{NH}[\vO]+2\Lag_{5}^{NH}[\vUp-\vOPx]-2\Lag_{5}^{H}[\vOP+2X\vOPx]+\Lag_{4}^{NH}[\vO\vOP^{2}-2\vOPp]+2\Lag_{4}^{H}[X\vO\vOP^{2}] \\
+2&\Lag_{3}^{NH}[X\vOP^{2}(\vUp-3\vOPx)-\vOP^{3}+2\vO\vOP\vOPp]-6\Lag_{3}^{H}[X\vOP^{2}(\vOP+2X\vOPx)]-6\Lag_{2}^{NH}[X\vOP^{2}\vOPp].
\end{split}
\label{eq:DisfLov}
\end{equation}
Then, using that $\Lag_{i}^{NH}$ are related with $\Lag_{i}^{H}$ \cite{Ezquiaga:2016nqo}, this result may indicate that we can obtain the full Horndeski theory from the disformal Lovelock theory, meaning that we can get here the quintic Horndeski Lagrangian that was missing in the disformal volume element or Einstein-Hilbert. Nevertheless, we are going to prove now that this is not the case, so that all the disformal Gauss-Bonnet contributions are indeed total derivatives. For doing that, we need to recall the exact forms that relates $\Lag_{i}^{NH}$ with $\Lag_{i}^{H}$ presented in \cite{Ezquiaga:2016nqo}. They correspond to 
\begin{align}
&\Lag_{6}^{NH}[\frac{3X}{2}G_{5}]=-\Lag_{5}^{NH}[3G_{5,\phi}]+\D\Lag_{(030)}^{D-1}[G_{5,X}]+3\D\Lag^{D-1}_{(110)}[G_{5}]\,, \\
&\Lag_{5}^{NH}[G_{4,X}]=-\Lag_{4}^{NH}[G_{4,\phi}]-\Lag_{5}^{H}[2XG_{4,X}+G_{4}]+\D\Lag_{(020)}^{D-1}[G_{4,X}]+\D\Lag^{D-1}_{(100)}[G_{4}]\,, \\
&\Lag_{4}^{NH}[G_{3}]=-2\Lag_{3}^{NH}[G_{3,\phi}]-2\Lag_{4}^{H}[XG_{3}]+2\D\Lag_{(010)}^{D-1}[G_{3}]\,, \\
&\Lag_{3}^{NH}[G_{2,X}]=-\Lag_{2}^{NH}[G_{2,\phi}]-\Lag_{3}^{H}[(G_{2}+2XG_{2,X})]+\D\Lag_{(000)}^{D-1}[G_{2}]\,. 
\end{align}
Then, it is easy to show that all the terms from (\ref{eq:DisfLov}) add up to form all the above exact forms. This follows
\begin{equation}
\begin{split}
\T{\Lag}_{(200)}=&\Lag_{6}^{NH}[1+X\vU]+2\Lag_{5}^{NH}[\vUp-\vOPx]-2\Lag_{5}^{H}[\vOP+2X\vOPx]+\Lag_{4}^{NH}[\vO\vOP^{2}-2\vOPp]+2\Lag_{4}^{H}[X\vO\vOP^{2}] \\
&+2\Lag_{3}^{NH}[X\vOP^{2}(\vUp-3\vOPx)-\vOP^{3}+2\vO\vOP\vOPp]-6\Lag_{3}^{H}[X\vOP^{2}(\vOP+2X\vOPx)]-6\Lag_{2}^{NH}[X\vOP^{2}\vOPp] \\
=&\Lag_{(200)}+\frac{2}{3}\D\Lag_{(030)}^{D-1}[\vUx]+2\D\Lag_{(110)}^{D-1}[\vU]-2\D\Lag_{(020)}^{D-1}[\vOPx]-2\D\Lag_{(100)}^{D-1}[\vOP] \\
&+2\D\Lag_{(010)}^{D-1}[\vO\vOP^{2}]-2\D\Lag_{(000)}^{D-1}[X\vOP^{3}]\,,
\end{split}
\end{equation}
which proves that the special disformal Gauss-Bonnet transforms to the Gauss-Bonnet plus total derivatives. 

Altogether, this result implies that one cannot generate the full Horndeski theory from the Lovelock action via disformal transformation. In order to get this mapping, it is necessary to relax some of the initial condition. For instance, it has been studied in the literature \cite{VanAcoleyen:2011mj} that Galileon actions can arise from Lovelock theory in five dimensions applying a standard Kaluza-Klein compactification, i.e. a diagonal metric in the extra dimension with no additional vector field. In that situation, the Gauss-Bonnet is no longer a topological term and it can generate dynamics.
%-------
%SEC. IV: HORNDESKI'S ORBITS
%-------
\section{Horndeski's Orbits}
\label{sec:Orbits}
%------
%FIGURE: Horndeski's Orbits
\begin{figure}[t!]
\vspace{-10pt}
\begin{center}
\includegraphics[width=.6\textwidth]{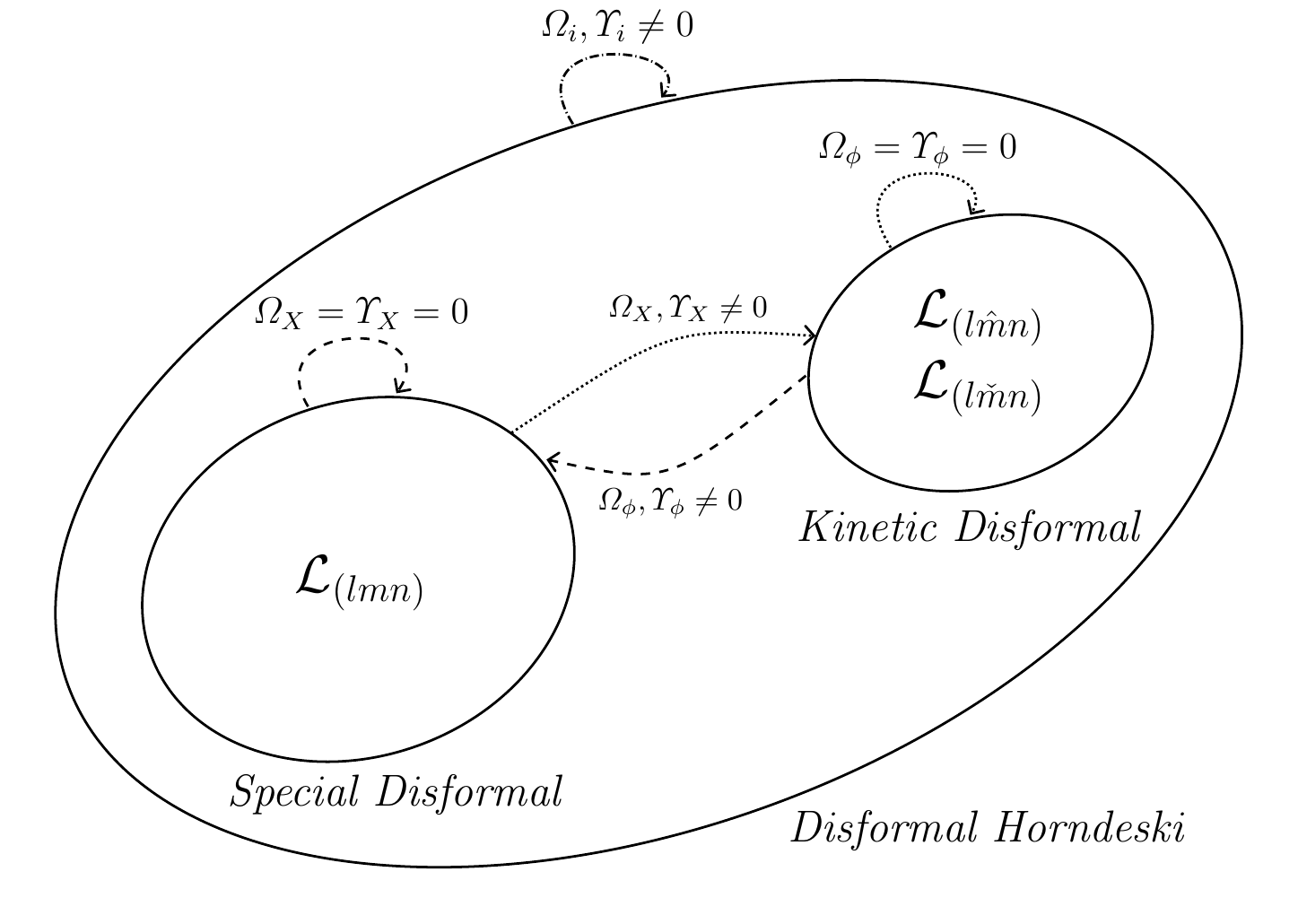}
\end{center}
\vspace{-15pt}
\caption{Diagram of Horndeski's orbits. Each closed solid line represents a different orbit, which is defined as a set of theories invariant under a certain type of disformal transformations. The arrows indicate the connections of different orbits through specific field redefinitions defined by the coefficients $\vO_{i}$ and $\vU_{i}$ (see Tab. \ref{tab:xi}). There are three main orbits: the special disformal ($\vOX=\vUX=0$), the kinetic disformal ($\vOP=\vUP=0$) and the disformal Horndeski ($\vO_{i},\vU_{i}\neq0$). The general basis of scalar-tensor Lagrangians $\Lag_{(lmn)}$ is defined in (\ref{eq:L}) and its extensions $\Lag_{(l\J{m}n)}$ and $\Lag_{(l\I{m}n)}$ are presented in Sec. \ref{subsec:ExtendedBasis}.}
\label{fig:HorndeskiOrbits}
\end{figure}
%------
In the previous section, we have computed the disformal transformation of Horndeski theory. When the disformal transformation is special, i.e. when there is no dependence on $X$ in the disformal coefficients and thus $\vOX=\vUX=0$, we obtain that the transformed theory is also Horndeski, as it was first found in Ref. \cite{Bettoni:2013diz}. This result can be extended to any Lagrangian of the form $\Lag_{(lmn)}$ since, for special disformal transformations, the building blocks do not introduce new elements. Therefore, the set $\Lag_{(lmn)}$ could be defined as the set of Lagrangians invariant under special disformal transformations. Following this logic, we denote this set of Lagrangians as the \emph{special disformal orbit}.

When the transformation is fully general, implying that there is a kinetic dependence in the disformal coefficients, new Lagrangians arise that are not part of the starting set. This defines the \emph{first Horndeski's orbit}, i.e. the set of theories that are disformally related to Horndeski theory but which do not belong to it. In this section, we are going to compute the disformal transformation of this first Horndeski's orbit. Our objective is to find if a \emph{second Horndeski's orbit} exits or the first one closes under disformal transformations. Of course, if one performs a disformal transformation on a disformal Horndeski theory, one will conclude that the final theory is also a disformal Horndeski theory. This is because disformal transformations can be composed so that one could define two consecutive transformations as only one with new coefficients\footnote{Being explicit, a disformal transformation will modify the vielbein as $\T{\theta}_{(1)}^{a}[\theta,\Psi]=\T{C}_{1}\theta^{a}+\T{D}_{1}\Psi^{a}$. If we perform a second one, then we obtain $\T{\theta}_{(2)}^{a}[\T{\theta}_{(1)},\T{\Psi}_{(1)}]=\T{C}_{2}\T{\theta}_{(1)}^{a}+\T{D}_{2}\T{\Psi}_{(1)}^{a}=(\T{C}_{2}\T{C}_{1})\theta^{a}+(\T{C}_{2}\T{D}_{1}+\T{D}_{2}\gamma_{1})\Psi^{a}$, where $\T{\Psi}_{(1)}^{a}=\gamma_{1}\Psi^{a}$ and $\gamma_{1}=(\T{C}_{1}-2X\T{D}_{1})^{-1}$. Accordingly, one could condense the two redefinitions in one by defining $\T{\theta}_{(12)}^{a}\equiv \T{C}_{12}\theta^{a}+\T{D}_{12}\Psi^{a}$ with $\T{C}_{12}=\T{C}_{2}\T{C}_{1}$ and $\T{D}_{12}=\T{C}_{2}\T{D}_{1}+\T{D}_{2}\gamma_{1}$. Thus, several disformal transformations can be composed, but they do not commute. In general the commutator of two disformal transformations reads $\T{\theta}_{(12)}^{a}-\T{\theta}_{(21)}^{a}=((\T{C}_{2}-\gamma_{2})\T{D}_{1}-(\T{C}_{1}-\gamma_{1})\T{D}_{2})\Psi^{a}$. Note that even a purely conformal and a purely disformal transformation do not commute.}. However, it is not a trivial question what the disformal transformation of the extended basis is. We are interested here in determining if new Lagrangians can be generated.

In order to compute the disformal transformation of the extended basis, we need to determine first how the extended building blocks transform. For the case of $\J{\Psi}^{a}$, which was defined in Sec. \ref{subsec:ExtendedBasis}, it turns out that it changes in a very similar way to the other building blocks $\Phi^{a}$ and $\Psi^{a}$, namely $\HPs_{\mathrm{disf}}^{a}=\beta_{\HPs}\HPs^{a}+\beta_{\BPh}\BPh^{a}+\beta_{\BPs}\BPs^{a}+\beta_{\Psi}\Psi^{a}$ (cf. Table \ref{tab:notation} for a summary of this notation). For the case of $\I{\Psi}^{a}=\pP\J{\Psi}^{a}$, we only need to compute in addition the transformation of $\pP$ that yields $\langle\T{\Phi}\rangle=\chi_{1}+\chi_{2}\pP$, where the precise form of these $\beta_{i}$ and $\chi_{i}$ coefficients is presented in Eqs. (\ref{eq:Disf11}-\ref{eq:DisfpP}) of Appendix \ref{app:Computations}. Some of these coefficients will have a dependence on second-derivative scalars such as $\pP$. However, once we plug the disformal building blocks into the disformal Lagrangian, we can eliminate these dependences thanks to the algebraic identities that relate different Lagrangians. These identities, which are summarized in Appendix \ref{subapp:AntisymId}, are a consequence of the antisymmetric structure of the Lagrangians. With them, we are able to express the disformal transformations of the extended basis as a linear combination of Lagrangians of the original and extended basis with coefficients that depend only in $\phi$ and $X$.

Interestingly, whenever we perform a kinetic disformal transformation, $\vOP=\vUP=0$ (cf. Table \ref{tab:xi}), the disformal extended building blocks generate only extended building blocks. As a result, if we perform a kinetic disformal transformation of an element of the extended basis, we will obtain another one of the same class. This means that this set closes under this type of field redefinitions defining the \emph{kinetic disformal orbit}.

Altogether, these results yield an important consequence, i.e. the disformal transformation of the extended basis of Lagrangians will not generate new terms outside of the extended and the original set. Consequently, the first Horndeski's orbit closes under disformal transformations, meaning that there is only one \emph{disformal Horndeski's orbit}. This allows us to classify the different sets of Lagrangians, constructed with our building blocks, with respect to their invariance under disformal transformations in the following manner: 
\begin{enumerate}[(a)] 
\item \emph{Special disformal orbit:} set of Lagrangians that is invariant under special disformal transformations, i.e. $\vOX=\vUX=0$. In four dimensions, this set corresponds to Horndeski theory, which is built with the Lagrangians $\Lag_{(lmn)}$.
\item \emph{Kinetic disformal orbit:} set of Lagrangians that is invariant under kinetic disformal transformations, i.e. $\vOP=\vUP=0$. In four dimensions, it is composed of $\Lag_{(0\J{1}0)}$, $\Lag_{(0\J{1}1)}$, $\Lag_{(0\J{2}0)}$, $\Lag_{(0\J{2}1)}$, $\Lag_{(0\I{1}0)}$ and $\Lag_{(0\J{1}1)}$.
\item \emph{Disformal Horndeski's orbit:} set of Lagrangians that is invariant under disformal transformations. It is formed by the special and kinetic disformal orbits. It defines also the group of Lagrangians that can be disformally related to Horndeski theory.
\end{enumerate}
This classification is also schematically presented in Fig. \ref{fig:HorndeskiOrbits}.

More importantly, since we have found a closed set under disformal transformation, any theory constructed with a Lagrangian that does {\em not} belong to this set cannot be disformally related. This means that if we construct a theory with a Lagrangian that is not contained in the disformal Horndeski orbit, this theory cannot be connected with Horndeski via disformal transformations. This is the case of one class of Extended Scalar-Tensor theories \cite{Crisostomi:2016czh,BenAchour:2016fzp}. When expressing this theory in our basis of Lagrangians, we encounter that the constraints in the parameters simplifies. We then find that, for instance, the class \textsuperscript{2}N-IIIi of quadratic Lagrangians in EST \cite{Crisostomi:2016czh} is characterized for having the Lagrangian $\Lag_{(01^{2}0)}$. This Lagrangian does not belong to the disformal Horndeski's orbits and thus cannot be related with Horndeski via general disformal transformations. We discover a similar situation for the \textsuperscript{3}N-II class of cubic theories \cite{BenAchour:2016fzp}. For completeness, we present the details of this calculation in Appendix \ref{subapp:EST}.

%-------
%SEC. V: EXTENDING (BEYOND) HORNDESKI THEORIES
%-------
\section{Extending (Beyond) Horndeski Theories}
\label{sec:ExtendedDisformal}

Up to now, we have investigated different aspects of disformal transformations. As stressed along this work, this kind of field redefinitions is characterized for being linear in the vielbein (metric) and for containing up to first derivatives of the scalar. Originally, Bekenstein himself refrained from considering transformations with higher derivatives of the scalar with the intention of avoiding higher derivatives in the matter EoM \cite{Bekenstein:1992pj}. However, it has been recently realized \cite{Zumalacarregui:2013pma,Gleyzes:2014dya,Langlois:2015cwa} that there are viable theories with higher derivative EoM. Therefore, it seems reasonable to at least reconsider this initial Ansatz in light of the present knowledge. In the following, we are going to introduce how disformal transformations could be generalized to include second derivatives of the scalar field. This will be an extension of some of the arguments of Ref. \cite{Zumalacarregui:2013pma}, where this kind of transformations were first proposed. Then, we will analyze if these transformations could be related with some scalar-tensor theories beyond Horndeski. As we will see, the differential form formalism will be very advantageous for this extension of the disformal transformations since it allows an easy generalization in terms of the other building blocks. 
%-------
%EXTENDED DISFORMAL TRANSFORMATIONS
%-------
\subsection{Extended Disformal Transformations}
\label{subsec:ExtendedDisformal}
%--
A field redefinition can be classified either by its tensor structure of by its order in derivatives \cite{Zumalacarregui:2013pma}. In components notation, a disformal transformation is built with a vector that can be identified with the first derivatives of the scalar $\phi_{,\mu}$. If one wants to add a tensor contribution to the transformation, one is forced to introduce second derivatives $\phi_{;\mu\nu}$. An analogous situation happens in the context of transformations of the vielbein. Accordingly, the simplest possibility to enlarge a disformal transformation consists on adding the 1-form that encodes second derivatives, $\Phi^{a}$, in the redefinition of the vielbein $\theta^{a}$. This yields
\begin{equation}
\T{\theta}^{a}=\T{C}\theta^{a}+\T{D}\Psi^{a}+\T{E}\Phi^{a}\,.
\label{eq:EPhivierbein}
\end{equation}
However, a field redefinition of this kind posses a serious drawback. An inverse map cannot be found with the same simple form of (\ref{eq:EPhivierbein}), since it must contain an infinite number of terms. This is because the new term $\Phi^{a}$ is formed with a tensor with two indices rather than two vector fields such as $\Psi^{a}$. Let us see how we arrive to this conclusion (cf. Appendix \ref{app:InverseExtended} for the explicit calculation). The transformed vielbein is defined as $\T{\theta}^{a}=\T{e}^{a}_{\  \mu}dx^{\mu}$. Its inverse is obtained from the condition $\T{e}^{a}_{\  \mu}\T{e}^{\mu}_{\  b}=\delta^{a}_{b}$, which is a direct consequence of the definition of the metric tensor (\ref{eq:disfMetric}). Now, if we postulate the inverse vielbein $\T{e}^{\mu}_{\  b}$ to be of the same form of (\ref{eq:EPhivierbein}) but with exchanged indices, i.e. $\T{e}^{\mu}_{\  b}=\T{\alpha}e^{\mu}_{\  b}+\T{\beta}\nabla^{\mu}\phi\nabla_{b}\phi+\T{\gamma}\nabla^{\mu}\nabla_{b}\phi$, we find that there is no solution. This is due to the terms with second derivatives in the vielbein and its inverse that, when contracted together, form a unique term quadratic in second derivatives with coefficient $\T{E}\cdot\T{\gamma}$. Therefore, in order to satisfy the condition $\T{e}^{a}_{\  \mu}\T{e}^{\mu}_{\  b}=\delta^{a}_{b}$, the coefficient $\T{\gamma}$ must be set to zero. Nevertheless, if this coefficient vanishes, the other ones must vanish too because there are no terms to compensate the second derivatives of $\T{E}\Phi^{a}$. Clearly, we arrive at the trivial solution. The only way to obtain an actual inverse is by introducing an infinite serie of powers of second derivatives. In that case, one can find the coefficients of the inverse recursively without difficulties.%
\footnote{A similar (although much simpler) problem appears when considering special disformal transformations with no $X$ dependence. In that case the contravariant metric includes a factor $\gD$ that depends explicitly on $X$, and hence looses the special property.}

Furthermore, one arrives at the same conclusion for any transformation containing higher powers of second derivatives contracted $\nabla_{\mu}\nabla^{\alpha_{1}}\phi\nabla_{\alpha_{1}}\nabla_{\alpha_{2}}\phi\cdots\nabla^{\alpha_{n-1}}\nabla_{\nu}\phi$. In the differential forms language, this is the case for any transformation containing the generalized 1-form $\lp\Phi^{n}\rp^{a}$ that is defined as a $n$-th contraction of second derivatives, cf. Tab. \ref{tab:notation} and Appendix \ref{app:HigherBasis} for more details. With this approach, there is a straightforward way to determine that the inverse exist, that is, showing that the determinant of the vielbein is not zero. The advantage is that the determinant is directly given by the wedge product of $D$ basic blocks $\T{\theta}^{a}$, as we did in (\ref{eq:disfDet}) for the disformal transformation. In general, the transformed Hodge dual basis will be given by
\begin{equation}
\begin{split}
\T{\theta}^{\star}_{\  a_{1}\cdots a_{k}}&=\sum_{j=0}^{D-k}\frac{1}{(D-k-j)!}\T{C}^{j}\T{E}^{D-k-j}\cdot\Phi^{a_{k+1}}\wedge\cdots\wedge\Phi^{a_{D-j}}\wedge\theta^{\star}_{\  a_{1}\cdots a_{D-j}} \\
&+\sum_{j=0}^{D-1-k}\frac{1}{(D-1-k-j)!}\T{C}^{j}\T{D}\T{E}^{D-1-k-j}\cdot\Phi^{a_{k+1}}\wedge\cdots\wedge\Phi^{a_{D-1-j}}\wedge\Psi^{a_{D-j}}\wedge\theta^{\star}_{\  a_{1}\cdots a_{D-j}}\,,
\end{split}
\label{eq:ExtendedHodge}
\end{equation}
which can be deduced in analogy to the well-known binomial formula. Noticeably, the main difference with respect to the disformal case, cf. (\ref{eq:disfHodge}), is that now we have all possible combination with the second derivative 1-form, since the wedge product of two or more $\Phi^{a}$ is non-zero, in contrast to the case of $\Psi^{a}$. Subsequently, we only need to particularize to $k=0$ to obtain 
\begin{equation}
\T{\theta}^{\star}=\T{\Lag}_{(000)}=\sum_{j=0}^{D}\frac{1}{(D-j)!}\T{C}^{j}\T{E}^{D-j}\Lag_{(0(D-j)0)}+\sum_{j=0}^{D-1}\frac{1}{(D-1-j)!}\T{C}^{j}\T{D}\T{E}^{D-1-j}\Lag_{(0(D-1-j)1)}\,,
\end{equation}
where the different Lagrangians are defined in (\ref{eq:L}). This means that the square root of the determinant of the metric $\T{\theta}^{\star}=\sqrt{-\T{g}}d^{D}x$ will be composed of second derivatives contracted up to a power equal to the dimension $D$. Generically, this expression will not be zero and the inverse will exist. Nevertheless, such inverse will be an infinite series, as it can be inferred from the inverse determinant.

Alternatively, one could avoid this issue with the inverse considering higher order basis elements $\lp\Psi^{mn}\rp^{a}$ in the transformations. These elements are formed of second derivatives contractions contracted with gradient fields, as explicitly shown in Appendix \ref{app:HigherBasis}. The generalized disformal transformation will then be 
\begin{equation}
\begin{split}
\tilde{\theta}^{a}=&\tilde{C}\theta^{a}+\tilde{D}_{nm}\left(\Psi^{mn}\right)^{a} \\
=&\tilde{C}\theta^{a}+\tilde{D}_{00}\Psi^{a}+\tilde{D}_{01}\BPh^{a}+\tilde{D}_{10}\BPs^{a}+\tilde{D}_{11}\HPs^{a}+\mathcal{O}(n,m>1)\,,
\end{split}
\label{eq:Evierbein}
\end{equation}
where in the first line we are assuming Einstein's summation convention and in the second one we are introducing the notation of Sec. \ref{sec:Disformal} summarized in Tab. \ref{tab:notation}. These field redefinitions can be generically denoted as \emph{extended disformal transformations}\footnote{In components notation, such a transformation would read $\tilde{g}_{\mu\nu}=Cg_{\mu\nu}+D_{nm}\phi^{,\rho}\left.\Phi^{m}\right._{\rho(\mu}\left.\Phi^{n}\right._{\nu)\gamma}\phi^{,\gamma}$, where a $n$-th power contraction of second derivatives is defined by $\left.\Phi^{n}\right._{ab}=\phi_{;az_{1}}\left.\phi^{;z_{1}}\right._{;z_{2}}\cdots\left.\phi^{;z_{n-1}}\right._{;b}$. There is a one-to-one correspondence between $D_{nm}$ and $\T{D}_{nm}$. For all coefficient but $D_{00}$ being zero, one recovers the usual disformal metric.}.
One should notice that the coefficient $\T{D}_{00}$ represents the previous disformal coefficient $\T{D}$. Also, every $\T{D}_{nm}$ coefficient will have a different mass dimension to compensate the extra powers of second derivatives. Subsequently, we can repeat the process of finding an inverse. The main difference is that when contracting two $\Psi^{nm}$ one does {\em not} get a term of higher order in second derivatives. Instead, one obtains $\lp\Psi^{mn}\rp^{a}_{\  b}\lp\Psi^{pq}\rp^{b}=\langle\Phi^{n+p}\rangle\lp\Psi^{mq}\rp^{a}$, which introduces an extra factor $\langle\Phi^{n+p}\rangle$. Thus, one can find an inverse. Nevertheless, its coefficients will be, in general, functions of scalars with $n$ powers of second derivatives $\langle\Phi^{n}\rangle$. This means that this kind of transformations (\ref{eq:Evierbein}) represent \emph{viable extensions} of the disformal transformations in which both the vielbein and its inverse are described with finite series of extended building blocks.

Before continuing, there are some points to remark about these extended disformal transformations. From the effective field theory perspective, every higher power term in the expansion will be suppressed by the cutoff scale of the theory. This implies that much below this cutoff scale an extended disformal transformation will effectively behave as a disformal one. Furthermore, whenever there is a second derivative of the scalar, we are introducing derivatives of the vielbein (or the metric in component notation). Consequently, the transformation is no longer linear in the vielbein. In addition, the fact that the inverse coefficients are also functions of second derivative scalars such as $\pP$ will have important consequences in the connection of these transformations with present scalar-tensor theories. We will analyze these possible relations in the next Sec. \ref{subsec:EST}. Lastly, let us point out that metrics with second derivatives of the scalar appear in various places. For instance, they correspond to the effective background metric for the propagation of gravitational waves in quintic Horndeski theory \cite{Bettoni:2016mij}. Moreover, it shows up as the typical replacement for the metric perturbations when applying the St\"uckulberg trick to theories with massive gravitons \cite{ArkaniHamed:2002sp}.
%-------
%IDENTIFYING EXTENDED ST
%-------
\subsection{Identifying Extended Horndeski theories}
\label{subsec:EST}
%--
Once you start with a well-defined scalar-tensor theory, it is possible to create a new healthy theory by applying a field redefinition. This field redefinition must be invertible and non-singular for the number of degrees of freedom to be preserved.%
\footnote{An interesting appliation of the contrary is mimetic gravity \cite{Chamseddine:2013kea} and its generalizations theories, which are based on a non-invertible conformal or disformal relation \cite{Arroja:2015wpa}}
Conversely, one can use a field redefinition to relate a novel theory with a previously studied one. In the context of disformal transformations, this avenue has been very productive. Previously, we have seen that a general disformal transformation of Horndeski theory introduce Lagrangians beyond the original setup. These Lagrangians constitute one of the two viable classes of Extended Scalar-Tensor (EST) theories \cite{Langlois:2015cwa,Crisostomi:2016czh,BenAchour:2016fzp} both at quadratic and cubic order. In fact, we show in Appendix \ref{subapp:EST} how to rewrite the constraints defining each sub-class of EST theories in terms of the our basis of Lagrangians. We find that they simplify in our basis, becoming clear that one sub-family corresponds directly to the disformal transformation of Horndeski theory. Still, as pointed out in Ref. \cite{deRham:2016wji}, the other class of EST theories could also be related to a previously studied theory via a-yet-to-be-found field redefinition. Accordingly, the question is: could this be achieved with an extended disformal transformation?

In the previous section, we have found that the field redefinition (\ref{eq:Evierbein}) is an invertible extension of a disformal transformation. Nevertheless, we have also discovered that its inverse contains coefficients which are functions of second derivative scalars such as $\pP$. This can be already studied in the lowest order extended disformal transformation
\begin{equation}
\T{\theta}^{a}=\T{C}(\phi,X)\theta^{a}+\T{D}(\phi,X)\Psi^{a}+\T{E}(\phi,X)\BPh^{a}\,.
\label{eq:ExtendedLowest}
\end{equation}
The effect of the inverse vielbein can be captured in the transformed gradient $\T{\nabla}_{a}\phi=\T{e}_{a}^{\  \mu}\nabla_{\mu}\phi$ or more conveniently in the transformed 1-form $\T{\Psi}^{a}=\T{\nabla}^{a}\phi\T{\D}\phi=\eta^{ab}\T{\nabla}_{b}\phi\D\phi$. For the above transformation, this building block becomes
\begin{equation}
\T{\Psi}^{a}=\frac{1}{\T{C}(\T{C}-2X\T{D}+\pP\T{E})}((\T{C}+\pP\T{E})\Psi^{a}+2X\T{E}\BPs^{a})\,.
\label{eq:ExtendedPsi}
\end{equation}
We observe that, in contrast with a disformal transformation, this transformation can increase the power of second derivative terms. Furthermore, it contains non-polynomial functions of $\pP$. This will happen too when transforming the other basic building block $\T{\Phi}^{a}=\T{\D}(\T{\nabla}^{a}\phi)$. As a consequence, a generic extended disformal transformation of a ``Galileon-like" Lagrangian $\Lag_{(lmn)}$ will lead to Lagrangians of the extended basis but with non-polynomial functions of $\pP$. Therefore, these transformations \emph{cannot} generate EST theories because these theories are constructed solely with monomials/polynomials of second derivative terms of different powers (see Appendix \ref{subapp:EST}). This is an important result and highlights the singularity of the recently proposed EST theories. 
%-------
%SEC. VI: CONNECTING GRAVITY THEORIES
%-------
\section{Connecting Gravity Theories}
\label{sec:ConnectingMG}

In addition to the extended disformal transformations presented in the previous section, we would like to discuss other possible routes to generalize disformal transformations. Specifically, one could consider adding fields with different spin combined to give the appropriate tensor structure \cite{Zumalacarregui:2013pma}. In the following, we will investigate field redefinitions in which an arbitrary number of scalars are included. Afterwards, we will study field redefinitions in which spin-1 and spin-2 fields are present. This exercise would serve us both to emphasize the intrinsic similarities among different gravity theories and to show the transversality of the differential forms approach. Finally, we will discuss how the  formalism of differential forms could be used in the future to make progress in the study of general Modified Gravity theories. We will concentrate in the case of multi-scalar-tensor theories, general vector-tensor theories and multi-tensor theories.
%---SPIN 0----
\subsection{Mixing with Spin-0 Fields}
\label{subsec:s0}
%--------
The simplest manner to enlarge a disformal transformation with extra fields would be to consider a transformation with $N$ different scalar fields $\phi_{A}$, where capital letters are indices of the internal field space $A=1,\cdots,N$. Such a transformation would read
\begin{equation}\label{eq:DisfMulti}
\T{\theta}^{a}=\T{C}\theta^{a}+\sum_{A=1}^{N}\T{D}_{A}\lp\Psi_{A}\rp^{a}\,,
\end{equation}
where $\lp\Psi_{A}\rp^{a}$ is the 1-form encoding the first derivatives associated to each field $\phi_{A}$. The coefficients of this transformation would be functions of scalars built with the fields and their first derivatives. Applying this kind of transformations, one would end up with a multi-scalar-tensor theory. The main difference with respect to the single scalar case is that now there are many more possible interactions. In particular, an interaction between two first derivative 1-forms $\lp\Psi_{A}\rp^{a}\wedge\lp\Psi_{B}\rp^{b}$ is not zero if they are built with different fields, i.e. if $A\neq B$. This kind of new interaction arises already when transforming the volume element (or cosmological constant). Restricting for simplicity to two scalar fields, one can easily generalize the redefinition of the Hodge dual basis (\ref{eq:disfHodge}) to
\begin{equation}
\T{\theta}^{\star}_{a_{1}\cdots a_{k}}=\T{C}^{D-k-2}\lp\T{C}^{2}\theta^{\star}_{a_{1}\cdots a_{k}}+\T{C}\lp\T{D}_{1}\lp\Psi_{1}\rp^{b}+\T{D}_{2}\lp\Psi_{2}\rp^{b}\rp\wedge\theta^{\star}_{a_{1}\cdots a_{k}b}+\T{D}_{1}\T{D}_{2}\lp\Psi_{1}\rp^{b_{1}}\wedge\lp\Psi_{2}\rp^{b_{2}}\wedge\theta^{\star}_{a_{1}\cdots a_{k}b_{1}b_{2}}\rp\,.
\label{eq:MultiDisfHodge}
\end{equation}
from which we can obtain the transformed volume element setting $k=0$ as  
\begin{equation}
\T{\theta}^{\star}=\T{\Lag}_{(000)}=\T{C}^{D-2}\lp\T{C}^{2}\Lag_{(000)}+\T{C}\T{D}_{1}\Lag_{(00(10))}+\T{C}\T{D}_{2}\Lag_{(00(01))}+\T{D}_{1}\T{D}_{2}\Lag_{(00(11))}\rp\,.
\end{equation}
Here, we are generalizing our single scalar-tensor basis $\Lag_{(lmn)}$ given in (\ref{eq:L}) to
\begin{equation}
\Lag_{(l(m_{1}\cdots m_{N})(n_{1}\cdots n_{N}))}=\bigwedge_{i=1}^{l}\mathcal{R}^{a_{i}b_{i}}\wedge\bigwedge_{J=1}^{N}\bigwedge_{j=1}^{m_{J}}\Phi_{J}^{\  c_{J,j}}\wedge\bigwedge_{K=1}^{N}\bigwedge_{k=1}^{n_{K}}\Psi_{K}^{\  d_{K,k}}\wedge\theta^{\star}_{~a_{1}b_{1}\cdots a_{l}b_{l}c_{1,1}\cdots c_{N,m}d_{1,1}\cdots d_{N,n}}\,.
\label{eq:LMulti}
\end{equation}
Note that when there are non scalar 1-forms we omit the additional parentheses in the subindex of the Lagrangian. Theories with this type of first derivative interactions fall into the class of multi-scalar theories minimally coupled to gravity studied in Ref. \cite{Damour:1992we}. Non-minimally coupled Horndeski-like theories have much richer interactions that includes second derivatives interactions. This leads to a more complex analysis. In the bi-scalar-tensor case, the most general second order equations of motion have been found but they have not been associated to a particular Lagrangian yet \cite{Ohashi:2015fma}. The bi-scalar results have also shown that the previous attempts to construct a multi-scalar-tensor theory were not fully general \cite{Padilla:2012dx,Kobayashi:2013ina}. A possible route to find general multi-scalar-tensor theories could be to apply the program of Ref. \cite{Ezquiaga:2016nqo} to theories constructed with this new basis of Lagrangians (\ref{eq:LMulti}). The advantage would be that the systematic structure of the differential forms language could be easily extended to a theory with additional building blocks.

%---SPIN 1----
\subsection{Mixing with Spin-1 Fields}
\label{subsec:s1}
%--------
More interestingly, one could mix fields with different spins as proposed in \cite{Zumalacarregui:2013pma}. Restricting to integer spins, one could consider adding a spin-1 field to the vielbein. In analogy with the disformal transformation, one could define
\begin{equation}\label{eq:DisfVect}
\T{\theta}^{a}=\T{C}\theta^{a}+\T{V}\A^{a}\,,
\end{equation}
where $\A^{a}=A^{a}\A=A^{a}A_{b}\theta^{b}$ is a 1-form encoding the vector field. Here, $\T{C}$ and $\T{V}$ would be functions of the modulus square of the vector field, which we parametrize for convenience with $-2X_{V}=A^{a}A_{a}$. Noticeably, one can recover a disformal transformation by going to the scalar limit $\A^{a}\rightarrow\Psi^{a}$. Consequently, we can benefit from all the machinery developed for disformal transformations. In this sense, we could introduce a 1-form encoding the first derivatives of the vector field $\lp\Phi_{V}\rp^{a}=\D A^{a}$, which would be analogous to the 1-form describing the second derivatives of the scalar field $\Phi^{a}=\D\nabla^{a}\phi$. With this dictionary in hand, one could translate all the calculations performed in Sec. \ref{sec:DisformalTheories} to obtain results in vector-tensor theories. In particular, if one starts from Lovelock's action and applies the vielbein redefinition (\ref{eq:DisfVect}) with constant coefficients, which is equivalent to a special disformal transformation, the resulting theory would correspond to the generalization of Proca theory in curved space \cite{Heisenberg:2014rta}. The only precaution we must have in the transliteration is that now a covariant derivative does not conmute with the vector field, i.e. $\lp\Phi_{V}\rp^{a}=\D A^{a}\neq \nabla^{a}\A$, in contrast to the case of the gradient field, i.e. $\Phi^{a}=\D\nabla^{a}\phi=\nabla^{a}\D\phi$. Therefore, there would be additional terms with respect to Horndeski Lagrangian. These terms can be parametrized by the 1-form $\F^{a}=\nabla^{a}\A-\D A^{a}$, which can be related with the usual Abelian 2-form field strength $\F=\frac{1}{2}F_{ab}\theta^{a}\wedge\theta^{b}$ via an interior product.\footnote{Specifically, the 2-form field strength is given by $\F=\D\A=\frac{1}{2}(\nabla_{a}A_{b}-\nabla_{b}A_{a})\theta^{a}\wedge\theta^{b}$. Thus, one realizes that $\F^{a}=\mathfrak{i}_{\delta^{a}}\F$, where $\delta^{a}\equiv\eta^{ab}\theta_{b}$ and $\mathfrak{i}_{X}$ is the interior product operator that maps $p$-forms onto $(p-1)$-forms by contracting the indices with the vector field $X$. One can find more details about this operation in Ref. \cite{Ezquiaga:2016nqo} or in any mathematical textbook such as \cite{nakahara2003geometry}.} With this additional building block, one could follow the same principles used in Ref. \cite{Ezquiaga:2016nqo} for scalar-tensor theories to build a general vector-tensor theory.

Field redefinitions of this kind have been studied in tensorial notation in \cite{Kimura:2016rzw}. They have also been used to screen the cosmological vector field \cite{BeltranJimenez:2013fca} and to study Weyl geometry as a vector-tensor theory \cite{Jimenez:2014rna,Jimenez:2015fva}. However, they do not represent the most general transformation since they do not incorporate derivatives of the vector field. In this respect, some field redefinitions with derivatives have been investigated for conformal transformations in \cite{EspositoFarese:2009aj} and for disformal transformations in \cite{Goulart:2013laa}. The key point is to construct a covariant field redefinition that involves derivatives of the vector without introducing derivatives of the metric. Thus, the transformation must be constructed in terms of antisymmetric combinations of covariant derivatives of the vector field. From the differential forms perspective, this could be achieved by including the aforementioned 1-form $\F^{a}$ in the vielbein redefinition
\begin{equation}\label{eq:DisfVectE}
\T{\theta}^{a}=\T{C}\theta^{a}+\T{V}\A^{a}+\T{W}\F^{a}\,,
\end{equation}
where the coefficients $\T{C}$, $\T{V}$ and $\T{W}$ could depend on the modulus of the vector field and scalars formed with contractions of the field strength. Interestingly, these transformations would have some structural similarities with the extended disformal transformations discussed in Sec. \ref{sec:ExtendedDisformal}. However, there is a main difference since this transformation is only first order in derivatives. By its own interest, the analysis of this kind of field redefinitions should be addressed elsewhere.

In addition, one could think on combining both scalars and vectors fields. For the simplest case, this was studied in the context of TeVeS \cite{Bekenstein:2004ne,Bruneton:2007si}, where the field redefinition that generates the scalar-vector-tensor interactions reads\begin{equation}
\T{\theta}^{a}=\T{C}(\phi)\theta^{a}+\T{V}(\phi)\A^{a}\,.
\end{equation}
This represents a generalization of the simple vector-tensor transformation (\ref{eq:DisfVect}) in which the coefficients are allowed to depend on an additional scalar field $\phi$. Another place where there is an interplay of spin-0, spin-1 and spin-2 fields is when a general dimensional reduction in the Kaluza-Klein framework is considered. In that case, the vector field arises from the non-diagonal components of the metric in the extra dimension. This would account for an extension of the analysis presented in Sec. \ref{subsec:Lovelock} in which the field redefinition is allowed to depend in $\theta^{a},\  \Psi^{a}$ and $\A^{a}$. This might constitute an interesting avenue to study scalar-vector-tensor theories.

%---SPIN 2----
\subsection{Mixing with Spin-2 Fields}
\label{subsec:s2}
%--------
Finally, one could think on adding extra spin-2 fields. For simplicity, we will restrict to just one additional spin-2 field, but the following construction can be easily generalized to several spin-2 fields. In the language of differential forms, adding additional spin-2 fields means adding extra vielbeins \cite{Hinterbichler:2012cn}. We will denote the usual gravitational vielbein, to which matter is coupled, as before $\theta^{a}$ and the second dynamical vielbein will be encoded in $\Theta^{a}$. Thus, a transformation that mixes both fields would be
\begin{equation}\label{eq:DisfTensor}
\T{\theta}^{a}=\T{C}\theta^{a}+\T{F}\Theta^{a}\,,
\end{equation}
where we set $\T{C}$ and $\T{F}$ to be constants.%
\footnote{This kind of vielbein redefinitions have been studied in the context of couplings of massive (bi-)gravity to matter \cite{deRham:2014naa,Noller:2014sta,deRham:2014fha}. In tensorial notation, the metric transforms as $\T{g}_{\mu\nu}=\T{C}^{2}g_{\mu\nu}+2\T{C}\T{F}g_{\mu\alpha}X^{\alpha}_{\  \nu}+\T{F}^{2}f_{\mu\nu}$, where $f_{\mu\nu}$ is the second metric and $X^{\mu}_{\  \alpha}X^{\alpha}_{\  \nu}=g^{\mu\alpha}f_{\alpha\nu}$. Note that this type of coupling to matter leads to a theory that is different from minimally coupled bigravity.}
However, this transformation will share the same issues with the inverse of the extended disformal transformation (\ref{eq:EPhivierbein}), i.e. it cannot be generically written in a finite, polynomial form. This will affect, for instance, to the definition of the curvature 2-form. One way out of this situation is to impose the symmetric vielbein condition \cite{Hinterbichler:2012cn}. Then, as we explicitly show in Appendix \ref{app:InverseS2}, a simple inverse can be found.  Alternatively, one may consider the fields as perturbations, such as in \cite{Noller:2014ioa}. Here, in order to keep the discussion simple and general, we are not going to specialize in any of those situations. Thus, we parametrize the transformed curvature as $\T{\R}^{ab}=\R^{ab}+\R_{\Omega}^{\  \  ab}$, where $\R_{\Omega}^{\  \  ab}=\D \Omega^{ab}+\Omega^{a}_{\  c}\wedge \Omega^{cb}$ contains the 2-form curvature associated to the second vielbein $\Theta^{a}$ and also derivative interactions of the two vielbeins, e.g. $\omega^{a}_{\  c}\wedge \Omega^{cb}$, where $\Omega^{ab}=\T{\omega}^{ab}-\omega^{ab}$ is the difference of the new connection and the original one.
 
As in the previous cases, we can forecast the structure of the theory by applying this field redefinition to Lovelock theory. Starting with the volume element, or the zero-th order Lovelock Lagrangian $\Lag_{(000)}$, one finds that all possible combinations of $\theta^{a}$ and $\Theta^{a}$ are present. This can be deduced from the transformation of the extended Hodge dual basis $\T{\theta}^{\star}_{\  a_{1}\cdots a_{k}}$ presented in (\ref{eq:ExtendedHodge}) by eliminating the disformal part and changing $\Phi^{a}\rightarrow\Theta^{a}$ and $\T{E}\rightarrow\T{F}$. We obtain, for a $D$ dimensional spacetime, that
\begin{equation}
\T{\Lag}_{(000)}=\T{\theta}^{\star}=\sum_{j=0}^{D}\frac{1}{(D-j)!}\T{C}^{j}\T{F}^{D-j}\Theta^{a_{1}}\wedge\cdots\wedge\Theta^{a_{D-j}}\wedge\theta^{\star}_{\  a_{1}\cdots a_{D-j}}\,. 
\end{equation}
Similarly for the Einstein-Hilbert Lagrangian, we find
\begin{equation}
\T{\Lag}_{(100)}=\T{\R}^{ab}\wedge\T{\theta}^{\star}_{\  ab}=\sum_{j=0}^{D-2}\frac{1}{(D-j-2)!}\T{C}^{j}\T{F}^{D-j-2}(\R^{ab}+\R_{\Omega}^{\  \  ab})\wedge\Theta^{a_{3}}\wedge\cdots\wedge\Theta^{a_{D-j-2}}\wedge\theta^{\star}_{\  aba_{3}\cdots a_{D-j-2}}\,. 
\end{equation}
In order to exemplify this general result, we particularize to four dimensions. Then, from the volume element we obtain the following interactions
\begin{equation}
\T{\Lag}_{(000)}=\T{F}^{4}\Theta^{\star}+\frac{1}{3!}\T{C}\T{F}^{3}\Theta^{a}\wedge\Theta^{b}\wedge\Theta^{c}\wedge\theta_{\  abc}^{\star}+\frac{1}{2!}\T{C}^{2}\T{F}^{2}\Theta^{a}\wedge\Theta^{b}\wedge\theta_{\  ab}^{\star}+\T{C}^{3}\T{F}\Theta^{a}\wedge\theta_{\  a}^{\star}+\T{C}^{4}\theta^{\star}\,, 
\end{equation}
which are nothing but the well-known symmetric polynomials appearing in massive gravity and bi-gravity \cite{Hassan:2011vm}. The last term is the cosmological constant term and the third one would lead to a mass term for the graviton if the second vielbein is fixed to be non-dynamical $\Theta^{a}=\delta^{a}_{\  \mu}dx^{\mu}$. On the other hand, the first Lovelock Lagrangian $\Lag_{(100)}$ transforms in four dimensions to
\begin{equation}
\T{\Lag}_{(100)}
=\R^{ab}\wedge\theta^{\star}_{ab}+\R^{ab}\wedge\theta^{c}\wedge\Theta^{\star}_{abc}+\R^{ab}\wedge\Theta^{\star}_{ab}+\R_{\Omega}^{\  \  ab}\wedge\theta^{\star}_{ab}+\R_{\Omega}^{\  \  ab}\wedge\Theta^{c}\wedge\theta^{\star}_{abc}+\R_{\Omega}^{\  \  ab}\wedge\Theta^{\star}_{ab}\,,
\end{equation}
where $\Theta^{\star}_{\  a_{1}\cdots a_{k}}$ is the Hodge dual basis constructed with the second vielbein $\Theta^{a}$. Here, the first term corresponds to the usual kinetic term of GR and the last one contains the other kinetic term of bi-gravity \cite{Hassan:2011zd} plus some derivative interactions. One should notice that the other terms also involve derivative interactions. Nevertheless, since this theory arises from a well-defined field redefinition, there are not additional ghost degrees of freedom. The question of the existence of a full non-linear theory with derivative interactions different from the one presented above, which arises from a field redefinition, is still open. There have been arguments claiming that this is not possible \cite{deRham:2013tfa,deRham:2015rxa}, although some new interactions were found in the pseudo-linear regime \cite{Hinterbichler:2013eza}.

Altogether, these results can be used to connect the vielbein formalism in massive gravity, bi-gravity or, generically, multi-gravity \cite{Hinterbichler:2012cn} with the differential form approach to scalar-tensor theories \cite{Ezquiaga:2016nqo}. Looking forward, one could implement the Hamiltonian vielbein formalism, already developed in multi-gravity, to investigate the number of degrees of freedom in general scalar-tensor theories. This would constitute a new avenue to count the physical degrees of freedom that should be addressed elsewhere. 
%-------
%SEC. VII: DISCUSSION
%-------
\section{Discussion}
\label{sec:Discussion}

Present and future data from experimental setups and astrophysical and cosmological observations demand a good understanding of alternative theories of gravity to test against General Relativity. In this wide landscape of models, scalar-tensor (ST) theories appear as the simplest modification of Einstein's theory of gravity since only one extra degree of freedom is incorporated. However, in the search of general ST theories, one needs to deal with quite complex and particular interactions to keep the theory theoretically consistent. In this sense, it is both instructive and practical to seek for connections between different theories in order to gain insights of their theoretical construction and possible simplifications in the calculations. This can be achieved by studying different field redefinitions.

In this work, we have investigated the role of field redefinitions in ST gravity. Our novel approach is based on the  formalism of differential forms developed in Ref. \cite{Ezquiaga:2016nqo}. This formalism naturally accounts for the specific interactions of general ST theories and allows for a systematic analysis of their viability. Since we are interested in ST theories with derivative interactions, we have considered field redefinitions linear in the spin-2 field but including derivatives of the spin-0 field. We began with transformations including first derivatives of the scalar, i.e. \emph{disformal} transformations. Then, we have analyzed \emph{extended disformal} transformations, which includes up to second derivatives of the scalar. Finally, we have explored field redefinitions with more DoF with \emph{different spin}, namely several spin-0 fields, spin-1 fields and spin-2 fields. All these cases represent an excellent example of the great economy of means of the differential forms language, which simplifies the calculations and clarifies the analysis.

In this framework, every field is defined in the tangent space. The fundamental object describing the geometry of space-time is the vielbein, which can be encoded in the 1-form $\theta^{a}=e^{a}_{\ \mu}dx^{\mu}$. Accordingly, we have first introduced a disformal transformation of the vielbein in the tangent space, i.e. $\T{\theta}^{a}=\T{C}\theta^{a}+\T{D}\Psi^{a}$. Interestingly, this transformation includes the basic building block containing first derivatives of the scalar $\Psi^{a}$ that was used to construct general ST Lagrangians in \cite{Ezquiaga:2016nqo}. With this transformation in hand, we have computed the disformal transformation of the connection 1-form $\omega^{a}_{\  b}$. This is essential to obtain the transformation of the other building blocks, the 1-form with second derivatives of the scalar $\Phi^{a}$ and the curvature 2-form $\R^{ab}$. Conveniently, the disformal 1-form connection (\ref{eq:disfConnection}) is formed by different terms constructed with the basis elements $\theta^{a}$, $\Psi^{a}$ and $\Phi^{a}$, combined with derivatives of the scalar. Each term can be parametrized with a coefficient $\vO_{i}$ or $\vU_{i}$ that indicates respectively if it is sourced by the disformal parameter $\T{C}$ or $\T{D}$, or by their dependence in the scalar field $i=\phi$ or its first derivatives $i=X$. A summary of these coefficients for each type of disformal transformation can be found in Tab. \ref{tab:xi}. This is an important result because it allows us to identify directly the origin of each Lagrangian in terms of the disformal coefficients. As a consequence, we can conclude that all new Lagrangians beyond Horndeski theory \cite{Horndeski:1974wa} will be generated by the kinetic dependence of the disformal transformation, i.e. by $\vOX$ and $\vUX$. This is due to the fact that, in the transformation of the original basis elements, the only new building blocks are provided by these coefficients. These additional elements, classified in Tab. \ref{tab:notation}, highlight the necessity to enlarge the basis of Lagrangians (\ref{eq:L}) to accommodate their disformal transformations.

Thanks to the clear mapping of each building block, the calculation of the disformal transformation of a given Lagrangian simplifies greatly. Along this article, we have performed a tour through different disformal theories. First, we have computed the full disformal Horndeski theory. The outcome of this calculation partially overlaps previous works, as we discuss in Sec. \ref{sec:DisformalTheories}. The interest of our result resides in both the novelty of the approach and the completeness of the calculation, since we obtain at once the most general transformation for the most general Horndeski theory. In addition, we have presented the complete disformal transformation of the beyond-Horndeski Lagrangians \cite{Gleyzes:2014dya}. We find that the invariance under purely disformal transformations of these Lagrangians can be understood in a very transparent way within the language of differential forms. Lastly, we have performed a special disformal transformation of Lovelock theory \cite{Lovelock:1971yv} in four dimensions. We explicitly show that the Gauss-Bonnet maps onto itself plus exact forms (total derivatives), as it should since it is a topological term. This is a good example of how our formalism simplifies the calculations and the importance of knowing the possible relations between different ST theories, which could be either algebraic identities or exact forms \cite{Ezquiaga:2016nqo}.

Moreover, we have identified the different \emph{Horndeski's orbits}, i.e. the different sets of Lagrangians that are invariant under certain types of disformal transformations (see Fig. \ref{fig:HorndeskiOrbits}). For that purpose, we have computed the disformal transformation of the new Lagrangians that arise in the disformal Horndeski theory. We encounter that these Lagrangians are invariant under kinetic disformal transformations, $\vOP=\vUP=0$, defining the kinetic disformal orbit. This is consistent with the results of Ref. \cite{Achour:2016rkg}. From the previous calculation, we can infer that Horndeski theory is invariant under special disformal transformations, $\vOX=\vUX=0$, thus defining the special disformal orbit. This result, which was known since Ref. \cite{Bettoni:2013diz}, can be easily understood within our formalism because the transformed building blocks of the original basis $\theta^{a}$, $\Psi^{a}$, $\Phi^{a}$ and $\R^{ab}$ only introduce new terms via kinetic dependence. Altogether, these two orbits combine to form a closed orbit, the disformal Horndeski orbit, which englobes all Lagrangians that can be disformally related to Horndeski.

It is interesting that these new Lagrangians that appear in the disformal Horndeski theory have a particular structure. This structure is triggered by the kinetic dependence of the conformal coefficient, i.e. by $\vOX$. It is possible to distinguish if the new Lagrangians are originated by the presence of a disformal term or just by the conformal factor. When there is only kinetic dependence in the disformal sector, $\vOX=0$ and $\vUX\neq0$, one recovers Horndeski theory plus the beyond-Horndeski Lagrangians $\Lag_{(021)}$ and $\Lag_{(031)}$. These combinations of non-Horndeski Lagrangians can be related with the degenerate ST theories or Extended Scalar-Tensor (EST) theories \cite{Langlois:2015cwa,Crisostomi:2016czh,BenAchour:2016fzp}. In fact, when the EST constraints are presented in our basis of Lagrangians, their form substantially simplifies, cf. Appendix \ref{subapp:EST} for the comparison. On the one hand, the class of EST theories disformally related to Horndeski directly reproduce the aforementioned combinations of non-Horndenski Lagrangians. On the other hand, the family of EST theories not related to Horndeski always include a Lagrangian of the extended basis that does not belong to the disformal Horndeski orbit. This is a clear sign that this class of theories cannot be disformally related to Horndeskit.

In addition to the analysis of disformal transformations, we have also explored generalizations of these field redefinitions. Following the proposal of Ref. \cite{Zumalacarregui:2013pma}, we have considered transformations with higher order in derivatives of the scalar and with new tensor structures, which can be accommodated by adding additional fields. The former can be easily achieved by including other 1-forms basis elements, apart from $\Psi^{a}$, in the vielbein redefinition. In this sense, the first choice has been to incorporate the basic building block with second derivatives $\Phi^{a}$. However, we find that this kind of term and its higher order powers, $\lp\Phi^{n}\rp^{a}$, are not convenient since the inverse vielbein would require an infinite series of terms. This problem is not present when the second derivatives are introduced via contractions with first derivatives, i.e. with the generalized building blocks $\lp\Psi^{mn}\rp^{a}$. This represents an example of a well-defined extended disformal transformations. To our knowledge, such transformations have not been previously studied in the literature. Thus, we have analyzed further the lowest order case, $m=0$ and $n=1$, in order to determine if this transformation can be used to relate Horndeski theory with the class of EST theories not disformally related to it. We find that they cannot be generically linked because the transformed Lagrangian would include non-polynomial functions of second derivative contractions such as $\pP$, and the EST Lagrangians are constructed solely with monomials/polynomials of second derivative terms. This result stresses the distinctiveness of the recently proposed EST theories.

Field redefinitions with different types of fields open an interesting avenue to connect gravity theories. In this work, we have presented how field redefinitions of the vielbein with several spin-0 fields, with spin-1 fields and spin-2 fields can be written in the language of differential forms, discussing their relation with different theories such as multi-scalar-tensor theories \cite{Damour:1992we}, generalized Proca theory \cite{Heisenberg:2014rta} or bi-gravity \cite{Hassan:2011zd}. We find that there are interesting lines of research for future progress. In the case of multi-scalar theories, one could apply the program of Ref. \cite{Ezquiaga:2016nqo} to systematically determine the set of Lagrangians with second order EoM. For theories with spin-1 fields, one could construct general vector-tensor actions substituting the scalar gradient 1-form $\Psi^{a}$ by a 1-form encoding the vector field $\A^{a}$. To be complete, one would also need to include a 1-form $\F^{a}$ related to the field strength of the field. This new element could be used to construct a field redefinition with first derivatives of the spin-1 field. Such transformation may play a similar role to disformal transformations in ST gravity but for vector-tensor theories. Lastly, one could use field redefinitions with several spin-2 fields to connect the vielbein formulation of multi-gravity theories \cite{Hinterbichler:2012cn} with the differential forms language of ST gravity \cite{Ezquiaga:2016nqo}. Along these lines, one could apply the Hamiltonian vielbein analysis, already developed in this kind of theories, to ST gravity. This would represent a new manner of counting the number of physical DoF in ST theories.

It is important to note that, in the analysis of field redefinitions with additional fields, we have focus our attention to the gravitational sector. Considering this sector alone, both the original and the transformed theories are dynamically equivalent. Thus, no new DoF propagates. In order to notice the presence of these additional fields, one needs a \textquotedblleft detector\textquotedblright. Such detector corresponds to the matter Lagrangian. If the transformed theory couples to the matter sector with a different vielbein than the field redefined one or with a different type of coupling than the original one, both theories will no longer be equivalent. As a consequence, additional propagating DoF could be probed. Of course, since the equivalence is broken, a proper study of the number of DoF and their stability should be performed. Inclusion of the matter Lagrangian and its transformations may be done in a future publication.\footnote{Here the difference between the first and second order actions may be relevant. That is, $S_1 =-m\int d\tau(-\dot X^\mu\dot X_\mu)^{1/2}$ and $S_2 =1/2 \int d\tau(e^{-1}\dot X^\mu\dot X_\mu-e m^2)$, where $\dot X^{\mu}\equiv dX^{\mu}(\tau)/d\tau$ and $e$ is the einbein (1D vielbein). Both are world-line reparametrization invariant and Poincare invariant in the target space, but the second one is valid even for massless particles, and also easier to implement in the Path Integral \cite{Polchinski:1998rq}. Moreover, $S_2$ can be generalized much more easily under general disformal transformations than $S_1$.}

Altogether, field redefinitions have been a very useful tool to understand gravity since the inception of the Jordan-Brans-Dicke theory. More recently, an extension of these ideas has yielded further understanding of the larger classes of Horndeski and beyond Horndeski theories. The elegance and compactness of the language of differential forms have allowed us to simplify, unify and generalize these results, providing a new tool to delineate and understand the landscape of alternative gravitational theories.
%--
%Acknowledgements
\subsection*{Acknowledgements}
%--
We would like to thank Jose Beltr\'{a}n and Kurt Hinterbichler for useful correspondence in particular topics of this article. This work is supported by the Research Projects of the Spanish MINECO, FPA2013-47986-03-3P and FPA2015-68048-C3-3P, and the Centro de Excelencia Severo Ochoa Program SEV-2012-0249. J.M.E. is supported by the FPU Grant No. FPU14/01618 and thanks the hospitality of Nordita during his visit. M.Z. thanks IFT-UAM-CSIC for hospitality during the completion of this work. This article is based upon work from COST Action CA15117 (CANTATA), supported by COST (European Cooperation in Science and Technology). The computations of the disformal transformations in tensorial notation have been checked using the xAct package for Mathematica \cite{Brizuela:2008ra,xAct}.
%----------------------------------------------------------------------------------
%----------------------------------------------------------------------------------
%---
%APPENDICES
%---
\appendix
%---
%HIGHER ORDER BASIS
%---
\section{Extended basis}
\label{app:HigherBasis}
Here, we extend the fundamental building blocks of our basis of scalar-tensor Lagrangians $\Lag_{(lmn)}$ allowing for contractions with vectors constructed with arbitrary number of first and second derivatives of the scalar field, as introduced in the Appendix B of Ref. \cite{Ezquiaga:2016nqo}. The generalizations of the 1-form encoding the second derivatives of the scalar $\Phi^{a}$ and the first derivative 1-form $\Psi^{a}$ are correspondingly
\begin{align} \label{eq:higher blocks}
&\left(\Phi^{n}\right)^{a}\equiv\left.\Phi^{n}\right.^{a}_{~b}\theta^{b} & &\mathrm{and} & &\lp\Psi^{mn}\rp^{a}\equiv\left.\Phi^{m}\right.^{a}_{~b}\phi^{,b}\phi^{,c}\left.\Phi^{n}\right._{cd}\theta^{d}
\end{align}
where $\left.\Phi^{n}\right._{ab}=\phi_{;az_{1}}\left.\phi^{;z_{1}}\right._{;z_{2}}\cdots\left.\phi^{;z_{n-1}}\right._{;b}$ is a $n$-th power contraction of second derivatives, $\phi_{;ab}=\nabla_{a}\nabla_{b}\phi$ is a second derivative and $\phi_{,a}=\nabla_{a}\phi=\partial_{a}\phi$ is a first derivative. With these new building blocks and imposing invariance under LLT, a generalized version of $\Lag_{(lmn)}$ was presented in \cite{Ezquiaga:2016nqo}. One should notice that not all combinations of (\ref{eq:higher blocks}) are possible. In fact, whenever there are two (or more) terms $\lp\Psi^{mn}\rp^{a}$ with the same $m$ or $n$, the total will be zero due to antisymmetry. A summary of this notation is placed in Tab. \ref{tab:notation}.
%--
\subsection{Lagrangians}
\label{subapp:Lag}

For the purpose of analyzing disformal transformations, it is only necessary to consider some of the terms of the extended basis (\ref{eq:higher blocks}), provided that a disformal transformation does not change the power of second derivatives, as it can be deduced of the transformation of the building blocks in Sec. \ref{subsec:DisformalBasis}. When dealing with quartic Horndeski, only $\lp\Psi^{10}\rp^{a}$, $\lp\Psi^{01}\rp^{a}$ and $\lp\Psi^{11}\rp^{a}$ are needed. The first two could be alternatively seen as contractions of $\Phi^{a}$ with the gradient field $\nabla^{a}\phi$, which correspond to the Lagrangians $\Lag_{(l\bar{m}n)}$ defined in Ref. \cite{Ezquiaga:2016nqo}. Thus, these terms are denoted by $\BPh^{a}\equiv\lp\Psi^{01}\rp^{a}$ and $\BPs\equiv\lp\Psi^{10}\rp^{a}$. To be complete, one needs to allow similar operations with the other terms, leading to $\Lag_{(\bar{l}mn)}$. Accordingly, the new term $\J{\Psi}^{a}\equiv\lp\Psi^{11}\rp^{a}$ could be interpreted as a contraction with $\nabla^{a}\nabla_{z}\phi\nabla^{z}\phi$. Lagrangians built with such a contraction will be denoted by a circumflex accent (an over hat) in the corresponding element, becoming $\Lag_{(l\hat{m}n)}$ or $\Lag_{(\hat{l}mn)}$ as e.g. in (\ref{eq:Lhat}). Since we only want to go up to cubic order, the component expressions of the relevant new Lagrangians are
\begin{align}
\Lag_{(0\J{1}0)}&=\J{\Psi}^{a}\wedge\theta^{\star}_{\  a}\,=\pPP\cdot\eta\,, \\
\Lag_{(0\J{1}1)}&=\J{\Psi}^{a}\wedge\Psi^{b}\wedge\theta^{\star}_{\  ab}\,=(-2X\pPP-\pP^{2})\cdot\eta\,, \\
\Lag_{(0\J{2}0)}&=\J{\Psi}^{a}\wedge\Phi^{b}\wedge\theta^{\star}_{\  ab}\,=(\pPP\tP-\pPPP)\cdot\eta\,, \\
\Lag_{(0\J{2}1)}&=\J{\Psi}^{a}\wedge\Phi^{b}\wedge\Psi^{c}\wedge\theta^{\star}_{\  abc}\,=(\pP(\pPP-\pP\tP)-2X(\pPP\tP-\pPPP))\cdot\eta\,,
\end{align}
where $\eta$ is the volume element in four dimensions $\eta=\theta^{\star}_{4D}=\theta^{1}\wedge\cdots\wedge\theta^{4}$.

In the case of quintic Horndeski, one should consider a further generalization, which corresponds to contractions with two second derivative terms. These Lagrangians will be represented with a Czech accent over the contracted terms, e.g. $\Lag_{(l\I{m}n)}$ or $\Lag_{(\I{l}mn)}$. The Lagrangians appearing in the calculations read
\begin{align}
\Lag_{(0\I{1}0)}&=\I{\Psi}^{a}\wedge\theta^{\star}_{\  a}\,=\pP\Lag_{(0\hat{1}0)}=\pP\pPP\cdot\eta\,, \\
\Lag_{(0\I{1}1)}&=\I{\Psi}^{a}\wedge\Psi^{b}\wedge\theta^{\star}_{\  ab}\,=\pP\Lag_{(0\hat{1}1)}=(-2X\pP\pPP-\pP^{3})\cdot\eta\,, 
\end{align}
where the extended basis element corresponds to $\I{\Psi}^{a}\equiv\nabla^{a}\nabla_{z}\phi\nabla^{z}\phi\nabla^{y}\phi\nabla_{y}\nabla_{x}\phi\lp\Psi^{01}\rp^{x}$.

If one has the intention of describing all possible linear combinations of Lagrangians with a given power of second derivatives, then, it is necessary to consider $\lp\Phi^{r}\rp^{a}$, defined in (\ref{eq:higher blocks}), as a building block. When there is only one higher power element of order $r$ in the Lagrangian, one could denote it with a super-index in the basis, i.e. $\Lag_{(lm^{r}n)}$. For the purpose of linking the differential form language used here and in \cite{Ezquiaga:2016nqo} with the notation of works related with Extended Scalar-Tensor (EST) theories \cite{Achour:2016rkg,Crisostomi:2016czh,BenAchour:2016fzp}, see Sec. \ref{subapp:EST}, we will need four Lagrangians of $\Lag_{(lm^{r}n)}$, i.e.
\begin{align}
\Lag_{(01^{2}0)}&=\lp\Phi^{2}\rp^{a}\wedge\theta^{\star}_{\  a}\,=\tPP\cdot\eta\,, \\
\Lag_{(02^{2}0)}&=\lp\Phi^{2}\rp^{a}\wedge\Phi^{b}\wedge\theta^{\star}_{\  ab}\,=(\tP\tPP-\tPPP)\cdot\eta\,, \\
\Lag_{(02^{2}1)}&=\lp\Phi^{2}\rp^{a}\wedge\Phi^{b}\wedge\Psi^{c}\wedge\theta^{\star}_{\  abc}\,=(2\pPPP-\pPP\tP-\pP\tPP-2X(\tP\tPP-\tPPP))\cdot\eta\,, \\
\Lag_{(01^{3}0)}&=\lp\Phi^{3}\rp^{a}\wedge\theta^{\star}_{\  a}\,=\tPPP\cdot\eta\,. 
\end{align}
In order to complete the correspondence, we have to introduce one Lagrangian more. Its novelty resides in the element $\lp\Psi^{12}\rp^{a}$. For shortness, when there is one such term, we will name the Lagrangian with an over dot $\Lag_{(l\dot{m}n)}$. The Lagrangians required is  
\begin{align}
\Lag_{(0\dot{1}0)}&=\lp\Psi^{12}\rp^{a}\wedge\theta^{\star}_{\  a}\,=\pPPP\cdot\eta\,,
\end{align}
In conclusion, using linear combinations of this set of Lagrangians, one can build any arbitrary action up to cubic order in second derivatives. We will use this fact to relate this basis with the one of EST theories in Appendix \ref{subapp:EST}.

%--
\subsection{Antisymmetric Identities}
\label{subapp:AntisymId}

Importantly, these new Lagrangians of the extended basis are not all linearly independent. There exist antisymmetric identities that relate them, in analogy with the results of \cite{Ezquiaga:2016nqo}. The difference is that now, since we are dealing with Lagrangians with higher power of second derivatives, the relations will include factors of the type of $\pP$. In general terms, these algebraic identities among Lagrangians can be written as
\begin{align}
&\Lag_{(l(\overline{m+1})n)}=\pP\Lag_{(lmn)}-2l\Lag_{(\J{l}mn)}-m\Lag_{(l\J{m}n)}+2Xn\Lag_{(l(\overline{m+1})(n-1))}\,, \label{eq:extanti1} \\
&\Lag_{(l\J{m}1)}=-2X\Lag_{(l\J{m}0)}-2l\Lag_{(\B{l}\J{m}0)}-\pP \Lag_{(l\B{m}0)}\,, \label{eq:extanti2} \\
&\Lag_{(l(\widehat{m+1})n)}=\pPP\Lag_{(lmn)}-2l\Lag_{(\I{l}mn)}-m\Lag_{(l\I{m}n)}-n\pP\Lag_{(l(\overline{m+1})(n-1))}\,.\label{eq:extanti3}
\end{align}
Subsequently, one can obtain any relation particularizing for the appropriate $l,\  m,\  n$. For instance, if we would like to simplify a second derivative factor acting in front of $\Lag_{(010)}$, one could use the following identities
\begin{align}
\pP\Lag_{(010)}&=\Lag_{(0\B{2}0)}+\Lag_{(0\J{1}0)}\,, \\
\pPP\Lag_{(010)}&=\Lag_{(0\J{2}0)}+\Lag_{(0\I{1}0)}\,, \\
\pP^{2}\Lag_{(010)}&=-2X(\Lag_{(0\J{2}0)}+\Lag_{(0\I{1}0)})-(\Lag_{(0\J{2}1)}+\Lag_{(0\I{1}1)})\,,
\end{align}
which are obtained from (\ref{eq:extanti1}), (\ref{eq:extanti3}) and a repeated action of (\ref{eq:extanti1}) respectively.

Lastly, let us emphasis that these antisymmetric identities are very useful for calculating the disformal transformation of a given Lagrangian. The reason is that some of the coefficients of the transformed building blocks $\vO_{i}$, $\vU_{i}$, $\lambda_{i}$ and $\alpha_{i}$ contain factors of $\pP$. Therefore, we can use these relations to rewrite the outcome in terms of Lagrangians of the original and extended basis (see Sec. \ref{subsec:ExtendedBasis}). 

%----
\subsection{Exact Forms}
\label{subapp:ExactForms}

However, we cannot only relate Lagrangians of a general basis with algebraic identities but also with exact forms. An exact form is the formal analog of a total derivative. For constructing them, we need to know first how a covariant, exterior derivative $\D$ acts on every building block. The exterior derivatives of the extended building blocks are
\begin{align}
&\D[\BPh^{a}]=\Phi^{a}\wedge\Phi^{z}\nabla_{z}\phi\,, \\
&\D[\BPs^{a}]=\lp\Phi^{2}\rp^{a}\wedge\D\phi+\nabla^{a}\Phi^{z}\wedge\Psi_{z}\,, \\
&\D[\HPs^{a}]=\lp\Phi^{2}\rp^{a}\wedge\Phi^{z}\nabla_{z}\phi+\nabla^{a}\Phi^{z}\nabla_{z}\phi\wedge\Phi^{y}\nabla_{y}\phi \,.
\end{align}
With this, one can construct general exact forms. Allowing for contractions with $\nabla^{e}\phi$ only, one obtains
\begin{equation}
\begin{split}
\D\B{\Lag}_{(lmn)}^{D-1}[G_{i}]=&\D\lp G_{i}\bigwedge_{i}\R^{a_{i}b_{i}}\wedge\bigwedge_{j}\Phi^{c_{j}}\bigwedge_{k}\Psi^{d_{k}}\wedge\theta^{\star}_{\  ea_{i}b_{i}c_{j}d_{k}}\nabla^{e}\phi\rp \\
&=G_{i,\phi}\Lag_{(lm(n+1))}-G_{i,X}\Lag_{(l(\overline{m+1})n)}+G_{i,\pP}\D(\pP)\nabla^{e}\phi\wedge[\Lag_{(lmn)}]_{e}\\
&+G_{i}\left(\Lag_{(l(m+1)n)}-m\Lag_{((\overline{l+1})(m-1)n)}-n\Lag_{(l(m+1)n)}\right)\,,
\end{split}
\end{equation}
where one could further extend this formula, if convenient, using that $\D[\pP]=2\lp\Phi^{2}\rp^{z}\nabla_{z}\phi+\nabla_{z}\phi\nabla^{z}\Phi^{y}\nabla_{y}\phi$. Alternatively, one can make use of the interior product of $\Phi^{a}$, which is nothing but a reduction to its component form, namely $\mathfrak{i}_{\nabla\phi}\Phi^{a}=\nabla^{a}\nabla_{z}\phi\nabla^{z}\phi$, to lower the order of the Lagrangian to a $(D-1)$-form, where $D$ is the dimension of the space-time. With the latter, one can build the following exact forms
\begin{equation}
\begin{split}
\D\J{\Lag}_{(lmn)}^{D-1}[G_{i}]=&\D\lp G_{i}\bigwedge_{i}\R^{a_{i}b_{i}}\wedge\mathfrak{i}_{\nabla\phi}\Phi^{e}\wedge\bigwedge_{j}\Phi^{c_{j}}\bigwedge_{k}\Psi^{d_{k}}\wedge\theta^{\star}_{\  ea_{i}b_{i}c_{j}d_{k}}\rp \\
&=G_{i,\phi}\Lag_{(l(\overline{m+1})n)}-G_{i,X}\Lag_{(l(\widehat{m+1})n)}+G_{i,\pP}\D(\pP)\nabla^{e}\nabla_{z}\phi\nabla^{z}\phi\wedge[\Lag_{(lmn)}]_{e} \\
&+G_{i}\lp\D(\nabla^{e}\nabla_{z}\phi\nabla^{z}\phi)\wedge[\Lag_{(lmn)}]_{e}-m\Lag_{((\widehat{l+1})(m-1)n)}-n\Lag_{(l(\overline{m+2})(n-1))}\rp\,.
\end{split}
\end{equation}
Equivalently to the case of the antisymmetric identities, previous section, we could construct any desired exact form just by setting a specific $l,\ m,\  n$. Again, these relations among Lagrangians have an utility when computing the disformal transformation of a scalar-tensor theory. Specifically, the exact forms are necessary for computing the transformation of Lagrangians including the curvature. This is because the disformal 2-form curvature $\T{\R}^{ab}$ contains terms with covariant exterior derivatives, cf. (\ref{eq:DisfCurvature}). With the above relations, one is able to rewrite the full disformal theory only with Lagrangians of the original and extended basis.

%--
\subsection{Relation with Extended Scalar-Tensor theories}
\label{subapp:EST}
As discussed in Sec. \ref{subsec:ExtendedBasis}, there are different possibilities to construct general scalar-tensor theories. Along this work, we have constructed Lagrangians using a basis based on the antisymmetry of the differential forms. On the other hand,
works related to the study of degenerate scalar-tensor theories \cite{Achour:2016rkg,Crisostomi:2016czh,BenAchour:2016fzp}, also named Extended Scalar-Tensor (EST) theories, write the action of the theory in terms of monomials with different powers of second derivatives of the scalar field. At quadratic order \cite{Crisostomi:2016czh}, the EST action $S_{EST}^{(2)}=\int d^{4}x\sqrt{-g}\sum_{i=1}^{5}a_{i}L_{i}^{(2)}$ is given by the following Lagrangians densities 
\begin{align}
L_{1}^{(2)}&=\tPP\,, & L_{2}^{(2)}&=\tP^{2}\,, & L_{3}^{(2)}&=\tP\pP\,, & L_{4}^{(2)}&=\pPP\,, &\mathrm{and} &  & L_{5}^{(2)}=\pP^{2}\,. \nonumber
\end{align}
This set of Lagrangian densities $L_{i}^{(2)}$ can be directly linked with our basis of Lagrangians, recalling that in our notation $-2X=\nabla_{\mu}\phi\nabla^{\mu}\phi$. In particular, those ones that are linearly independent and quadratic in second derivatives can be rewritten as
\begin{align}
&L_{1}^{(2)}\cdot\eta=\Lag_{(01^{2}0)}\,, \\
&L_{2}^{(2)}\cdot\eta=\Lag_{(020)}+\Lag_{(01^{2}0)}\,, \\
&L_{3}^{(2)}\cdot\eta=-\frac{1}{2}\Lag_{(021)}-X\Lag_{(020)}+\Lag_{(0\hat{1}0)}\,, \\
&L_{4}^{(2)}\cdot\eta=\Lag_{(0\hat{1}0)}\,, \\
&L_{5}^{(2)}\cdot\eta=-\Lag_{(0\hat{1}1)}-2X\Lag_{(0\hat{1}0)}\,,
\end{align}
where $\eta$ is the volume element and the component expression of these Lagrangians is given in Sec. \ref{subapp:Lag}. Noticeably, the Lagrangian $\Lag_{(01^{2}0)}$ does not appear in the disformal transformations. Nevertheless, it is needed to form a complete basis of Lagrangians at quadratic order. As we will see, it will characterize those degenerate quadratic scalar-tensor theories that cannot be related to Horndeski with a disformal transformation.

The corresponding Lagrangians densities $L_{i}^{(3)}$ of the cubic EST theory $S_{EST}^{(3)}=\int d^{4}x\sqrt{-g}\sum_{i=1}^{10}b_{i}L_{i}^{(3)}$ \cite{BenAchour:2016fzp} read
\begin{align}
L_{1}^{(3)}&=\tP^{3}\,, & L_{2}^{(3)}&=\tP\tPP\,, & L_{3}^{(3)}&=\tPPP\,, & L_{4}^{(3)}&=\tP^{2}\pP\,, & L_{5}^{(3)}&=\tP\pPP\,, \nonumber \\
L_{6}^{(3)}&=\tPP\pP\,, & L_{7}^{(3)}&=\pPPP\,, & L_{8}^{(3)}&=\pPP\pP\,, & L_{9}^{(3)}&=\tP\pP^{2}\,, &L_{10}^{(3)}&=\pP^{3}\,. \nonumber  
\end{align}
In the same manner, they can be easily related with our extended basis of Lagrangians through
\begin{align}
&L_{1}^{(3)}\cdot\eta=\Lag_{(030)}+3\Lag_{(02^{2}0)}+\Lag_{(01^{3}0)}\,, \\
&L_{2}^{(3)}\cdot\eta=\Lag_{(02^{2}0)}+\Lag_{(01^{3}0)}\,, \\
&L_{3}^{(3)}\cdot\eta=\Lag_{(01^{3}0)}\,, \\
&L_{4}^{(3)}\cdot\eta=\Lag_{(0\J{2}0)}-2X\Lag_{(02^{2}0)}-\Lag_{(02^{2}1)}+\Lag_{(0\dot{1}0)}-\frac{2}{3}X\Lag_{(030)}-\frac{1}{3}\Lag_{(031)}\,, \\
&L_{5}^{(3)}\cdot\eta=\Lag_{(0\J{2}0)}+\Lag_{(0\dot{1}0)}\,, \\
&L_{6}^{(3)}\cdot\eta=-\Lag_{(0\J{2}0)}-2X\Lag_{(02^{2}0)}-\Lag_{(02^{2}1)}+\Lag_{(0\dot{1}0)}\,, \\
&L_{7}^{(3)}\cdot\eta=\Lag_{(0\dot{1}0)}\,, \\
&L_{8}^{(3)}\cdot\eta=\Lag_{(0\I{1}0)}\,, \\
&L_{9}^{(3)}\cdot\eta=-2X\Lag_{(0\J{2}0)}-\Lag_{(0\J{2}1)}+\Lag_{(0\I{1}0)}\,, \\
&L_{10}^{(3)}\cdot\eta=-2X\Lag_{(0\I{1}0)}-\Lag_{(0\I{1}1)}\,,
\end{align}
where, again, the component form of the Lagrangians can be found in Sec. \ref{subapp:Lag}. One should notice that only six of these Lagrangians are necessary for constructing the disformal Horndeski theory. The other four, $\Lag_{(02^{2}0)}$, $\Lag_{(02^{2}1)}$, $\Lag_{(0\dot{1}0)}$ and $\Lag_{(01^{3}0)}$, are only needed to have a complete basis up to cubic order in second derivatives. The appearance of these Lagrangians will be the smoking gun for a theory that cannot be disformally related to Horndeski theory.

Interestingly, with this mapping between the different notations, we are now able to determine how the EST constraints look in our approach. Starting with the quadratic case \cite{Crisostomi:2016czh}, it was shown in Ref. \cite{deRham:2016wji} that only two sub-classes propagate three healthy degrees of freedom. The first one, denoted \textsuperscript{2}N-I, leads to the following theory
\begin{equation}
\begin{split}
\Lag_{EST}^{(2)\mathrm{N-I}}&=\Lag_{4}^{H}[f_{2}]-\frac{1}{2}(f_{2}+2Xa_{1})g_{2}(2\Lag_{(020)}-3g_{2}\Lag_{(0\J{1}0)}) \\
&-\frac{2f_{2}a_{3}-a_{1}(a_{1}-3Xa_{3}+f_{2,X})}{2(f_{2}+2Xa_{1})}(\Lag_{(021)}-g_{2}\Lag_{(0\J{1}1)})+\frac{f_{2}a_{3}-a_{1}(a_{1}-Xa_{3}+f_{2,X})}{2(f_{2}+2Xa_{1})}\Lag_{(021)}\,, 
\end{split}
\end{equation}
where the $a_{i}$'s are the coefficients in front of each $L_{i}^{(2)}$ and we have defined $g_{2}\equiv\frac{(a_{1}+Xa_{3}+f_{2,X})}{f_{2}+2Xa_{1}}$. Notably, the EST constraints exhibit a much simpler form in this basis of Lagrangians. Moreover, it is also evident that the above expression shares the same structure of the quadratic terms of a disformal quartic Horndeski theory, cf. (\ref{eq:DisfH4}) and substitute $\vOX$ for $g_{2}$. In fact, there is a one-to-one correspondence between both theories. This shows that this \textsuperscript{2}N-I class of theories just represents a disformal quartic Horndeski theory.

The other quadratic EST theory is the sub-family classified as \textsuperscript{2}N-IIIi. When translated to the differential form formalism, the most important feature is that the constraints enforce the Lagrangian $\Lag_{(01^{2}0)}$ to be present. Such a Lagrangian cannot be generated by a disformal transformation of quartic Horndeski. It cannot be produced either by a disformal transformation of the new Lagrangians generated in a disformal transformation as shown in Sec. \ref{sec:Orbits}. Therefore, this highlights that the \textsuperscript{2}N-IIIi class of theories cannot be related to previously known theories via disformal transformations.

For the cubic case \cite{BenAchour:2016fzp}, there are also two sub-families. The first one, \textsuperscript{3}N-I, yields\footnote{We remind the reader that in our notation $-2X\equiv\nabla_{\mu}\phi\nabla^{\mu}\phi$ and $\Lag_{5}^{H}[G_{5}]=-2(G_{5}G^{ab}\Phi_{ab}-\frac{1}{6}G_{5,X}(\tP^{3}-3\tP\tPP+2\tPPP))$.}
\begin{equation}
\begin{split}
\Lag_{EST}^{(3)\mathrm{N-I}}&=\Lag_{5}^{H}[f_{3}]+\frac{1}{2}Xb_{1}g_{3}(2\Lag_{(030)}-6g_{3}\Lag_{(0\J{2}0)}+3g_{3}^{2}\Lag_{(0\I{1}0)}) \\
&-\frac{1}{6}b_{4}(2\Lag_{(031)}-6g_{3}\Lag_{(0\J{2}1)}+3g_{3}^{2}\Lag_{(0\I{1}1)})\,, 
\end{split}
\end{equation}
where now we define $g_{3}\equiv\frac{(3b_{1}-2Xb_{4}-f_{3,X})}{3Xb_{1}}$. Again, the EST constraints greatly simplifies in this formalism. Clearly, this theory has a form equal to the terms of cubic order in a disformal quintic Horndeski theory, cf. (\ref{eq:DisfH5}) and exchange $\vOX$ for $g_{3}$. As for the quadratic case, there is a bijective mapping between both theories that makes transparent that the class of theories \textsuperscript{3}N-I describes a disformal quintic Horndeski theory.

The other possible cubic theory, \textsuperscript{3}N-II, is characterized for the presence of $\Lag_{(02^{2}0)}$, $\Lag_{(02^{2}1)}$ and $\Lag_{(0\dot{1}0)}$. Importantly, such Lagrangians cannot be derived from a disformal transformation of quintic Horndeski or its new disformally related Lagrangians. Consequently, these \textsuperscript{3}N-II theories represent scalar-tensor theories that cannot be related disformally to Horndeski theory.

%---
%FULL COMPUTATIONS
%---
\section{Full Disformal Building Blocks}
\label{app:Computations}

In this appendix we include the concrete form of the coefficients of the disformal building blocks. They will be functions of the disformal coefficients $\T{C}$ and $\T{D}$ through the $\vO_{i}$ and $\vU_{i}$ parameters. For the case of $\T{\Phi}^{a}$, cf. (\ref{eq:DisfPhi}), the coefficients are
\begin{align}
\lPh&=\gD\vO\,, \label{eq:l0} \\
\lT&=-\frac{1}{2}\pP\gD\vO\vOX-X\gD\vOP\,, \label{eq:l1} \\
\lPs&=-\frac{1}{4}\pP\gD(\vOX\vU+2\vUX)-\gD(\vOP-X\vUP)\,, \label{eq:l2} \\
\lBPh&=\frac{1}{2}\gD\vO(\vOX-\vU-2X\vUX)\,, \label{eq:l3} \\
\lBPs&=\frac{1}{2}\gD(\vOX-2X\vUX)\,, \label{eq:l4}
\end{align}
where the subindex indicates the building block associated to each $\lambda_{i}$. Noticeably, there are also $\pP$ factors in the above definitions. This is a hint pointing out that there will be terms generated beyond the basis $\Lag_{(lmn)}$. These new terms appear through the kinetic dependence of the disformal transformation, i.e. via $\vOX$ and $\vUX$.

Accordingly, the coefficients of the disformal curvature 2-form $\T{\R}^{ab}$, cf. (\ref{eq:DisfCurvature}), read
\begin{align}
&\alpha_{\B{R}}=-\vU\,, \label{eq:a1}\\
&\alpha_{\Phi\Phi}=\vU+\frac{1}{2}X\vU^{2}\,, \label{eq:a2}\\
&\alpha_{\BPh\Phi}=(\frac{1}{2}\vU^{2}-\vUx)\,, \label{eq:a3}\\
&\alpha_{\BPs\Phi}=(\frac{1}{2}\vOX\vU-\vO\vUX)\,, \label{eq:a4}\\
&\alpha_{\Phi\Psi}=(\vUp-\frac{1}{2}\vOP\vU-\frac{1}{2}\pP\vU(\vUX+\frac{1}{2}\vOX\vU))\,, \label{eq:a5}\\
&\alpha_{\Phi\theta}=-\vO(\vOP+\frac{1}{2}\pP\vOX\vU)\,, \label{eq:a6}\\ 
&\alpha_{\BPh\theta}=(\vOPx-\frac{1}{2}\vOP(\vOX+\vU)-\frac{1}{4}\pP\vOX\vU(\vOX+\vU))\,, \label{eq:a7}\\
&\alpha_{\BPs\theta}=(\vOP(X\vUX-\frac{1}{2}\vOX)+\frac{1}{2}\pP\vOX(\vO\vUX-\frac{1}{2}\vOX\vU))\,, \label{eq:a8}\\
&\alpha_{\Psi\theta}=(\frac{1}{2}\vOP^{2}-\vOPp+\frac{1}{2}\pP\vOP(\vOX\vU+\vUX)+\frac{1}{4}\pP^{2}\vOX\vU(\vUX+\frac{1}{2}\vOX\vU)-\frac{1}{2}\pPP\vOX\vUX)\,, \label{eq:a9}\\
&\alpha_{\theta\theta}=(\frac{1}{2}X\vOP^{2}+\frac{1}{2}\pP\vO\vOP\vOX+\frac{1}{8}\pP^{2}(1+\vO)\vOX^{2}\vU-\frac{1}{4}\pPP\vOX^{2})\,, \label{eq:a10}\\ 
&\alpha_{\HPs\theta}=\frac{1}{2}\vOX^{2}\,, \label{eq:a11}\\
&\alpha_{\HPs\Psi}=\frac{1}{2}\vOX\vUX\,, \label{eq:a12}\\
&\alpha_{\vec{\Phi}\theta}=\frac{1}{2}\vOX\vU\,, \label{eq:a13}\\
&\alpha_{\BPs\BPh}=-\frac{1}{2}\vU\vUX\,, \label{eq:a14}\\ 
&\B{\alpha}_{\Phi\theta}=\vOX\,, \label{eq:a15}\\
&\B{\alpha}_{\Phi\Psi}=\vUX\,, \label{eq:a16}\\
&\B{\alpha}_{\pP\theta}=-\frac{1}{2}\vOX\vU\,, \label{eq:a17}
\end{align}
where, again, the subindex indicates the building blocks forming a 2-form associated to each coefficient $\alpha_{i}$. The last three coefficients have an over-bar to indicate that they are different from the rest because they are defined inside an exterior derivative. Noticeably, most of the terms are functions of $\vOX$ and $\vUX$

In addition, we can apply a similar procedure for the extended basis presented in Sec. \ref{subsec:ExtendedBasis}. The additional building blocks that we need to transform are $\J{\Psi}^{a}\equiv\lp\Psi^{11}\rp^{a}$ and $\I{\Psi}^{a}\equiv\pP\J{\Psi}^{a}$. For that purpose, we compute how second derivatives transform alone, without forming a 1-form. We obtain
\begin{equation} \label{eq:Disf2deriv}
\T{\nabla}^{a}\T{\nabla}_{b}\phi=\B{\lambda}_{\Phi}\nabla^{a}\nabla_{b}\phi+\B{\lambda}_{\theta}\delta^{a}_{\  b}+\B{\lambda}_{\Psi}\nabla^{a}\phi\nabla_{b}\phi+\B{\lambda}_{\BPh}\nabla^{a}\phi\nabla_{b}\nabla_{z}\phi\nabla^{z}\phi+\B{\lambda}_{\BPs}\nabla^{a}\nabla_{z}\phi\nabla^{z}\phi\nabla_{b}\phi\,,
\end{equation}
where the coefficients are given by
\begin{align}
\B{\lambda}_{\Phi}&=\lambda_{\Phi}\gC\,, \label{eq:bl0} \\ 
\B{\lambda}_{\theta}&=-\frac{1}{2}\gD\gC\lp2X\vOP+\pP\vO\vOX\rp\,, \label{eq:bl1} \\
\B{\lambda}_{\Psi}&=-\frac{1}{4}\gD\gC\lp4(\vOP-X\vUp)-\pP\vO(\vU(\vU-\vOX)+2(\vO-2)\vUX)\rp\,, \label{eq:bl2} \\
\B{\lambda}_{\BPh}&=\B{\lambda}_{\BPs}=\frac{1}{2}\gC\lambda_{\Phi}(\vOX-\vU-2X\vUX)\,. \label{eq:bl3} 
\end{align}
In order to compute this calculation, we have used that $\T{\nabla}_{a}\phi=\gD\nabla_{a}\phi$, which is a consequence of the fact that a partial derivative does not change but the vielbein does. One should notice also that the above transformation shares the same structure as $\T{\Phi}^{a}$, cf. (\ref{eq:DisfPhi}), but the coefficients are not equal since we are only transforming now the components. With this result, one can easily conclude, as anticipated in Sec. \ref{sec:Orbits}, that
\begin{equation} \label{eq:Disf11}
\J{\Psi}_{\mathrm{disf}}^{a}=\beta_{\HPs}\HPs^{a}+\beta_{\BPs}\BPs^{a}+\beta_{\BPh}\BPh^{a}+\beta_{\Psi}\Psi^{a}
\end{equation}
where $\beta_{\HPs}=\gD^{2}\epsilon_{1}\B{\epsilon}_{1}$, $\beta_{\BPs}=\gD^{2}\epsilon_{1}\B{\epsilon}_{2}$, $\beta_{\BPh}=\gD^{2}\epsilon_{2}\B{\epsilon}_{1}$, $\beta_{\Psi}=\gD^{2}\epsilon_{2}\B{\epsilon}_{2}$ and
\begin{align}
\epsilon_{1}&=\B{\lambda}_{\Phi}-2X\B{\lambda}_{\BPh}=\gD^{2}\gX^{-1}\,, \label{eq:e1} \\ 
\B{\epsilon}_{1}&={\lambda}_{\Phi}-2X{\lambda}_{\BPh}=\T{C}\epsilon_{1}\,, \label{eq:ne1} \\
\epsilon_{2}&=\B{\lambda}_{\theta}-2X\B{\lambda}_{\Psi}+\pP\B{\lambda}_{\BPs}=\gD\B{\epsilon}_{2}-\frac{1}{2}X\gC\gX^{-1}\lambda_{\Phi}\vU\pP  \,, \label{eq:e2} \\
\B{\epsilon}_{2}&={\lambda}_{\theta}-2X{\lambda}_{\Psi}+\pP{\lambda}_{\BPs}=X\gC(\vOP-2X\vUp)\,. \label{eq:ne2} 
\end{align}
Accordingly, only $\epsilon_{2}$ is a function of $\pP$ apart from $\phi$ and $X$.

Furthermore, we can easily compute the disformal transformation of the other building block $\I{\Psi}^{a}$. We just need to contract the disformal second derivatives (\ref{eq:Disf2deriv}) with disformal gradients $\T{\nabla}^{a}\phi$ to obtain
\begin{equation} \label{eq:DisfpP}
\langle\T{\Phi}\rangle=\gD^{2}(\epsilon_{1}\pP-2X\epsilon_{2})=\chi_{1}+\chi_{2}\pP\,,
\end{equation}
where $\chi_{1}=-2X^{2}\gD^{3}\gC(\vOP-2X\vUp)$ and $\chi_{2}=\gD^{3}\gX^{-1}\vO$. Subsequently, we have to combine this result with (\ref{eq:Disf11}) to obtain $\I{\Psi}^{a}_{\mathrm{disf}}$, which will be again a linear combination of extended building blocks. In this linear combinations, some of the coefficients will be functions of $\pP$. One can note that terms of the original basis, i.e. $\Psi^{a}$ and $\Phi^{a}$, are only sourced by the dependence in $\phi$ via $\vOP$ and $\vUP$ in both $\HPs^{a}$ and $\IPs^{a}$. This is a sign that the Lagrangians of the extended basis will be invariant under kinetic disformal transformations.

Finally, let us emphasis that when computing a disformal transformation of a given Lagrangian with one of these extended building blocks, one can use the antisymmetric identities presented in Appendix \ref{subapp:AntisymId} to eliminate any dependence of the coefficients in second derivative scalars such as $\pP$. This allows to present the transformed Lagrangian as a linear combination of Lagrangians of the extended basis with coefficients depending only on $\phi$ and $X$.

%---
%SUMMARY DISFORMAL
%---
\section{Complete Disformal Horndeski Theory}
\label{app:DisformalHorndeski}

In this appendix we summarize the general disformal transformation of the full Horndeski action. For that task, we need to rewrite every Lagrangian in terms of Horndeski ones, when possible. This includes using the identities derived in Ref. \cite{Ezquiaga:2016nqo} to express the $\Lag_{i}^{NH}$, which are given by
\begin{align}
\Lag_{2}^{NH}[E_{2}]=&E_{2}\Lag_{(001)}\,, \label{eq:LNH2} \\
\Lag_{3}^{NH}[E_{3}]=&E_{3}\Lag_{(011)}\,, \label{eq:LNH3} \\
\Lag_{4}^{NH}[E_{4}]=&E_{4}\Lag_{(101)}+E_{4,X}\Lag_{(021)}\,, \label{eq:LNH4} \\
\Lag_{5}^{NH}[E_{5}]=&E_{5}\Lag_{(111)}+\frac{1}{3}E_{5,X}\Lag_{(031)}\,, \label{eq:LNH5} \\
\Lag_{6}^{NH}[E_{6}]=&E_{6}\Lag_{(200)}+2E_{6,X}\Lag_{(120)}+\frac{1}{3}E_{6,XX}\Lag_{(040)}\,, \label{eq:LNH6} 
\end{align}
terms of $\Lag_{i}^{H}$. Once this manipulation is performed, such a theory will be compound of the following Lagrangians
\begin{equation}
\begin{split}
\T{\Lag}^{H}&=\sum_{i=2}^{5}\Lag_{i}^{H}[\B{G}_{i}]+F_{4}\Lag_{(021)}+H_{4}(2\Lag_{(020)}-3\vOX\Lag_{(0\J{1}0)})+I_{4}(\Lag_{(021)}-\vOX\Lag_{(0\J{1}1)}) \\
&+H_{5}(2\Lag_{(030)}-6\vOX\Lag_{(0\J{2}0)}+3\vOX^{2}\Lag_{(0\I{1}0)})+I_{5}(2\Lag_{(031)}-6\vOX\Lag_{(0\J{2}1)}+3\vOX^{2}\Lag_{(0\I{1}1)})
\end{split}
\end{equation}
where the coefficients of the Horndeski Lagrangians are given by
\begin{align}
\B{G}_{2}=&\T{G}_{2}\T{C}^{3}\gD^{-1}-2X\T{G}_{3}\T{C}^{2}(\vOP+X\vUp)+3X\T{G}_{4}\T{C}(\gD^{-1}\vOP^{2}+2\T{C}\vOPp)+12X^{3}\T{G}_{4,X}\gD^{-2}\gX\vOP\vUp \\
+&6X^{2}\T{G}_{5,\phi}\vO\vOP^{2}+4X^{3}\T{G}_{5,X}\gX\vO^{-1}\vOP^{2}(1-3\vOp)+2X\int\T{E}_{3,\phi}dX\,, \nonumber \\
\B{G}_{3}=&\T{G}_{3}\T{C}^{2}\vO-\T{G}_{4}\gD^{-2}(\vOP(\T{C}(\vO^{2}+\vO+1)-X\vU)-2X\vUp)+4X\T{G}_{4,\phi}\T{C}\vU(\T{C}-X\T{D}) \\
-&{2X\T{G}_{4,X}\gD^{-1}\gX\T{C}(\vOP+2X\vUp)}+4X(\T{G}_{5}\vO_{,\phi\phi}-\T{G}_{5,\phi}(\vO^{2}\vOP-\vOp)-\int(\T{G}_{5}\vOx)_{\phi\phi}dX) \nonumber \\
-&2X^{2}\T{G}_{5,X}\gX\vOP(\vOP-4\vOp)-\int\T{E}_{3}dX\,, \nonumber \\
\B{G}_{4}=&\T{G}_{4}\T{C}\gD^{-1}-2X\T{G}_{5}\vOp+2X\int(\T{G}_{5,\phi}\vOx+\T{G}_{5}\vO_{,X\phi})dX\,, \\
\B{G}_{5}=&\T{G}_{5}\vO-\int\T{G}_{5}\vOx dX
\end{align}
and the rest by
\begin{align}
F_{4}&=\frac{1}{4}\T{C}(\T{C}\T{G}_{4}(\vOX\vU+2\vUX)-2\T{G}_{4,X}(2\T{D}(2\vO-1)+{\T{C}\gX(2X\vOX\vU+(1-2\vO)\vU+(3-2\vO)2X\vUX)}) \\
&-\frac{1}{2}X(\vOX\vU+2\vUX)(\vO^{2}\T{G}_{5,\phi}-{2X^{2}\gX\vUp\T{G}_{5,X}})\,, \nonumber  \\
H_{4}&=-\frac{1}{2}\T{C}^{2}\vOX(\T{G}_{4}\vO-{2X\gX\T{G}_{4,X}})+X\vO\vOX(\vO^{2}\T{G}_{5,\phi}-{2X^{2}\gX\vUp\T{G}_{5,X}})\,, \\
I_{4}&=\frac{1}{2}\T{C}(\T{G}_{4}(X\T{D}\vOX\vU-2\T{C}(\vOX\vU+\vUX))+\T{G}_{4,X}(4X\T{D}\vU+\T{C}\gX((2\vO-1)\vOX-2X\vU^{2}-4X(\vO-2)\vUX)) \\ 
&+\frac{1}{2}(\vO((1+\vO(3\vO-2))\vOX+4X\vO\vUX)\T{G}_{5,\phi}-{2X^{2}\gX((3\vO-2)\vOX+4X\vUX)\vUp\T{G}_{5,X}})\,, \nonumber \\
H_{5}&=\frac{1}{6}{X\vO^{2}\gX\vOX\T{G}_{5,X}}\,, \\
I_{5}&=\frac{1}{12}X{\vO\gX(\vOX\vU+2\vUX)\T{G}_{5,X}}\,. 
\end{align}
All these coefficients are functions of $\vO_{i}$, $\vU_{i}$ and $\gamma_{i}$, which depend only in the disformal coefficients $\T{C}$ and $\T{D}$, cf. (\ref{eq:x1}-\ref{eq:x4}) and below the equations. The coefficient $\T{E}_{3}$ in the integrals is given by
\begin{align}
\T{E}_{3}=&\frac{1}{2}\T{G}_{3}\T{C}^{2}\vO(3\vOX+(\vO+1)\vU-\vO(\vOX-2X\vUX))-\T{G}_{4}\gD^{-2}(\vOP\vU(1+\vO)+3\vO^{2}\vOPx+\vUp) \nonumber  \\
+&\T{G}_{4,\phi}\T{C}(3\T{C}\vO\vOX-2(\T{C}-X\T{D})\vU)-\T{G}_{4,X}\T{C}(3\T{C}\vOP-\gD^{-1}\gX(\vOP+2X\vUp)(1-3X\vOX))  \\
-&2(\T{G}_{5}\vO_{,\phi\phi}-\int(\T{G}_{5}\vOx)_{\phi\phi}dX-\T{G}_{5,\phi}(\vO^{2}\vOP(1-3X\vOX)-\vOp))+X\T{G}_{5,X}\gX(\vOP-4\vOp)(1-3X\vOX)  \,. \nonumber
\end{align}
Notice that in order to compare this result with the literature it is necessary to include the proper notation. In particular, we are using $-2X\equiv\nabla_{\mu}\phi\nabla^{\mu}\phi$ and $\Lag_{5}^{H}=-2(G_{5}G^{ab}\Phi_{ab}-\frac{1}{6}G_{5,X}(\tP^{3}-3\tP\tPP+2\tPPP))$. Moreover, in order to map with the disformal metric transformation, $\T{g}_{\mu\nu}=Cg_{\mu\nu}+D\phi_{,\mu}\phi_{\nu}$, one just need to define
\begin{align}
\T{C}&=\sqrt{C} \,, \label{eq:cJ}\\
\T{D}&=\frac{\sqrt{C}-\sqrt{C-2XD}}{2X} \,. \label{eq:dJ}
\end{align}
Remarkably, with this full result, one can obtain the transformed Horndeski theory of any particular disformal transformation. One only needs to particularize the value of the coefficients $\vO_{i}$ and $\vU_{i}$, cf. Table \ref{tab:xi} for a list of all possibilities.

%---
%INVERSE VIELBEIN
%---
\section{Inverse Vielbein}
\label{app:InverseVeilbein}

A necessary condition for a field redefinition to preserve the number of DoF is to be invertible. In the context of this work, this implies that an inverse vielbein $e_{a}^{\  \mu}$ must exist. Such inverse can be computed either from
\begin{align}
& e^{a}_{\  \mu}e_{a}^{\  \nu}=\delta^{\mu}_{\nu} & &\mathrm{or} & &e^{a}_{\  \mu}e_{b}^{\  \mu}=\delta^{a}_{b}\,. \nonumber
\end{align} 
For instance, let us start by obtaining the inverse vielbein of a disformal transformation. The disformal vielbein can be directly read from the definition of the transformation
\begin{equation}
\T{\theta}^{a}=\T{e}^{a}_{\  \mu}dx^{\mu}=\T{C}\theta^{a}+\T{D}\Psi^{a}=(\T{C}e^{a}_{\mu}+\T{D}\nabla^{a}\phi\nabla_{\mu}\phi)dx^{\mu}.
\end{equation}
Then, postulating that the inverse vielbein has a similar structure $\T{e}_{a}^{\  \mu}=\T{A} e_{a}^{\  \mu}+\T{B} \nabla_{a}\phi\nabla^{\mu}\phi$, one has to solve the equation
\begin{equation}
\T{e}^{a}_{\  \mu}\T{e}_{a}^{\  \nu}=(\T{C}e^{a}_{\  \mu}+\T{D}\nabla^{a}\phi\nabla_{\mu}\phi)(\T{A} e_{a}^{\  \nu}+\T{B} \nabla_{a}\phi\nabla^{\nu}\phi)=\T{C}\T{A}\delta^{\nu}_{\mu}+(\T{C}\T{B}+\T{D}\T{A}-2X\T{D}\T{B})\nabla^{\nu}\phi\nabla_{\mu}\phi=\delta^{\nu}_{\mu}\,,
\end{equation}
where in the second equality we have used the properties of the original vielbein $e^{a}_{\  \mu}$. It is straightforward to conclude that the equation can be solved and that the coefficients of the disformal, inverse vielbein are
\begin{align}
& \T{A}=\T{C}^{-1}=\gC & &\mathrm{and} & &\T{B }=-\T{C}^{-1}(\T{C}-2X\T{D})^{-1}\T{D}=-\gC\gD\T{D}\,. \nonumber
\end{align} 
%--
%EXTENDED DISFORMAL
%--
\subsection{Extended Disformal Transformations}
\label{app:InverseExtended}
%--
One could applied the same logic to the extended disformal transformation $\T{\theta}^{a}=\T{C}\theta^{a}+\T{D}\Psi^{a}+\T{E}\Phi^{a}$ defined in (\ref{eq:EPhivierbein}). Considering only the conformal and extended disformal parts, one could postulate the inverse vielbein to be $\T{e}_{a}^{\  \mu}=\T{A}e_{a}^{\  \mu}+\T{B}\nabla_{a}\nabla^{\mu}\phi$, which should be a solution of
\begin{equation} \label{eq:ex1}
\T{e}^{a}_{\  \mu}\T{e}_{a}^{\  \nu}=(\T{C}e^{a}_{\  \mu}+\T{E}\nabla^{a}\nabla_{\mu}\phi)(\T{A} e_{a}^{\  \nu}+\T{B} \nabla_{a}\nabla^{\nu}\phi)=\T{C}\T{A}\delta^{\nu}_{\mu}+(\T{C}\T{B}+\T{E}\T{A})\nabla^{\nu}\nabla_{\mu}\phi+\T{E}\T{B}\nabla^{\nu}\nabla_{a}\phi\nabla^{a}\nabla_{\mu}\phi=\delta^{\nu}_{\mu}\,.
\end{equation}
However, this equation only admits the trivial solution $\T{A}=\T{B}=0$ since the last term is quadratic in second derivatives and no other term can compensate it. This will be a generic feature of any transformation containing $\Phi^{a}$ or its higher order powers $\lp\Phi^{n}\rp^{a}$. Consequently, the inverse will necessary be an infinite series of powers of second derivatives, so that every term can be compensated order by order. 

On the other hand, if we consider the other type of extended disformal transformation defined in Sec. \ref{subsec:ExtendedDisformal}, e.g. $\T{\theta}^{a}=\T{C}\theta^{a}+\T{E}\BPh^{a}$ and postulates $\T{e}_{a}^{\  \mu}=\T{A}e_{a}^{\  \mu}+\T{B}\nabla_{a}\phi\nabla_{z}\phi\nabla^{z}\nabla^{\mu}\phi$, one encounters that
\begin{equation}
\T{e}^{a}_{\  \mu}\T{e}_{a}^{\  \nu}=(\T{C}e^{a}_{\  \mu}+\T{E}\phi^{a}\phi_{z}\nabla_{z}\nabla_{\mu}\phi)(\T{A} e_{a}^{\  \nu}+\T{B}\phi_{a}\phi_{z}\nabla^{z}\nabla^{\nu}\phi)=\T{C}\T{A}\delta^{\nu}_{\mu}+(\T{C}\T{B}+\T{E}\T{A}+\pP\T{E}\T{B})\phi^{\nu}\phi^{z}\nabla_{z}\nabla_{\mu}\phi=\delta^{\nu}_{\mu}\,
\end{equation}
can be solved. The only difference with respect to the disformal case is that now the coefficients of the inverse vielbein will depend in second derivatives of the scalar via $\pP$.
%--
%SPIN-2
%--
\subsection{Vielbeins with several Spin-2 fields}
\label{app:InverseS2}
Another interesting situation is when different vielbeins are combined. Let us work with $\T{\theta}^{a}=\T{C}\theta^{a}+\T{F}\Theta^{a}$ such as in (\ref{eq:DisfTensor}), where $\Theta^{a}=E^{a}_{\  \mu}dx^{\mu}$. The first guess would be to define the inverse vielbein as $\T{e}_{a}^{\  \mu}=\T{A}e_{a}^{\  \mu}+\T{B}E_{a}^{\  \mu}$. Accordingly, we will have to solve
\begin{equation}
\T{e}^{a}_{\  \mu}\T{e}_{a}^{\  \nu}=(\T{C}e^{a}_{\  \mu}+\T{F}E^{a}_{\  \mu})(\T{A} e_{a}^{\  \nu}+\T{B}E_{a}^{\  \nu})=(\T{C}\T{A}+\T{F}\T{B})\delta^{\nu}_{\mu}+\T{C}\T{B}e^{a}_{\  \mu}E_{a}^{\  \nu}+\T{F}\T{A}E^{a}_{\  \mu}e_{a}^{\  \nu}=\delta^{\nu}_{\mu}\,,
\end{equation}
where we have used that $E^{a}_{\  \mu}E_{a}^{\  \nu}=\delta^{\mu}_{\nu}$. Unfortunately, we arrive at a similar situation to the one of (\ref{eq:ex1}). Thus, in general, the inverse will not have a simple form. The exception is if we impose that $e^{a}_{\  \mu}E_{a}^{\  \nu}-E^{a}_{\  \mu}e_{a}^{\  \nu}=0$. Then, by setting $\T{C}\T{B}=-\T{F}\T{A}$, one can find the following solution
\begin{align}
& \T{A}=(\T{C}^{2}-\T{F}^{2})^{-1}\T{C} & &\mathrm{and} & &\T{B }=-(\T{C}^{2}-\T{F}^{2})^{-1}\T{F} \,. \nonumber
\end{align} 
Interestingly, the condition imposed is nothing but the symmetric vielbein condition \cite{Hinterbichler:2012cn}, which is a dynamical constraint that arises in multi-gravity theories with Einstein-Hilbert kinetic terms. It can be use, for instance, to show the equivalence of the metric and vielbein formalism of bi-gravity. 
%---
%END APPENDICES
%---
%--- 
%BIBLIOGRAPHY
\bibliographystyle{h-physrev}
\bibliography{FieldRedefDiffFormsBib}
%---

%-----
%END
%----- 
\end{document}